\documentclass[preprint,12pt]{elsarticle}

\usepackage{amssymb}
\usepackage{hyperref}
\hypersetup{colorlinks=true,linkcolor=blue,urlcolor=blue}
\usepackage{amsmath,mathtools}
\usepackage{multicol}
\usepackage{tabularx}
\usepackage{rotating}
\usepackage{url}
\usepackage{float}
\usepackage{xcolor}
\usepackage{multirow}
\usepackage[ruled,vlined]{algorithm2e}
\usepackage{bbm}
\newcommand\given[1][]{\:#1\vert\:}
\usepackage{etoolbox}
\patchcmd{\pprintMaketitle}%
{\vskip36pt} 
{\vskip18pt} 
{}
{}

\usepackage[linguistics]{forest, adjustbox}
\usepackage{dsfont}
\usepackage{bm}
\usepackage{graphicx}
\usepackage{subcaption}
\usepackage{mwe}
\usepackage{adjustbox}

\journal{Computational Statistics and Data Analysis}

\begin{document}

\begin{frontmatter}

\title{GP-BART: a novel Bayesian additive regression trees approach using Gaussian processes}


\author[inst1]{Mateus Maia\texorpdfstring{\corref{cor1}}{}}

\author[inst1]{Keefe Murphy}
    
\author[inst1,inst2]{Andrew C. Parnell}

\affiliation[inst1]{organization={Hamilton Institute and Department of Mathematics and Statistics},
            addressline={\\Maynooth University}, 
            city={Maynooth},
            country={Ireland}}
            
\affiliation[inst2]{organization={Insight Centre for Data Analytics},
            addressline={Maynooth University}, 
            city={Maynooth},
            country={Ireland}}

\cortext[cor1]{\emph{Email address:} \texttt{mateus.maiamarques.2021@mumail.ie} (Mateus Maia)}

\begin{abstract}
The Bayesian additive regression trees (BART) model is an ensemble method extensively and successfully used in regression tasks due to its consistently strong predictive performance and its ability to quantify uncertainty. BART combines ``weak'' tree models through a set of shrinkage priors, whereby each tree explains a small portion of the variability in the data. However, the lack of smoothness and the absence of an explicit covariance structure over the observations in standard BART can yield poor performance in cases where such assumptions would be necessary. The Gaussian processes Bayesian additive regression trees (GP-BART) model is an extension of BART which addresses this limitation by assuming Gaussian process (GP) priors for the predictions of each terminal node among all trees. The model's effectiveness is demonstrated through applications to simulated and real-world data, surpassing the performance of traditional modeling approaches in various scenarios.
\end{abstract}




\end{frontmatter}


\section{Introduction}
\enlargethispage{0.25\baselineskip}
\label{sec:sample1}
Bayesian additive regression trees \citep[BART;][]{chipman2010bart} is a probabilistic machine learning model that has proved successful in both regression and classification settings \citep{zhao2018bayesian,zhang2020application,janizadeh2021novel}. Effectively, BART is a non-parametric Bayesian regression approach which learns through sums of trees \citep{chipman1998bayesian}, where each terminal node contribution is constrained by a regularising prior distribution. Given a vector of predictors $\mathbf{x}_{i}=\left(x_{i1},\ldots,x_{ip}\right)$, the target function $f(\mathbf{x}_i)$ is obtained by aggregating the small contributions of each tree, which is similar in flavour to the small step updates of gradient boosting algorithms \citep{friedman2001greedy}. 

Considering a univariate response and training observations denoted as $\{\mathbf{x}_{i},y_{i}\}_{i=1}^n$, the standard BART model is given by
\[y_{i} \given \mathbf{x}_i \sim \mbox{N} \left( \sum_{t=1}^{T}h\left(\mathbf{x}_{i};\mathcal{T}_{t},\mathbf{L}_{t}\right),\tau^{-1}\right),\]
\noindent where the function $h$ assigns a sampled value $\mu_{t \ell}$ to $\mathbf{x}_{i}$ within terminal node $\ell$ of the tree $\mathcal{T}_t$ across all $T$ trees and the vector $\mathbf{L}_{t} = \left(\mu_{t1},\ldots, \mu_{tb_t} \right)$ collects~the sampled mean parameters from the $b_{t}$ terminal nodes in tree $\mathcal{T}_t$. Here, $\mbox{N}(\cdot)$ denotes the normal distribution and $\tau$ is a residual precision term. In standard BART, terminal node parameters $\mu_{t \ell}$ are assigned a $\mbox{N}(\mu_{\mu}, \tau_{\mu}^{-1})$ prior, where the hyperparameters are selected to shrink the influence of each tree.

Our novel GP-BART method modifies the standard BART by using the function $g$ (replacing $h$) which assigns a vector of sampled values $\boldsymbol{\psi}_{t\ell}$ to the~$n_{t\ell}$ observations in node $\ell$ of tree $\mathcal{T}_t$, rather than the single value $\mu_{t\ell}$ used by BART. This is achieved by assuming a Gaussian process (GP) prior over each terminal node with constant mean $\mu_{t \ell}$ and a covariance function whose parameters are defined at the tree level.

In recent years, several extensions and modifications to the original BART model have been proposed to cover different types of data and assumptions \citep{hill2020bayesian}. To deal with the lack of smoothness, \citet{linero2018bayesian} presented a soft version of the BART model by advocating probabilistic split rules at the tree-building stage. \citet{starling2020bart} presented a BART extension, also incorporating GPs, which guarantees smoothness over a single target covariate by applying Gaussian process priors for each terminal node over the targeted variable. \citet{prado2021bayesian} proposed model trees BART that considers piecewise sums of linear functions at the terminal node level instead of piecewise sums of constants, adding flexibility. Our GP-BART considers GP models at the terminal node level, and can be seen as a piecewise sum of GPs which are inherently smooth.

Notably, our GP-BART approach is coherent with previous work of \citet{linero2017review}, who identified that the BART model is itself a GP, conditional on the tree structures, with a non-parametrically learned covariance matrix whereby each element is described by the proportion of times the two corresponding design points are allocated to the same terminal nodes across all trees. \citet{linero2017review} further showed that as $T\rightarrow \infty$, BART becomes a GP unconditionally. Therefore, it is natural to assume GP priors over the terminal nodes directly to circumvent the need for large $T$. More specifically, \citet{linero2017review} also shows that the implied kernel under this relation between BART and GPs is a function of the $\mathcal{L}_1$ distances between design points (similar results were also found in \cite{balog2016mondrian}). Following this, it is natural to allow kernels of other types, especially ones defined by different distance metrics. Here, we employ node-specific anisotropic exponentiated-quadratic kernels relying on squared Euclidean distances. Though these are parameterised, this enables covariance structures, more flexible than the one implied by the standard BART, to be learned non-parametrically when $T>1$, which would be too difficult to pre-specify under a single GP, or even a sum of GPs without tree splits.

The treed Gaussian process \citep[tGP;][]{gramacy2008bayesian} is another treed approach to GPs which defines all hyperparameters of a single GP at the terminal node level, thereby making it possible to incorporate non-stationarity into the model by varying the residual precision parameter across terminal nodes. However, to deal with the changing dimensions of the parameter space associated with growing and pruning a tree, this model requires the use of a reversible jump \cite{green1995reversible} algorithm, which comes with increased computational costs. Our GP-BART can also be seen as an additive ensemble of these treed GPs; though we define our priors and associated hyperparameters differently, the additive nature of the sum of GPs is shown here to yield superior performance. Finally, another example of previous work combining BART and GPs is provided by \citet{wang2022local}, who use node-level GPs differently, as an extrapolation strategy for improving BART's predictions for exterior points outside the range of the training data. The authors describe their approach as a `GPed tree', in contrast to the `treed GP' of \citet{gramacy2008bayesian}, and by extension GP-BART's ensemble of treed GPs.

We envisage our novel GP-BART framework being particularly suited for spatial data where smoothness in space is expected for certain covariate combinations, and thus useful in situations where GPs are commonly used \citep[e.g.,][]{banerjee2008gaussian,gelfand2016spatial,andugula2017gaussian,xie2018integrated}. As well as GPs, we introduce a further novelty to~\mbox{allow} for rotated splits. Traditional tree-based models can be interpreted as hyper-rectangles since each node is given in parallel-axis directions. This behavior leads to a staircase decision boundary which can inhibit the model's ability to approximate true boundaries. \citet{garcia2007nonlinear} propose non-linear projections of the tree models used in ensemble approaches to overcome this limitation, while \citet{menze2011oblique} describe an oblique forest model which selects optimal oblique directions using linear discriminant analysis. More recently, \citet{blaser2016random} proposed random rotation ensembles where the direction of rotation is selected randomly, yielding a more general decision boundary. In the GP-BART framework, the incorporation of random projections on various directions allows for splitting rules that are not limited to exclusively parallel axes. This flexibility enables the tree search algorithm to explore a broader sample space of the tree distribution, aiming to mitigate the issue of poor mixing \citep{wu2007bayesian}. The rotation moves can also be interpreted as another way to represent and model complex interactions among variables and should not be seen as strictly restricted to spatial features.

The remainder of this paper is structured as follows. Section \ref{sec:gp_bart} describes the GP-BART model, with mathematical formulations and key specifications. Section \ref{sec:gp_bart_sample} contains the sampling algorithm and describes prediction settings and uncertainty estimation. Sections \ref{sec:simulations} and \ref{sec:real_data} provide comparisons between GP-BART and other methods in simulated and real-data benchmarking scenarios, respectively. Finally, Section \ref{sec:discussion} presents conclusions regarding the proposed algorithm, some limitations, and potential future work.  We note that an implementation of our method is available in the \textsf{R} package \texttt{gpbart}, which is written in C\texttt{++} and available at:     \url{https://github.com/MateusMaiaDS/gpbart}, with which all results were obtained.


\section{Gaussian processes Bayesian additive regression trees}\label{sec:gp_bart}

For simplicity, we begin with the notation for a single tree model. Let $\mathcal{T}_1$ be a binary splitting tree with $b_1$ terminal nodes and let $\mathbf{G}_1\!=\! (\{\mu_{11},\boldsymbol{\phi}_{1},\nu\},\ldots,\linebreak{}\{\mu_{1b},\boldsymbol{\phi}_{1},\nu\})$ denote the sets of parameters associated with each terminal node's GP. Each GP, denoted by $\mathcal{GP}_{1\ell}(\boldsymbol{\mu}_{1\ell},\boldsymbol{\Omega}_{1 \ell}(\boldsymbol{\phi}_1, \nu))$, is characterised by a constant mean vector $\boldsymbol{\mu}_{1\ell}\!=\!(\mu_{1\ell},\ldots,\mu_{1\ell})$ and a covariance function $\boldsymbol{\Omega}_{1 \ell}(\boldsymbol{\phi}_1, \nu),\linebreak{}\:\forall\:\ell = 1,\ldots,b_1$, where $\boldsymbol{\phi}_1 = \{\phi_{11}, \dots, \phi_{1p^{{\star}}} \} \in \mathbb{R}^{p^{{\star}}}$  and $\nu$ are, respectively, the vector of length parameters and precision parameters of the chosen stationary kernel. Notably, this parameterisation allows for variable-specific length parameters $\phi_{1j}\:\forall\:j=1,\ldots,p^{{\star}}$, {where $p^\star\le p$ is the number of continuous predictors,} under which the kernel is still stationary but no longer isotropic.\enlargethispage{0.5\baselineskip}

In the standard BART, since the trees follow a binary structure, each new node is determined by split rules of the form $\smash{\{\mathbf{x}^{(j)} \leq c_{\mathbf{x}^{(j)}} \}} $ vs. $\{ \mathbf{x}^{(j)} > c_{\mathbf{x}^{(j)}} \}$ for continuous predictors, where $c_{\mathbf{x}^{(j)}}$ is a scalar uniformly sampled from the range of a specific covariate $\mathbf{x}^{(j)}$ in the matrix $\mathbf{X}$ of training set predictors. {Dummy variables are typically used to represent categorical predictors, which yields rules of the form} $\smash{\{ \mathbf{x}^{(j)} \in d_{\mathbf{x}^{(j)}} \}}$ vs. $\{ \mathbf{x}^{(j)} \notin d_{\mathbf{x}^{(j)}}\}$, where $d_{\mathbf{x}^{(j)}}$ denotes one of the variable's possible outcome levels.

For a single tree $\mathcal{T}_1$ with $b_1$ terminal nodes, the model is written as $y_{i}\given \mathbf{x}_i \sim \mbox{N}(g(\mathbf{x}_{i};\mathcal{T}_{1},\mathbf{G}_{1}),\tau^{-1})$, 
where the function $g$ assigns the predicted values $\boldsymbol{\psi}_{1\ell}$ from $\mathcal{GP}_{1 \ell}$ to the observations belonging to terminal node $\ell$. The description of the tree structure for GP-BART, which generalises the above to allow for rotated splitting rules, is deferred to Section \ref{sec:tree_structure}.

Expanding such a model into a sum-of-trees structure is achieved via
\[y_{i}\given \mathbf{x}_i \sim \mbox{N}\left(\sum_{t=1}^{T}g\left(\mathbf{x}_{i};\mathcal{T}_{t},\mathbf{G}_{t}\right), \tau^{-1}\right),
\]
where the parameters $\mathbf{G}_{t} = (\{\mu_{t1},\boldsymbol{\phi}_t,\nu\},\ldots,\{\mu_{tb_t},\boldsymbol{\phi}_t,\nu\})$ now characterise the terminal node GPs of each tree $\mathcal{T}_t$, now denoted by $\mathcal{GP}_{t \ell}(\boldsymbol{\mu}_{t \ell},\boldsymbol{\Omega}_{t \ell}(\boldsymbol{\phi}_t,\nu)),\linebreak{}\forall\:\ell=1,\ldots,b_t$, where $\boldsymbol{\mu}_{t \ell}=(\mu_{t \ell},\ldots,\mu_{t \ell})$ is again a constant vector, $\boldsymbol{\phi}_t = \left\{\phi_{t1}, \dots, \phi_{tp} \right\} \in \mathbb{R}^{p}$ is now specific to each tree, in addition to each variable, and $g$ now assigns the predicted values $\boldsymbol{\psi}_{t\ell}$ from $\mathcal{GP}_{t\ell}$. The GP-BART model can be interpreted as a piecewise sum of non-linear GPs whereby each of the $T$ trees will make a small contribution to the overall $\mathbb{E}[y_i\given \mathbf{x}_i]$, whereas BART can be interpreted as a less flexible piecewise sum of constants. Consequently, GP-BART typically requires fewer trees than the standard BART model.


As in standard BART, we require prior distributions for the tree structure and terminal node parameters; i.e., $(\mathcal{T}_{1},\mathbf{G}_{1}),\ldots, (\mathcal{T}_{t},\mathbf{G}_{t})$. We assume $\nu$ is fixed and select the following shrinkage priors assuming independence between trees and terminal nodes:
\begin{align}
    \begin{split}
    \pi\left(\left(\mathcal{T}_{1},\mathbf{G}_{1}\right),\ldots, \left(\mathcal{T}_{t},\mathbf{G}_{t}\right), \tau \right) &= \pi\left(\tau\right)\prod_{t=1}^{T} \pi \left(\mathcal{T}_{t},\mathbf{G}_{t} \right)\label{eq:full_join_model}\\
    &= \pi\left(\tau\right)\prod_{t=1}^{T} \pi \left( \mathbf{G}_{t}\given \mathcal{T}_{t} \right)  \pi\left(\mathcal{T}_{t}\right),
    \end{split}
\shortintertext{where}
    \mathclap{\pi\left(\mathbf{G}_{t}\given \mathcal{T}_{t}\right) = \pi \left(\boldsymbol{\phi}_{t}\right) \prod_{\ell=1}^{b_{t}} \pi\left(\boldsymbol{\psi}_{t \ell} \given \mu_{t \ell},\mathcal{T}_{t},\boldsymbol{\phi_{t}},\nu\right)\pi\left(\mu_{t \ell} \given \mathcal{T}_{t}\right).}    \label{eq:terminal_node_posterior}
\end{align}
We follow \cite{chipman2010bart} in our selection of priors for $\mathcal{T}_t$ and $\tau$ and adopt data-driven priors for the node-level $\boldsymbol{\mu}_{t\ell}$ in such a way that considerable probability is assigned around the range of the observed $\mathbf{y}$ given the induced prior from the sum of GPs. Associated hyperparameters are omitted from Equations \eqref{eq:full_join_model} and \eqref{eq:terminal_node_posterior}, for brevity, but we now fully define each prior in turn.

\subsection{The tree {structure}}
\label{sec:tree_structure}

The prior $\pi\left(\mathcal{T}_{t}\right)$ is specified {following} the standard setting given by \citep{chipman1998bayesian}{, with slight modifications to incorporate the rotated splitting rules.} Thus, the tree prior distribution is implicitly defined by a generating stochastic process. In the standard BART algorithm, the tree generation is initialised with a root node. Thereafter, the structure is learned via grow, prune, change, and swap moves. New trees are proposed by growing a new terminal node, removing a pair of terminal nodes, changing the split rule for an internal node, or swapping the split rules for a pair of internal nodes, where the type of move is chosen at random. Each proposed tree is then accepted or rejected via Metropolis-Hastings (MH); see \cite{chipman1998bayesian} for further details. Notably, the swap move is not incorporated by GP-BART due to computational complexity, as shown by \citet{kapelner2013bartmachine}, and the tendency of GP-BART to yield shallower trees, for which proposing such swap moves would not be feasible.

We also introduce two modified moves, termed 
``grow-rotate'' and ``change-rotate'', as replacements for the original ``grow'' and ``change'' moves, in order to enhance the predictive performance over standard BART. {We begin by successively selecting a pair of covariates $j$ and $j^\prime$ among the set of $p$ possible covariates in $\mathbf{X}$ at random with equal probabilities. Then, one of these predictors $\mathbf{x}^{(.)}$ is sampled, again with equal probability. The rotated splitting rules are restricted to the case where the covariate selected from the pair is continuous}. For each rotated split, an angle $\theta$ is {sampled with equal probability} from a predefined grid of $20$ equally spaced values within the interval $[0,\pi]$. To rotate both predictors with respect to $\theta$, it is possible to transform the  original coordinate system to $(\mathbf{x}^{(j)}_{r},\mathbf{x}^{(j^\prime)}_{r})$ by multiplying $\left(\mathbf{x}^{(j)},\mathbf{x}^{(j^\prime)}\right)$  by the rotation matrix
\begin{equation}\bm{\mathcal{R}}\left(\theta\right)={\begin{pmatrix}\cos \theta &-\sin \theta \\\sin \theta &\cos \theta \\\end{pmatrix}}.
\end{equation}
{Then, w}ithin the projected feature space, a split rule is selected by sampling a cutpoint from a uniform distribution $c_{\mathbf{x}_{r}^{(.)}} \sim \mbox{Uniform}(a_{\mathbf{x}_{r}^{(.)}},b_{\mathbf{x}_{r}^{(.)}})$, where $a_{\mathbf{x}_{r}^{(.)}}$ and $b_{\mathbf{x}_{r}^{(.)}}$represent the minimum and maximum values of the transformed selected split variable $\smash{\mathbf{x}_{r}^{(.)}}$ within the branch. These rules in the projection space correspond to rotated rules in the original space. If $\theta$ is chosen from the set $\boldsymbol{\theta}_0=\left\{0,\pi/2,\pi\right\}$, the rotation direction remains originally axis-aligned, effectively returning to the standard BART splitting rules with a univariate cutpoint as per Section \ref{sec:gp_bart}. Thus, the standard BART moves can be viewed as a specific case of their projected counterparts. Indeed, in cases where axis-aligned splits are sufficient, proposed projections at $\theta \in \boldsymbol{\theta}_0$ tend to be accepted instead of any other $\theta$ direction. 

{As stated, the above applies only when the selected covariate from the~pair is continuous.} While we omit categorical variables from the GPs, we do allow them to be used to form splitting rules. 
{If the sampled covariate within the pair is categorical, the angle $\theta$ is irrelevant; we assume $\theta \in \boldsymbol{\theta}_0$ and need not sample it.} When the selected covariate is binary, the splitting rule is simply a partition of its levels. {However, we make a further modification when~the sampled pair contains a nominal variable. We use} the reparameterisation suggested by \citeauthor{wright2019splitting} to identify optimal cutpoints for such predictors in treed methods (see \citep{wright2019splitting, ranger} for more details). This ultimately leads to split rules of the form ${\{}\mathbf{x}^{(.)} \in \{\mathcal{S}\}{\}}$ {vs. $\{\mathbf{x}^{(.)} \notin \{\mathcal{S}\}\}$}, where~$\{\mathcal{S}\}$ denotes a subset of the levels of the given covariate {and also allows for continuous covariates to be rotated with respect to nominal ones. Otherwise,} GP-BART utilises the default moves from BART {outlined in Section \ref{sec:gp_bart} when $p=1$}.

{To summarise, the prior for $T_{t}$ can be  divided into five aspects; namely, (i) the distribution on the pair of candidate splitting variables at each interior node, (ii) the distribution on the selected splitting variable, conditioned on the chosen pair, (iii) the distribution on the rotation angle $\theta$, given the selected variable, and
(iv) the distribution on the splitting cutpoint, conditional on the chosen pair, variable, and angle. For these four aspects, the relevant priors coincide with the equiprobable discrete proposal distributions described above. Furthermore, (v) the prior} probability of an individual node at depth $d=0,1,2,\ldots$ being non-terminal is controlled by the hyperparameters $\alpha$ and $\beta$ through
\begin{equation}
     \Pr\left(\mbox{non-terminal node}\right) \propto \alpha \left(1 + d \right)^{-\beta},\; \alpha \in \left(0,1\right),\; \beta \in \left\lbrack0,\infty\right).
     \label{eq:nonterminal_prob}
\end{equation}
The tree prior $\pi\left(\mathcal{T}_{t}\right)$ is then given by a product of the probabilities of each node, since Equation \eqref{eq:nonterminal_prob} assumes independence between nodes. Following some evaluation of alternative parameterisations, we fix the default values $\alpha=0.95$ and $\beta=2$, as per the standard BART \citep{chipman2010bart}. The proposal distribution for a new tree is described by a discrete sample of the possible grow-rotate, change-rotate, and prune moves, with respective probabilities of $0.3$, $0.4$, and $0.3$. These probabilities align with those associated with the standard moves in \texttt{bartMachine} \cite{kapelner2013bartmachine}, whereby the $0.1$ probability of a swap move in the original BART \citep{chipman2010bart} is equally reapportioned to the grow-rotate and change-rotate moves{, without modifying the prior probability of the prune move.}

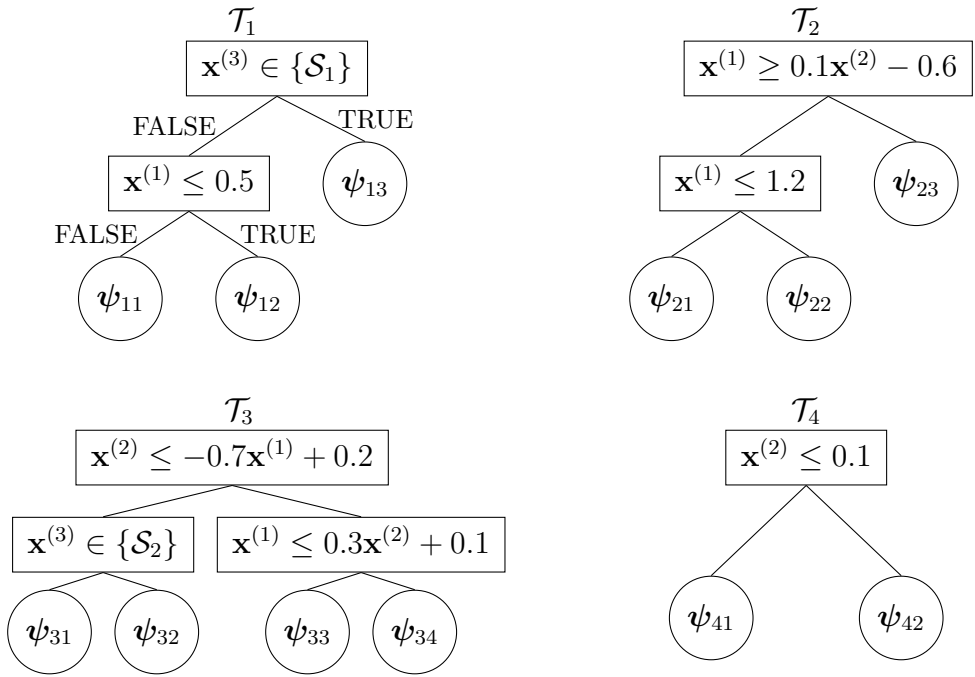
\begin{figure}[H]

\captionsetup[subfigure]{labelformat=empty}
        \centering
        \begin{subfigure}[t]{0.45\textwidth}
            \centering
            \caption[Network2]%
            {{ \normalsize $\mathcal{T}_{1}$}}
            \begin{forest}
for tree={
    grow=south, draw, minimum size=3ex, 
    inner sep=3pt, 
    s sep=7mm,
    l sep=6mm
    }
[$\:\mathbf{x}^{(3)} \in \{\mathcal{S}_1\}\:$,
    [$\:\mathbf{x}^{(1)} \leq 0.5\:$, edge label={node[midway, font=\footnotesize, left]{\:FALSE\:}}
        [$\boldsymbol{\psi}_{11}$, circle,  edge label={node[midway,left, font=\footnotesize]{\:FALSE\:}},]
        [$\boldsymbol{\psi}_{12}$, circle,  edge label={node[midway,right, font=\footnotesize]{\:TRUE\:}},]
    ]
    [$\boldsymbol{\psi}_{13}$, circle,  edge label={node[midway,right, font=\footnotesize]{\:TRUE\:}},]
]
\end{forest}
            \label{fig:mean and std of net14}
        \end{subfigure}
        \hfill
        \begin{subfigure}[t]{0.45\textwidth}   
            \centering 
                        \caption[]%
            {{\normalsize $\mathcal{T}_{2}$}}    
\begin{forest}
for tree={
    grow=south, draw, minimum size=3ex, inner sep=3pt, s sep=7mm, l sep=6mm
        }
[$\:\mathbf{x}^{(1)} \geq 0.1 \mathbf{x}^{(2)} - 0.6\:$,
    [$\:\mathbf{x}^{(1)} \leq 1.2\:$,
        [$\boldsymbol{\psi}_{21}$, circle]
        [$\boldsymbol{\psi}_{22}$, circle]
    ]
    [$\boldsymbol{\psi}_{23}$, circle]
]
\end{forest}
            \label{fig:mean and std of net34}
        \end{subfigure}
        \vskip\baselineskip
        \begin{subfigure}[t]{0.45\textwidth}  
            \centering 
            \caption[]%
            {{\normalsize $\mathcal{T}_{3}$}}
\begin{forest}
for tree={
    grow=south, draw, minimum size=3ex, 
    inner sep=3pt, 
    s sep=3mm,
    l sep=2mm
        }
[$\:\mathbf{x}^{(2)} \leq -0.7 \mathbf{x}^{(1)} + 0.2\:$,
    [$\:\mathbf{x}^{(3)} \in \{\mathcal{S}_2\}\:$,
        [$\boldsymbol{\psi}_{31}$, circle]
        [$\boldsymbol{\psi}_{32}$, circle]
    ]
        [$\:\mathbf{x}^{(1)} \leq 0.3\mathbf{x}^{(2)} + 0.1\:$,
        [$\boldsymbol{\psi}_{33}$, circle]
        [$\boldsymbol{\psi}_{34}$, circle]
    ]
]
\end{forest}
            \label{fig:mean and std of net24}
        \end{subfigure}
        \hfill
        \begin{subfigure}[t]{0.45\textwidth}   
            \centering 
            \caption[]%
{{\normalsize $\mathcal{T}_{4}$}}  
            \begin{forest}
for tree={
    grow=south, draw, minimum size=3ex, inner sep=3pt,
    s sep=14mm,
    l sep=12mm
        }
[$\:\mathbf{x}^{(2)} \leq 0.1\:$,
        [$\boldsymbol{\psi}_{41}$, circle]
        [$\boldsymbol{\psi}_{42}$, circle]
]
\end{forest}
\end{subfigure}
\caption[]
        {Graphical representation of four example trees from a GP-BART model. The splitting rules in each tree can take the form of a univariate cutpoint for continuous covariates (subject to $\theta \in \boldsymbol{\theta}_0$), a subset of factor levels for categorical covariates, or rotated split rules obtained by {random projections of a pair of covariates, provided the selected covariate from the pair is continuous}. Gaussian process priors are assumed for the predicted values for each terminal node in each tree, such that $\boldsymbol{\psi}_{t\ell}\sim\mathcal{GP}_{t\ell}$ \emph{a priori}.}
        \label{fig:graphical_bart}
\end{figure}

Figure \ref{fig:graphical_bart} summarises the main idea of our proposed statistical model, highlighting the modified terminal node priors and the rotated splitting rules, via four examples of regression trees within the ensemble. Here, there are two continuous predictors, $\mathbf{x}^{(1)}$ and $\mathbf{x}^{(2)}$, and $\mathbf{x}^{(3)}$ is categorical, with the sets $\{\mathcal{S}_1\}$ and $\{\mathcal{S}_2\}$ being subset of its levels. Notably, some split rules from trees $\mathcal{T}_{2}$ and $\mathcal{T}_{3}$ are obtained by projecting a randomly sampled non-parallel axis direction $\theta$ onto the pair $(\mathbf{x}^{(1)},\mathbf{x}^{(2)})$, resulting in rotated splitting rules.
 
\subsection{The prior on the Gaussian processes}\label{section:mu_gp_prior}

The main contribution of the GP-BART model is to define
\begin{equation}
\boldsymbol{\psi}_{t \ell}\given\mathcal{T}_{t},\mu_{t \ell},\boldsymbol{\phi}_{t},\nu \sim  \mbox{MVN} \left( \boldsymbol{\mu}_{t \ell}=\mu_{t \ell}\mathbf{1}_{n_{t \ell}}, \boldsymbol{\Omega}_{t \ell} \right)\label{psi_kappa}
\end{equation}
\noindent as a GP prior over the set of $n_{t \ell}$ observations belonging to terminal node $\ell$ of tree $\mathcal{T}_t$, where $\mathbf{1}_{n_{t \ell}}$ is a vector of ones of length $n_{t \ell}$, such that the mean vector is constant. Here, $\boldsymbol{\Omega}_{t\ell} \in \mathbb{R}^{n_{t \ell}} \times \mathbb{R}^{n_{t \ell}}$
is specified as a node-specific, stationary, anisotropic matrix of exponentiated-quadratic covariance terms, with its $(i,k)$-th element given by
\begin{equation}
    \nu^{-1}  \exp\left\{-\frac{1}{2}\sum_{j=1}^{p}\frac{ \left(    {x}^{(j)}_{i}-{x}^{(j)}_{k}\right)^{2}}{\phi_{tj}^{2}}\right\}.
    \label{eq:covariance_matrix_label}
\end{equation}
We normalise all predictors {to the $[0,1]$ range} to improve the numerical stability of the kernel. Notably, the trees themselves are unaffected by this, as the rules governing their structure are invariant to monotone transformations.
We set $\mu_{t \ell}\given \mathcal{T}_{t}, \sim \mbox{N}\left(\mu_{\mu}, \tau_{\mu}^{-1} \right)$ to exploit conjugacy and enable all $\mu_{t \ell}$ para\-meters to be marginalised out. Hence, Equation \eqref{psi_kappa} can be redefined~as 
\[\boldsymbol{\psi}_{t \ell}\given\mathcal{T}_{t},\boldsymbol{\phi}_{t},\nu,\mu_{\mu},\tau_{\mu} \sim  \mbox{MVN}\left(\mu_{\mu}\mathbf{1}_{n_{t\ell}}, \tau_{\mu}^{-1} \mathbf{1}_{n_{t\ell}}\mathbf{1}_{n_{t\ell}}^{\top} + \boldsymbol{\Omega}_{t \ell} \right),\]
in order to encourage better mixing. We adopt this likelihood formulation throughout and provide further details in \ref{sec::appendix_a}.

\citet{chipman2010bart} showed that the induced prior distribution on $\mathbb{E}[y_{i}\given\mathbf{x}_{i}]$ over all $T$ trees in a BART model allows for some expert knowledge to be incorporated about the contribution of each tree which can help to guide the choices of hyperparameter values. However, the presence of the GP priors on $\boldsymbol{\psi}_{t\ell}$ in GP-BART yields a different induced prior which we write as
\[\mathbb{E}\left[y_{i}\given\mathbf{x}_i\right] \sim \mbox{N}\left(T\mu_{\mu}, T\left(\nu^{-1} + \tau_{\mu}^{-1}  \right) \right).\]
\noindent Following the \citeauthor{chipman2010bart} approach, the key idea is to select the hyper\-parameters such that $\mathbb{E}[y_{i}\given\mathbf{x}_{i}]$ is between $y_{\min}$ and $y_{\max}$ with high probability. The confidence interval for $\mathbb{E}[y_{i}\given\mathbf{x}_{i}],\:\forall\:{i}=1,\ldots,n$, has boundaries
\[\begin{cases}
T\mu_{\mu} - k\sqrt{T}\left(\nu^{-1}+\tau^{-1}_{\mu}\right)^{1/2} = y_{\min}\\
T\mu_{\mu} + k\sqrt{T}\left(\nu^{-1}+\tau^{-1}_{\mu}\right)^{1/2} = y_{\max}
\end{cases}\]%
for a chosen $k$. We adopt $k=2$, which represents an approximate $95\%$  confidence interval. Following \citeauthor{chipman2010bart}, we re-scale $\mathbf{y}$ such that $y_{\min}= -0.5$ and $y_{\max} = 0.5$, set $\mu_{\mu} = 0$ and hence set the precision parameters to
\[\nu = \tau_{\mu} =  8k^{2}T,\]
in order to balance the contribution of both parameters.

Though $\nu$ and $\tau_{\mu}$ are both referred to as precision parameters, their roles and interpretations differ, with $\nu$ and $\tau_{\mu}$ -- both of which are fixed rather than estimated -- being the parameters that control the precision of the GPs and the $\mu_{t \ell}$ parameters, respectively. As we increase the number of trees $T$, the scale $\nu^{-1}$ of each GP decreases, regularising the model by setting the contribution of each GP to be small. Likewise, the precision of the $\mu_{t \ell}$ parameter is proportional to the number of trees, shrinking the mean of each terminal node as more tree components are added into the model. Setting both parameters in this way reduces the chance of only one single tree dominating the model.

\subsubsection{The prior on the length parameter}
\label{sec::prior_length_section}

{As shown in Equation \eqref{eq:covariance_matrix_label}, $\boldsymbol{\phi}_{t}$ controls the rate of decay with respect to the $\mathcal{L}_2$ distances between pairs of design points,
such that larger values of $\phi_{tj}$ will quickly decrease the contribution of variables which are uncorrelated with the true generation function $f(\mathbf{x}_{i})$. Thus, to enable the use of automatic relevance determination (ARD) over the variables used in the GPs while balancing computational considerations, we derive a discrete prior for the length parameter $\phi_{tj}$ for a given tree $t$ and covariate $j$ from a mixture of gamma distributions:
\begin{align}
\begin{split}
\kappa &\times \mbox{Ga}\left( a_{\phi_1} = 3, d_{\phi_1} = 2.5  \right) +\\
\left(1-\kappa\right) &\times \mbox{Ga} \left(a_{\phi_2} = 5000, d_{\phi_2} = 100\right),
\end{split}
\label{eq:prior_phi}
\end{align}
where $\mbox{Ga}(a,d)$ denotes a gamma distribution with expectation $a/d$. The two components govern smaller and larger values of $\phi_{tj}$, respectively, and we set the mixture weight $\kappa$ to $0.3$ throughout. 

Ultimately, we define a discrete prior for $\pi(\phi_{tj})$, with support given by $\mathbf{S}_\phi = \left\{0.1, 0.5, 1, 2, 3, 4, 50 \right\}$ in order to reflect the high-probability regions of the mixture in Equation \eqref{eq:prior_phi} and the fact that the precise magnitude of $\phi_{tj}$ is only important for smaller values. Furthermore, the prior probabilities are specified to be proportional to $d_\phi(k)$,  where $d_{\phi}(k)$ is the density of the mixture of gamma distributions in Equation \eqref{eq:prior_phi} evaluated at $k\in \mathbf{S}_\phi$. This leads to probabilities of $\Pr(\phi_{tj} = k)=(0.022, 0.206, 0.236, 0.035, 0.014, 0.002, 0.485)$, which reflect the fact that, \emph{a priori}, we expect each variable to have an equal chance of contributing meaningfully to the GPs. The $\phi_{tj}$ sampling processes is also done using MH, with the proposal distribution for new parameters given  by an equiprobable discrete distribution which reflects the support of our induced discrete prior $\pi(\phi_{tj})$ --- i.e., each value in $\mathbf{S}_\phi$ is sampled with equal probability --- and helps to avoid spurious length parameter values. 

The aforementioned normalisation of each predictor in $\mathbf{X}$ also aids the elicitation of this prior, by minimising the range of $\phi_{tj}$ and ensuring all covariates are on the same scale. Furthermore, the discrete proposal reduces the computational burden, as we can partially pre-compute all possible covariance functions. We calculate the fraction in the exponent of Equation \eqref{eq:covariance_matrix_label} for each length parameter value in $\mathbf{S}_\phi$, using all $n$ observations of the continuous covariates, and thereafter obtain $\boldsymbol{\Omega}_{t\ell}(\boldsymbol{\phi}_t,\nu)$ by appropriately utilising the quantities relevant to the sampled $\phi_{t1},\ldots,\phi_{tp^{\star}}$ values and subset of observations $\mathbf{X}_{(t\ell)}$ belonging to the corresponding terminal node.}

\subsection{The prior on the residual precision}

A conjugate gamma distribution $\tau \sim \mbox{Ga}(a_{\tau}, d_{\tau})$ is assumed for the residual precision parameter. To select the~\mbox{hyperparameters}, we follow \citet{chipman2010bart} in setting the shape $a_\tau$ and rate $d_\tau$ such that $\Pr\left(\tau \geq \hat{\tau}_{OLS}\right) = \eta_{\tau}$,
where $\eta_{\tau}$ is a high-probability value (we typically use~$\eta_{\tau} = 0.9$) and $\hat{\tau}_{OLS}$ is the precision calculated from an ordinary linear regression of $\mathbf{y}$ against the same set of predictors $\mathbf{X}$. The intuition behind this estimation strategy comes from the idea that, given the non-linearity of the GP and the piecewise additive component from BART, we can be optimistic that the precision of the model is greater than that of a linear model.

\section{Computational algorithms for inference and prediction}\label{sec:gp_bart_sample}

Given the observed $\mathbf{y}$, the posterior distribution for the trees and their parameters is given by
\begin{equation}\pi\left(\left(\mathcal{T}_{1},\mathbf{G}_{1}\right),\ldots, \left(\mathcal{T}_{T},\mathbf{G}_{T}\right), \tau \given \mathbf{y}\right).
\label{eq:posterior_gp_conditionals}
\end{equation}
We define the notation of a generic set $\mathcal{M}_{-t}$ as the the set of all $\mathcal{M}_{1},\ldots,\mathcal{M}_{T}$ elements except $\mathcal{M}_{t}$, such that $\mathcal{T}_{-t}$ corresponds to the set of $T-1$ trees except $\mathcal{T}_{t}$ with respective terminal node parameters $\mathbf{G}_{-t}$. The key feature necessary to sample from Equation \eqref{eq:posterior_gp_conditionals} is the ``Bayesian backfitting'' algorithm of \citet{hastie2000bayesian}, which enables iterative sampling of the $t$-th tree and its parameters. \citeauthor{hastie2000bayesian} showed that the distribution $ \pi\left(\mathcal{T}_{t},\mathbf{G}_{t}\given \mathcal{T}_{-t},\mathbf{G}_{-t},\tau, \mathbf{y}\right))$ can be rewritten in terms of the partial residuals
\begin{equation}
        \mathbf{R}_{t} = \left( \mathbf{r}_{t 1}, \dots, \mathbf{r}_{tb_{t}}  \right) \equiv \mathbf{y} - \sum_{r\neq t}^{T}  g\left(\mathbf{X};\mathcal{T}_{r},\mathbf{G}_{r}\right).\label{eq:residuals_def}
\end{equation}
\noindent The general structure of the sampler is thus given by:
\begin{align*}
1\colon \qquad &\mathcal{T}_{1}\given\mathbf{R}_{1},\boldsymbol{\phi}_{1},\nu,\tau_{\mu},\tau\\
2\colon \qquad &\boldsymbol{\psi}_{11},\ldots,\boldsymbol{\psi}_{1 b_1}\given \mathcal{T}_{1},\mathbf{R}_{1},\boldsymbol{\phi}_{1}, \nu,\tau_{\mu}, \tau\\
3\colon \qquad &\boldsymbol{\phi}_{1}\given \mathcal{T}_{1},\mathbf{R}_{1},\nu,\tau_{\mu},\tau\\
& \vdots \\
3T-2\colon \qquad &\mathcal{T}_{T}\given \mathbf{R}_{T},\boldsymbol{\phi}_{T},\nu,\tau_{\mu}, \tau\\
3T-1\colon \qquad &\boldsymbol{\psi}_{T1},\ldots,\boldsymbol{\psi}_{Tb_T}\given \mathcal{T}_{T},\mathbf{R}_{T},\boldsymbol{\phi}_{T}, \nu,\tau_{\mu},\tau\\
3T\colon \qquad &\boldsymbol{\phi}_{T}\given \mathcal{T}_{T},\mathbf{R}_{T}, \nu,\tau_{\mu},\tau\\[\medskipamount]
3T+1\colon \qquad &\tau\given  (\mathcal{T}_{1},\mathbf{G}_{1}) \ldots, (\mathcal{T}_{T},\mathbf{G}_{T}),a_{\tau},d_{\tau},\mathbf{y}.
\end{align*}%
\indent The algorithm is initialized with $T$ stumps (i.e., trees with a single root node), with all mean parameters $\mu_{t1} = 0$ and all length parameters $\phi_{tj}$ sampled from {the discrete proposal distribution described in Section \ref{sec::prior_length_section}}. Additionally, the residual precision parameter $\tau$ is sampled from its prior distribution. For stumps, only the grow-rotate move is proposed. Thereafter, once trees have reached sufficient depth $d=1$, new trees $\mathcal{T}^{\star}_{t}$ are sequentially proposed by randomly selecting one of the three available moves: grow-rotate, change-rotate, and prune, and then accepted or rejected according via MH. 

Though these moves modify the tree depth, $\mathbf{G}_t$ only changes dimension with respect to the means $\mu_{t1},\ldots,\mu_{tb_t}$, since $\nu$ is fixed and $\boldsymbol{\phi}_t$ is specified at the tree level. Consequently, this does not affect the sampling of $\mathcal{T}_t$, since all $\mu_{t\ell}$ parameters are marginalised out, thereby yielding a tractable tree posterior proportional to $\pi\left(\mathcal{T}_{t}\right) \pi\left(\boldsymbol{\phi}_{t}\right)\pi\left(\mathbf{R}_t\given \mathcal{T}_{t},\boldsymbol{\phi}_{t},\nu, \tau_{\mu},\tau\right)$, which does not depend on any varying-dimensional parameters at the terminal node level. The predicted values in each terminal node are updated by a Gibbs sampling scheme, with the associated full conditional distribution given by
\begin{equation}
    \boldsymbol{\psi}_{t \ell}\given   \ldots \sim \mbox{MVN} \left( \boldsymbol{\mu }_{\mathcal{GP}_{t \ell}} , \boldsymbol{\Sigma}_{\mathcal{GP}_{t \ell}}  \right) ,
    \label{eq:posterior_gp_distribution}
\end{equation}
where
\begin{align*}
    \boldsymbol{\mu}_{\mathcal{GP}_{t \ell}} &= \boldsymbol{\Lambda}_{t\ell}^\top \left(\tau^{-1}\bm{\mathcal{I}}_{n_{t \ell}} + \boldsymbol{\Lambda}_{t\ell}\right)^{-1}\mathbf{r}_{t \ell},\\
    \boldsymbol{\Sigma}_{\mathcal{GP}_{t \ell}} &= \boldsymbol{\Lambda}_{t\ell} - \boldsymbol{\Lambda}_{t\ell}^\top \left(\tau^{-1}\bm{\mathcal{I}}_{n_{t \ell}} + \boldsymbol{\Lambda}_{t\ell} \right)^{-1}\boldsymbol{\Lambda}_{t\ell},
\end{align*}
with $\boldsymbol{\Lambda}_{t\ell} =\tau_{\mu}^{-1} \mathbf{1}_{n_{t\ell}}\mathbf{1}_{n_{t\ell}}^{\top} + \boldsymbol{\Omega}_{t \ell}$ and $\bm{\mathcal{I}}_{n_{t \ell}}$ being an identity matrix of the indicated dimension.

Lastly, we sample the length parameters $\phi_{tj}\:\forall\:(j=1,\ldots, p^{{\star}}, t=1,\ldots,T)$ {from their discrete proposal distribution} using MH steps. Once all $T$ trees are updated, the precision parameter is sampled using a Gibbs step, with the full conditional given by
\begin{equation}
      \tau\given  \ldots \sim \mbox{Ga} \left(\frac{n}{2}+a_{\tau}, \frac{1}{2}\left(\mathbf{y}-\hat{\mathbf{y}})^\top\left(\mathbf{y}-\hat{\mathbf{y}}\right)\right) + d_{\tau} \right),
    \label{eq:sample_tau_posterior}
\end{equation}
where $\hat{\mathbf{y}} \equiv \sum_{t = 1}^{T} g(\mathbf{X}; \mathcal{T}_t, \mathbf{G}_{t})$ represents the sum of the predictions $\boldsymbol{\psi}_{t\ell}$ across all terminal nodes from all sampled trees. 

\subsection{Algorithm specifications and initialisation}\label{sec:hyper_init}

We set the number of trees $T$ to have a default value of $20$, since we require fewer trees than BART due to the inherent non-linearity of the GPs and achieved reasonable predictive performance in various scenarios demonstrated in Sections \ref{sec:simulations} and \ref{sec:real_data} using this value. Alternatively, this quantity could be selected via cross-validation, though the computational cost of doing so may be prohibitive. 

Employing the rotated splitting rules or using the standard moves from BART is also a setting of the model that can be toggled by the user, as well as which variables are included in the GPs themselves. All variables are allowed to form splitting rules, since it improves the model's prediction in general, especially for spatial data. If the rotated splits are deemed unnecessary, the sampler will not accept them and favour splits with $\theta \in \boldsymbol{\theta}_0$. By default, if there is no strict prior knowledge about the covariates, GP-BART includes all continuous variables in the GPs. Though a more parsimonious model could be achieved if the variables used in the GPs are merely a subset of those used to construct the trees, we do not consider this further here. 

We present the full structure of the GP-BART sampler in Algorithm \ref{alg:gp_bart_algorithm}, where the matrix of covariates $\mathbf{X}$ and response vector $\mathbf{y}$  from the training set enter as inputs. Trees, partial residuals, and hyperparameters are then initialised. For each MCMC sample, a proposed tree $\mathcal{T}^{\star}_{t}$ is accepted, if it is valid and contains no empty terminal nodes, with probability $\gamma^{\star} (\mathcal{T}_{t}, \mathcal{T}_{t}^{\star})$. {The novel aspects of the tree prior we introduce under GP-BART (i.e., priors over the pair of candidate splitting variables and the rotation angle $\theta$) cancel out in the MH acceptance ratio. Consequently, the ratio of priors $\pi\left(\mathcal{T}_t^\star\right)/\pi\left(\mathcal{T}_t\right)$ in $\gamma^\star\left(\mathcal{T}_t,\mathcal{T}^\star\right)$ and the transition probabilities $q(\cdot)$ for all moves remain unchanged from the formulations given by \citet{linero2018bayesian}}. The remaining parameters are sampled using Equations \eqref{eq:posterior_gp_distribution}--\eqref{eq:sample_tau_posterior}.

A standard number of iterations $N_{\text{MCMC}} = 3500$, of which the first $N_{\text{burn}}=1500$ are discarded, was found to yield a sufficient number of samples to reliably characterise the posterior in all applications herein. This was verified through examination of the convergence of posterior samples of $\tau$. Though the algorithm is computationally onerous given the matrix inversions associated with the use of GPs, we stress that such operations are of the order $\mathcal{O}(n_{t \ell}^3)$ within a given terminal node, rather than $\mathcal{O}(n^3)$ as they would be under a single GP. Further details of the computational performance of our algorithm in the context of a simulation study are deferred to Section \ref{sec::computationalcost}.\medskip

\begin{algorithm}[H]
\bgroup
\leftskip -7pt
\SetKwInput{KwInput}{Input}
\SetKwInput{KwOutput}{Output}
\SetKwInput{KwInit}{Initialise}
\SetAlgoLined
\KwInput{$\mathbf{X}$, $\mathbf{y}$, $T$, $N_{\text{MCMC}}$, $N_{\text{burn}}$, and all hyperparameters of the priors.}
 \KwInit{$T$ tree stumps with $\mu_{t1}=0\,\forall\,t$, $\phi_{tj}\,\forall\,(t,j)$ drawn {with equal probability from $\mathbf{S}_\phi$}, and $\tau$ drawn from its $\mbox{Ga}(a_{\tau},d_{\tau})$ prior.}
 \For{iterations $m$ from $1$ to $N_{\text{MCMC}}$}{
  \For{trees $t$ from $1$ to $T$}{
   Calculate the partial residuals $\mathbf{R}_t$ via Equation \eqref{eq:residuals_def}\;
   Propose a new tree $\mathcal{T}_{t}^{\star}$ by a grow-rotate, change-rotate, or prune move\;
   Accept and update $\mathcal{T}_{t}=\mathcal{T}_{t}^{\star}$ with probability 
   \[\gamma^{\star}\left(\mathcal{T}_{t},\mathcal{T}_{t}^{\star}\right) = \min \left\{1, \frac{\pi \big(\mathbf{R}_{t} \given  \mathcal{T}^{\star}_{t},\boldsymbol{\phi}_{t},\nu,\tau_{\mu},\tau \big) \pi\big(\mathcal{T}^{\star}_{t}\big){q\big(\mathcal{T}^{\star}_{t}\rightarrow\mathcal{T}_{t}\big)}}{\pi \big(\mathbf{R}_{t} \given  \mathcal{T}_{t},\boldsymbol{\phi}_{t},\nu,\tau_{\mu},\tau \big) \pi\big(\mathcal{T}_{t}\big){q\big(\mathcal{T}_{t}\rightarrow\mathcal{T}^{\star}_{t}\big)}}\right\}.
   \]
   
   \For{terminal nodes $\ell$ from $1$ to $b_{t}$}{
    Update $\boldsymbol{\psi}_{t \ell}$ via Equation \eqref{eq:posterior_gp_distribution}.
   }{
   \For{continuous predictors $j$ from $1$ to $p^{\star}$ used in the GPs}{
   Update ${\phi}_{tj}$ using MH.
   }}
   }{
     Update $\tau$ via Equation \eqref{eq:sample_tau_posterior}}.
 }
 \KwOutput{Samples from $\pi\left( (\mathcal{T}_{1},\mathbf{G}_{1}),\ldots, (\mathcal{T}_{T},\mathbf{G}_{T}), \tau \given \mathbf{y}\right)$.}
\caption{GP-BART sampling algorithm\label{alg:gp_bart_algorithm}}
\egroup
\end{algorithm}


\subsection{Prediction in GP-BART}

The trees in GP-BART models can provide out-of-sample predictions for a set of $n^\star$ new observations 
$\mathbf{X}^{\star}$. For a given terminal node $\ell$ in tree $\mathcal{T}_t$ for a particular MCMC sample, the joint posterior distribution of the node-level training predictions and the node-level test predictions is given by

\[\left(\begin{array}{@{}c|c@{}}
           \mathbf{\boldsymbol{\psi}}_{t \ell}&\multirow{2}{*}{$\cdots$\negmedspace}\\
           \boldsymbol{\psi}_{t \ell}^{\star}&
         \end{array}\right)\!\sim\mbox{MVN}\! \begin{pmatrix} 
         \begin{bmatrix}\mathbf{0}_{n_{t\ell}}\\ \mathbf{0}_{n^\star_{t\ell}}\end{bmatrix}\!,\!
         \begingroup
         \setlength\arraycolsep{1.5pt}
           \begin{bmatrix} 
           \boldsymbol{\Lambda}_{t \ell}
           & \boldsymbol{\Lambda}_{t \ell}^{\star} \\
          \boldsymbol{\Lambda}_{t \ell}^{\star\top} & \boldsymbol{\Lambda}_{t \ell}^{\star\star}
           \end{bmatrix}\endgroup\\
         \end{pmatrix}\!,\]
\noindent 
with $\boldsymbol{\Lambda}^{\star}_{t \ell} \in \mathbb{R}^{n^{\star}_{t \ell}} \times \mathbb{R}^{n_{t \ell}}$ and $\boldsymbol{\Lambda}_{t \ell}^{\star\star} \in \mathbb{R}^{n^{\star}_{t \ell}} \times \mathbb{R}^{n^{\star}_{t \ell}}$. Here, $n_{t\ell}$ and $n_{t\ell}^\star$ denote the number of observations assigned to terminal node $\ell$ of tree $\mathcal{T}_t$ for the training samples and new data, respectively. This posterior predictive distribution can be conditioned with respect to $\boldsymbol{\psi}_{t\ell}$ to yield
\[
    \boldsymbol{\psi}^{\star}_{t \ell}\given \boldsymbol{\psi}_{t \ell}, \mathbf{X}_{(t \ell)},\mathbf{X}^{\star}_{(t \ell)},\ldots \sim \mbox{MVN} \left( \boldsymbol{\mu}_{\mathcal{GP}^{\star}_{t \ell}} , \boldsymbol{\Sigma}_{\mathcal{GP}^{\star}_{t \ell}}
    \right) ,
\]%
where $\boldsymbol{\mu}_{\mathcal{GP}^{\star}_{t \ell}} = \boldsymbol{\Lambda}_{t \ell}^{\star \top} \boldsymbol{\Lambda}_{t \ell}^{-1}\boldsymbol{\psi}_{t \ell}$ and $\boldsymbol{\Sigma}_{\mathcal{GP}^{\star}_{t \ell}} = \boldsymbol{\Lambda}^{\star\star}_{t \ell} - \boldsymbol{\Lambda}_{t \ell}^{\star \top} \boldsymbol{\Lambda}_{t \ell}^{-1}\boldsymbol{\Lambda}^{\star}_{t \ell}$.

Ultimately, the function $g^\star$ assigns the vector $\boldsymbol{\mu}_{\mathcal{GP}^{\star}_{t \ell}}=\mathbb{E}(\boldsymbol{\psi}_{t\ell}^\star\given\ldots)$ to the associated new observations $\mathbf{X}_{(t \ell)}^\star$ on a \emph{per-iteration} basis, such that the estimates from GP-BART are given by 
\begin{equation}
    \hat{\mathbf{y}}^{\star^{(m)}} = \mbox{N}\left(\sum_{t=1}^{T}g^\star\left(\mathbf{X}^{\star}; \mathcal{T}^{(m)}_{t}, \mathbf{G}^{(m)}_{t}\right), \hat{\tau}^{-1(m)}\bm{\mathcal{I}}_{n^\star}\right),
    \label{eq:posterior_prediction}
\end{equation}
\noindent where $m$ indexes the draws from the posterior distribution after the burn-in iterations. The overall prediction $\bar{y}_i^\star$
for a new observation $\mathbf{x}^{\star}_{i}$ is then given by the average of the estimates $\hat{y}_{i}^{\star^{(1)}} ,\ldots, \hat{y}_{i}^{\star^{(M)}}$; i.e., $\bar{y}_i^\star=\frac{1}{M}\sum_{m=1}^{M}\hat{{y}}_{i}^{\star^{(m)}}$. Posterior samples from Equation \eqref{eq:posterior_prediction} can also be used to quantify the uncertainty in the predictions. For instance, with some large number $Q$ of draws \emph{per posterior sample}, the endpoints of a $(1-\alpha)\%$ prediction interval for a predicted value $\bar{{y}}_{i}^\star$ can be obtained from the upper and lower $\alpha/2$ quantiles of $(\hat{y}_{i}^{\star^{(11)}} ,\ldots, \hat{y}_{i}^{\star^{(M1)}}),\ldots,(\hat{y}_{i}^{\star^{(1q)}} ,\ldots, \hat{y}_{i}^{\star^{(MQ)}})$.

\section{Simulation studies}\label{sec:simulations}

In this Section, we present simulation studies to evaluate the performance of GP-BART from several different perspectives. In Section \ref{sec:benchmarking} we primarily aim to assess the efficacy of incorporating the rotated splitting rules and the GPs themselves for data with explicit spatial components, whereas in Section \ref{sec:friedman} we first aim to assess the ARD associated with the {equiprobable discrete} prior on the tree-varying, variable-specific length parameters $\phi_{tj}$ {described in Section \ref{sec::prior_length_section}}. An evaluation of the computational burden is also provided in Section \ref{sec::computationalcost}.

\subsection{Benchmarking experiments}\label{sec:benchmarking}

\begin{figure}[H]
     \centering
     \includegraphics[width=\textwidth]{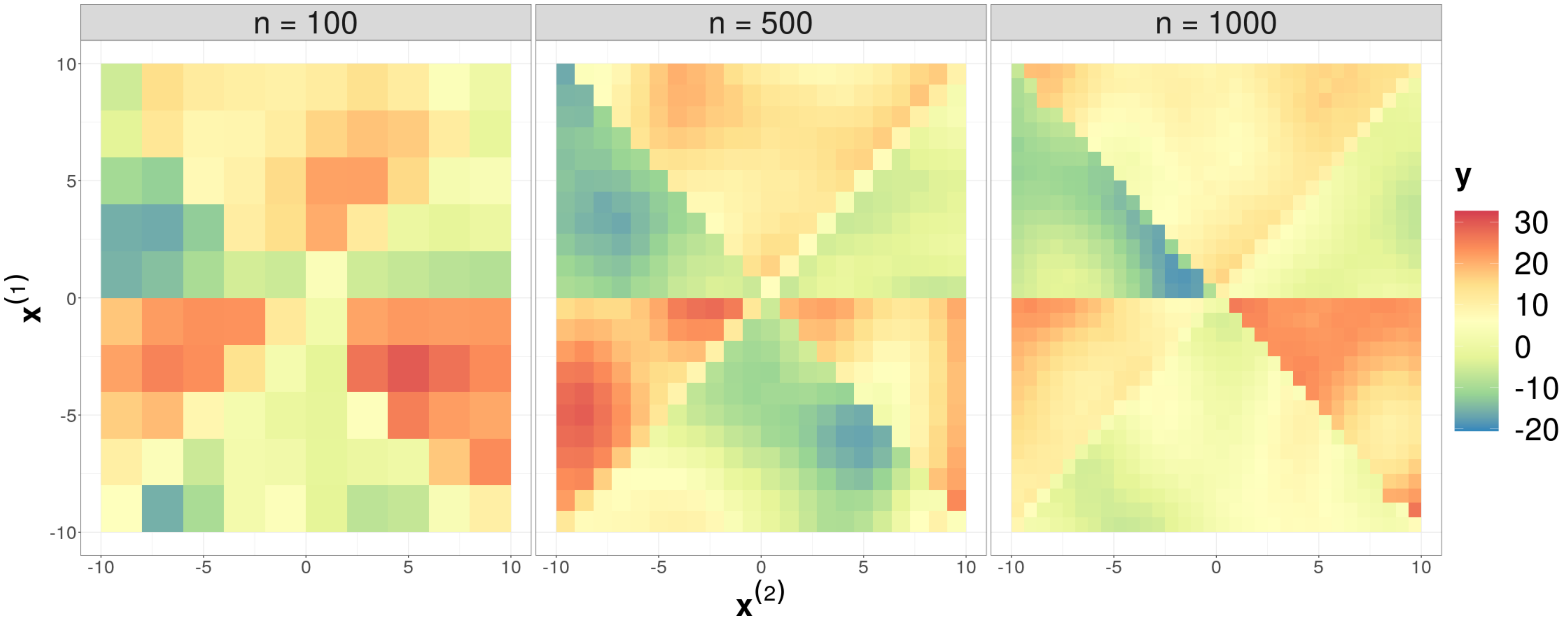}
     \caption{Simulated data with $n=\left\{100,500,1000 \right\}$ observations, respectively.}
     \label{fig:simulated_data}
 \end{figure}

In these experiments, the simulated data are composed by a summation of trees with two terminal nodes, built using the variables $\mathbf{X} = (\mathbf{x}^{(1)},\mathbf{x}^{(2)}$). These covariates are simulated such that each predictor is generated from a uniform grid between $-10$ and $10$. The values associated with each terminal node follow a multivariate normal distribution with specific mean and covariance parameters. We generate the response~variable via
%
\begin{align}
\begin{split}
\mathbf{y} &=\phantom{+}\left\lbrack\left( \boldsymbol{\mu}_{11} + \mathbf{s}_{11}\right)_{\mathbbm{1} \left(\mathbf{x}^{(1)}\,\leq\, \mathbf{x}^{(2)}\right)}\right.+\:\left.\left( \boldsymbol{\mu}_{12} + \mathbf{s}_{12} \right)_{\mathbbm{1}\left(\mathbf{x}^{(1)}\,>\,\mathbf{x}^{(2)}\right)}\right\rbrack\\[0.25ex]
&\phantom{=\:}+\left\lbrack\left( \boldsymbol{\mu}_{21} + \mathbf{s}_{21} \right)_{\mathbbm{1} \left(\mathbf{x}^{(1)}\,\leq\,-\mathbf{x}^{(2)}\right)}\right.+\:\left.\left(\boldsymbol{\mu}_{22} + \mathbf{s}_{22} \right)_{\mathbbm{1}\left(\mathbf{x}^{(1)}\,>\,-\mathbf{x}^{(2)}\right)}\right\rbrack\\[0.25ex]
&\phantom{=\:}+\left\lbrack\left( \boldsymbol{\mu}_{31} + \mathbf{s}_{31} \right)_{\mathbbm{1} \left(\mathbf{x}^{(1)}\,\leq\,0\right)}\right.+\:\left.\left( \boldsymbol{\mu}_{32} + \mathbf{s}_{32}  \right)_{\mathbbm{1}\left(\mathbf{x}^{(1)}\,>\,0\right)}\right\rbrack + \boldsymbol{\varepsilon},\label{eq:benchmarkDGP}
\end{split}
\end{align}
with number of trees $T=3$, each with two terminal nodes. The node-specific mean parameters $\boldsymbol{\mu}_{t\ell}$ are all constant vectors of the form $(\mu_{t\ell},\ldots,\mu_{t\ell})$, with respective values given by $\mu_{11} = -10$, $\mu_{21} = 0$, $\mu_{31} = 10$,  $\mu_{12} = 5$, $\mu_{22} = 20$, and $\mu_{32} = -15$. Multivariate normal spatial terms $\mathbf{s}_{t\ell} \sim \mbox{MVN} (\mathbf{0}_{n_{t\ell}}, \boldsymbol{\Omega}_{t\ell}(\boldsymbol{\phi}_t=\mathbf{3}_{n_{t\ell}},\nu=0.1))$ are added within each terminal node. Residual noise terms $\boldsymbol{\varepsilon} \sim \mbox{MVN}(\mathbf{0}_{n},\tau^{-1}\bm{\mathcal{I}}_{n})$ are also added. {Results obtained with residual precision values of $\tau=\left\{1,0.1, 0.01 \right\}$ lead to similar conclusions in that GP-BART still shows the best performance in terms of prediction accuracy and uncertainty calibration. For brevity, we show the data and results for $\tau=10$ here only and defer the other results to \ref{sec::appendix_b}.} Figure \ref{fig:simulated_data} shows the simulated data surfaces for data sets of size $n=\left\{100,500,1000\right\}$, respectively, highlighting the different partitioning behaviour and smoothness within each data set.
We compare the performance of our GP-BART model to other tree-based methods, namely BART \citep{chipman2010bart}, SoftBART \citep{linero2018bayesian}, and tGP \citep{gramacy2008bayesian}, as well as the universal kriging model \citep{cressie2015statistics} and latent Gaussian models using integrated nested Laplace approximations \citep[INLA;][]{lindgren2015bayesian}. We evaluate the results using $5$ repetitions of $5$-fold cross-validation; each fold is treated as a test set and prediction accuracy and uncertainty calibration are quantified using the root-mean-square error (RMSE) and the continuous ranked {probability score} (CRPS) \cite{gneiting2007strictly}, respectively, over all folds within a given repetition.
 
\begin{figure}[H]
     \centering
     \includegraphics[width=\textwidth]{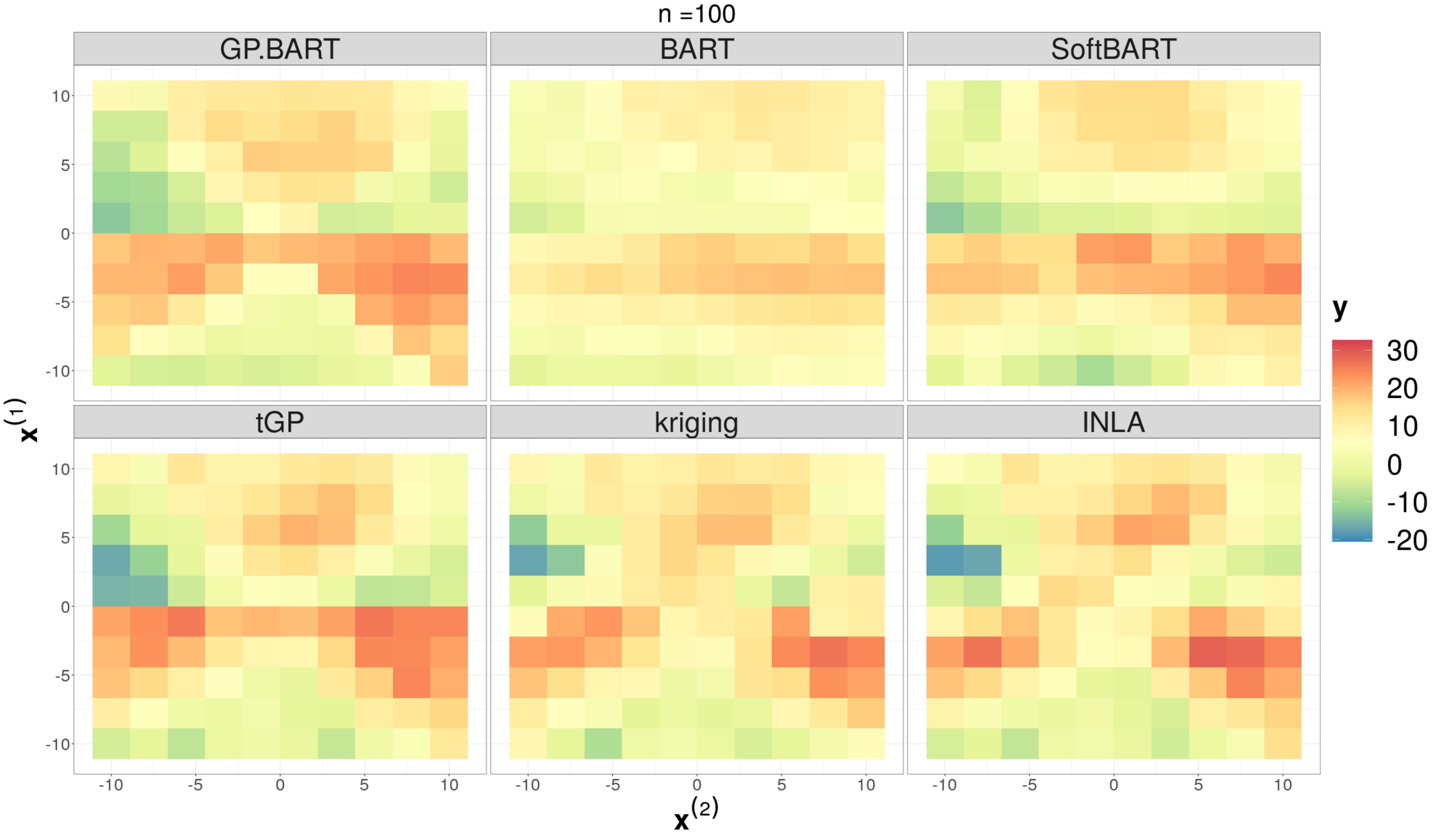}
     \caption{Predicted surfaces for the simulated scenario with $n=100$ observations from the first panel of Figure \ref{fig:simulated_data} using different methods over one randomly chosen test repetition.}
     \label{fig:sim_surface_100}
 \end{figure}

\begin{figure}[H]
     \centering
     \includegraphics[width=\textwidth]{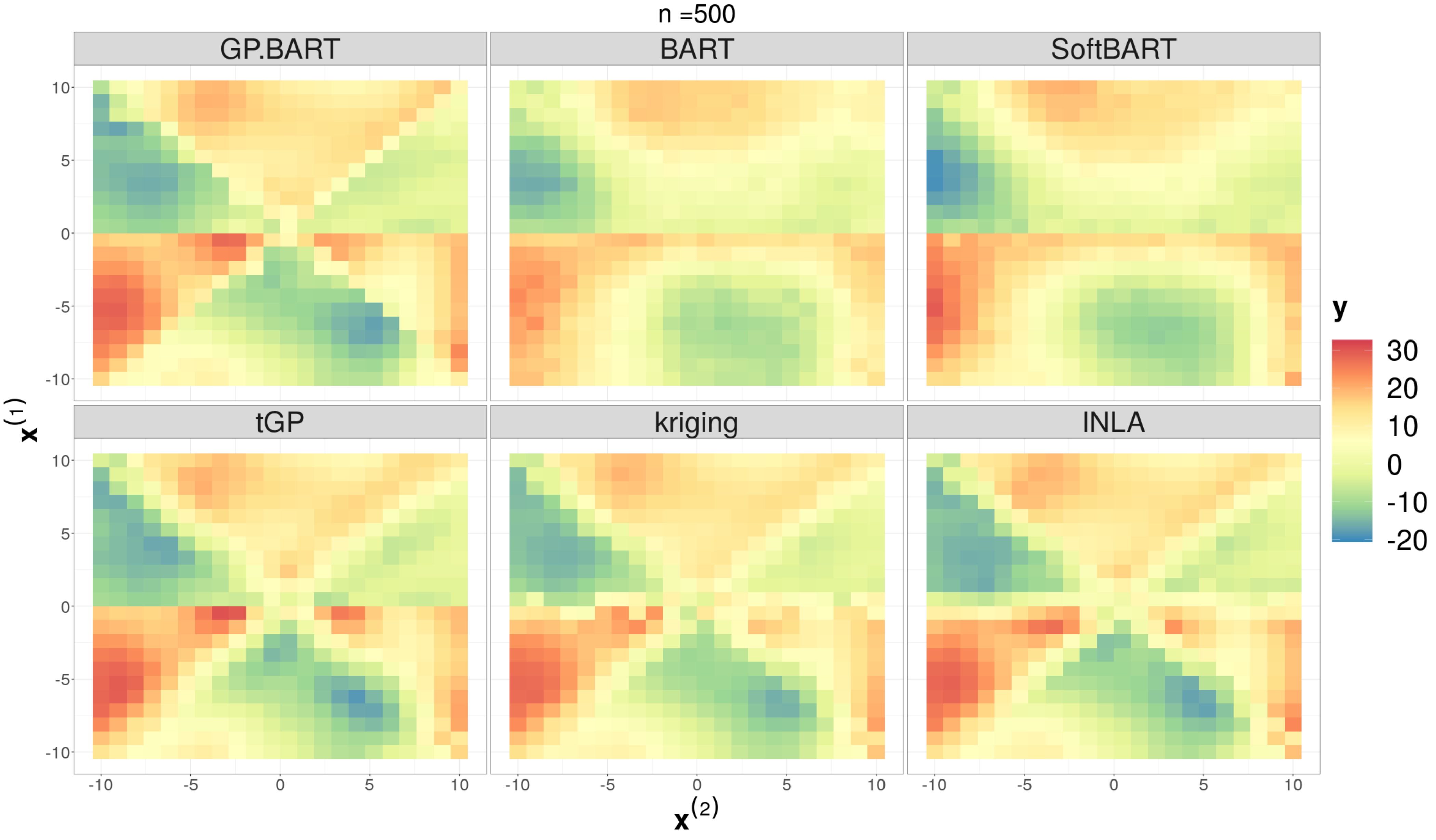}
     \caption{Predicted surfaces for the simulated scenario with $n=500$ observations from the second panel of Figure \ref{fig:simulated_data} using different methods over one randomly chosen test repetition.}
     \label{fig:sim_surface_500}
 \end{figure}

\begin{figure}[H]
     \centering
    \includegraphics[width=\textwidth]{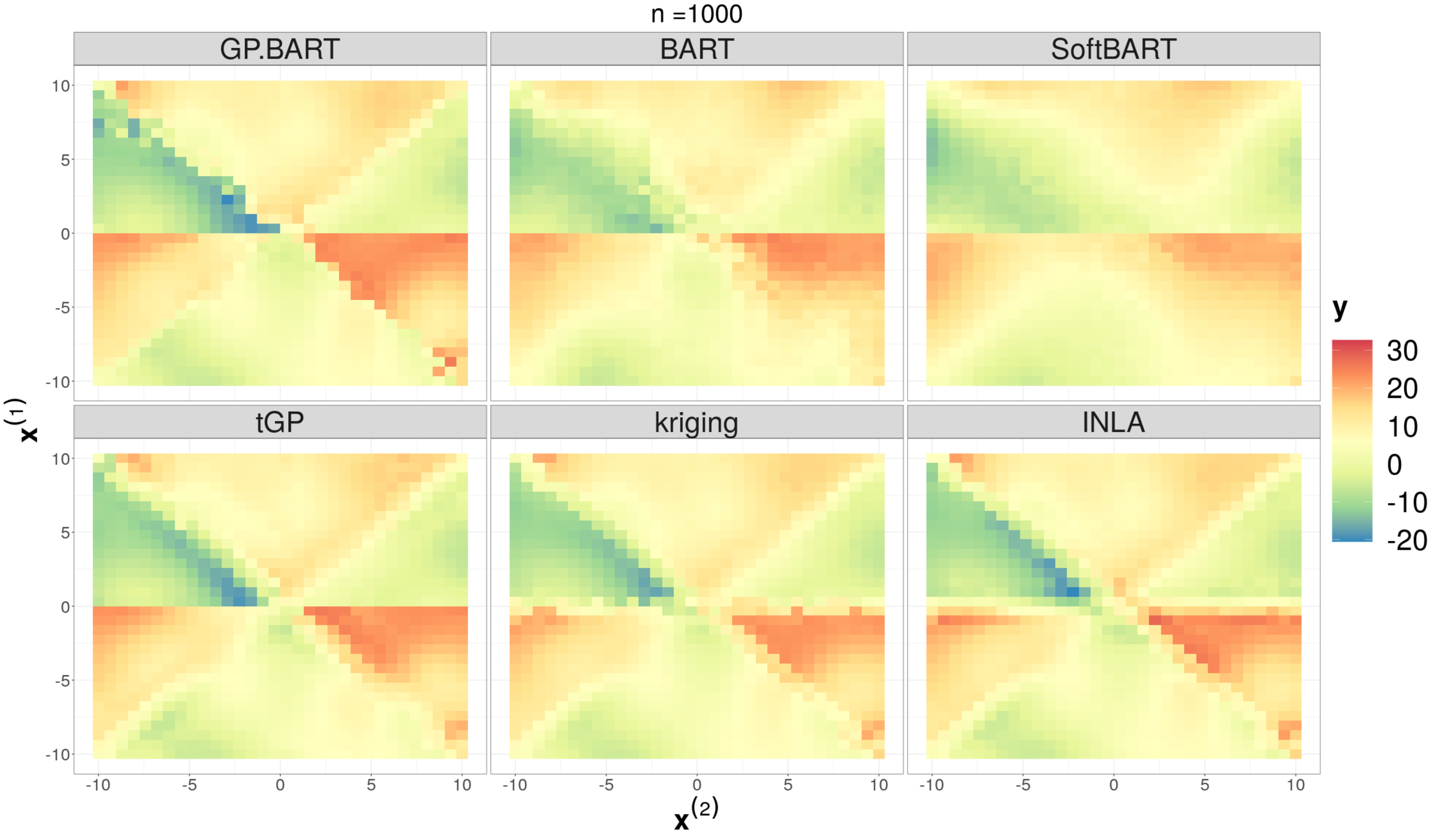}
     \caption{Predicted surfaces for the simulated scenario with $n\!=\!1000$ observations from~the third panel of Figure \ref{fig:simulated_data} using different methods over one randomly chosen test repetition.}
     \label{fig:sim_surface_1000}
 \end{figure}

The models are fitted using the \textsf{R} packages \texttt{BART} \cite{bart2021package}, 
\texttt{SoftBart} \cite{linero2022softbart},
\texttt{tgp} \cite{gramacy2010categorical}, \texttt{fields} \cite{fields2021package}, and \texttt{INLA} \cite{lindgren2015bayesian}, with their default settings. All hyperparameters for the GP-BART model were specified using their default values and settings previously described in Sections \ref{sec:gp_bart} and \ref{sec:gp_bart_sample}. To qualitatively compare the methods, we analyse the prediction surface generated by each algorithm for the data sets of size $n = \{100,500,1000\}$, shown in Figure \ref{fig:simulated_data}, using predictions over the test sets in the repeated $5$-fold setting. The corresponding plots are provided in Figures \ref{fig:sim_surface_100}, \ref{fig:sim_surface_500}, and \ref{fig:sim_surface_1000}, respectively. In each case, results from one randomly chosen repetition of the repeated $5$-fold cross-validation are used to construct the plots.
Though the provided plots indicate clear differences between each model type, each model's behaviour is similar across the sample sizes. GP-BART's prediction surfaces appear most similar to the original data shown in Figure \ref{fig:simulated_data} in each case. Indeed, GP-BART successfully identifies diagonal partitions due to its rotated splits, while BART, SoftBART, and tGP only produce splits parallel to the axes. Though BART and SoftBART uncover differences among the terminal node regions nonetheless, {their predictions are less accurate than their competitors by virtue of spatial dependence not being explicitly accounted for by these two methods. In addition, GP-BART can produce smoother surfaces than BART, as the nature of the original algorithm inherently involves the summation of stepwise-constant functions.} The tGP, kriging, and INLA predictions capture the spatial features well, but their failure to identify the partitions results in blurred prediction surfaces in areas where the data splits. {Therefore, we emphasise that the proposed model takes advantage of the benefits of rotated splits, explicitly defined spatial dependence assumptions, and the inherent smoothness from the GPs.}

\begin{figure}[H]
     \centering
     \includegraphics[width=\textwidth]{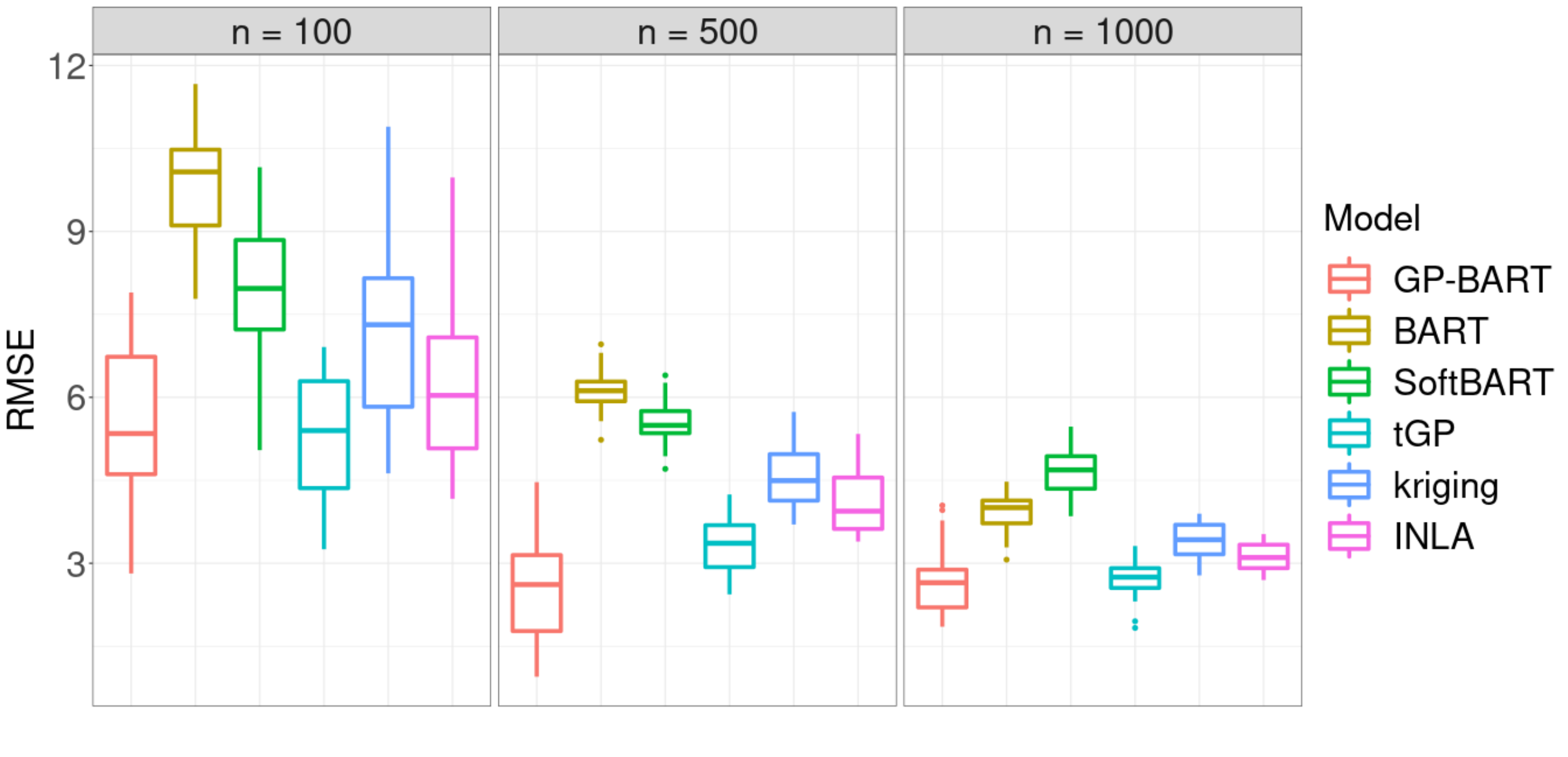}
     \caption{Comparisons between the RMSE obtained by the competing models for the simulated data using $10$-fold cross validation over different sample sizes. {These results show that GP-BART tends to deliver the lowest median RMSE, as it encompasses assumptions of spatial dependence, smoothness, and allows for rotated splits.}}
     \label{fig:sim_rmse_boxplot}
 \end{figure}

\begin{figure}[H]
     \centering
     \includegraphics[width=\textwidth]{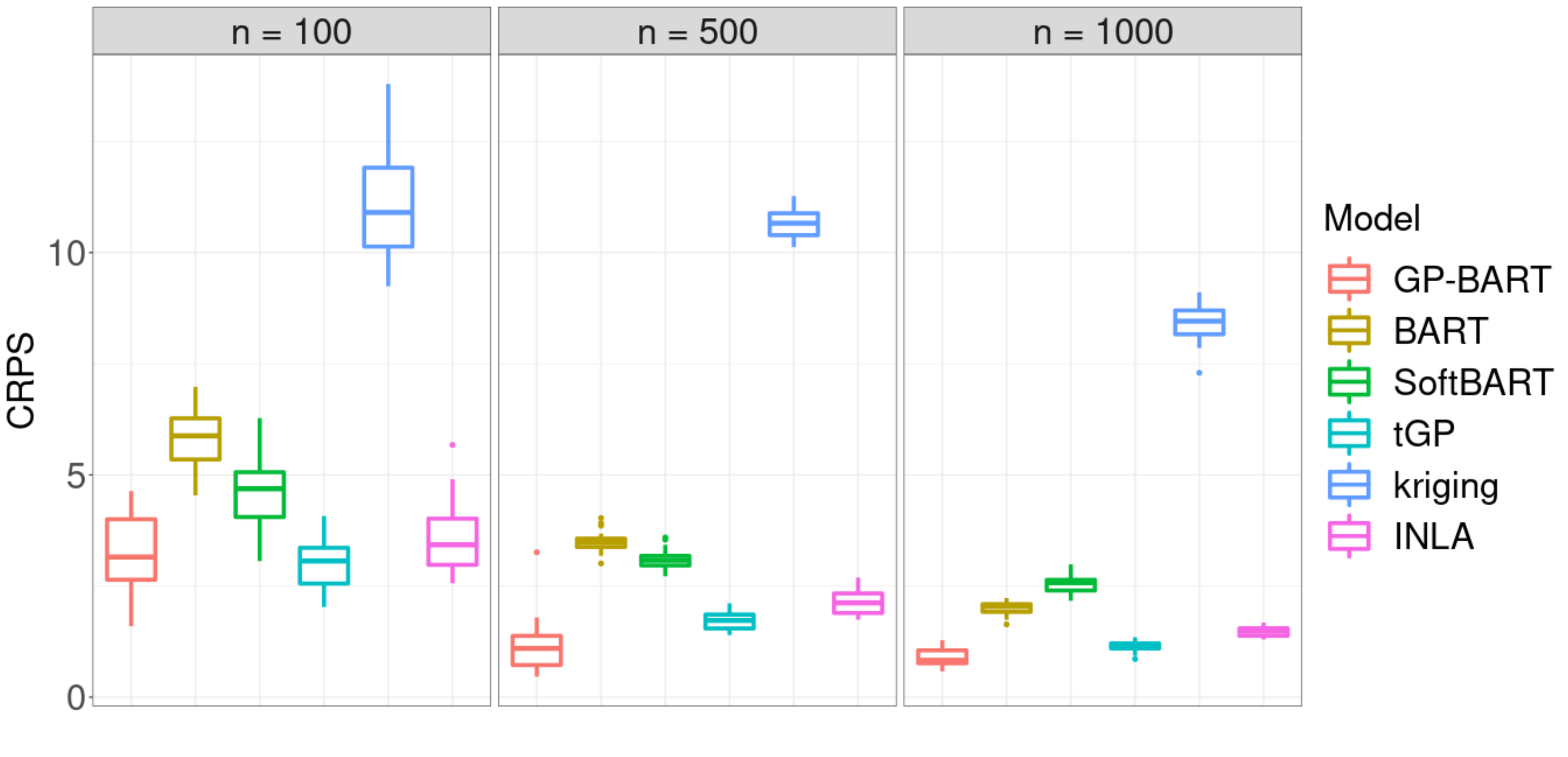}
     \caption{Comparisons between the CRPS values obtained by the competing models for the simulated data using $10$-fold cross validation over different sample sizes. {These results show that GP-BART tends to deliver the lowest median CRPS scores, as it encompasses assumptions of spatial dependence, smoothness, and allows for rotated splits.}}
     \label{fig:sim_pi_boxplot}
 \end{figure}

A quantitative comparison is shown in the boxplots in Figure \ref{fig:sim_rmse_boxplot}, which reflect the previous qualitative interpretations. Here, GP-BART presents substantially lower RMSE than its competitors, particularly for smaller $n$. We assess uncertainty calibration by examining boxplots of CRPS scores in Figure \ref{fig:sim_pi_boxplot}. These results show that the GP-BART model  presents the lowest CRPS values among all methods. Thus, considering both metrics jointly, GP-BART's performance in terms of prediction accuracy and uncertainty quantification is superior to the other models considered.

\begin{figure}[H]
    \centering
    \includegraphics[width=\textwidth]{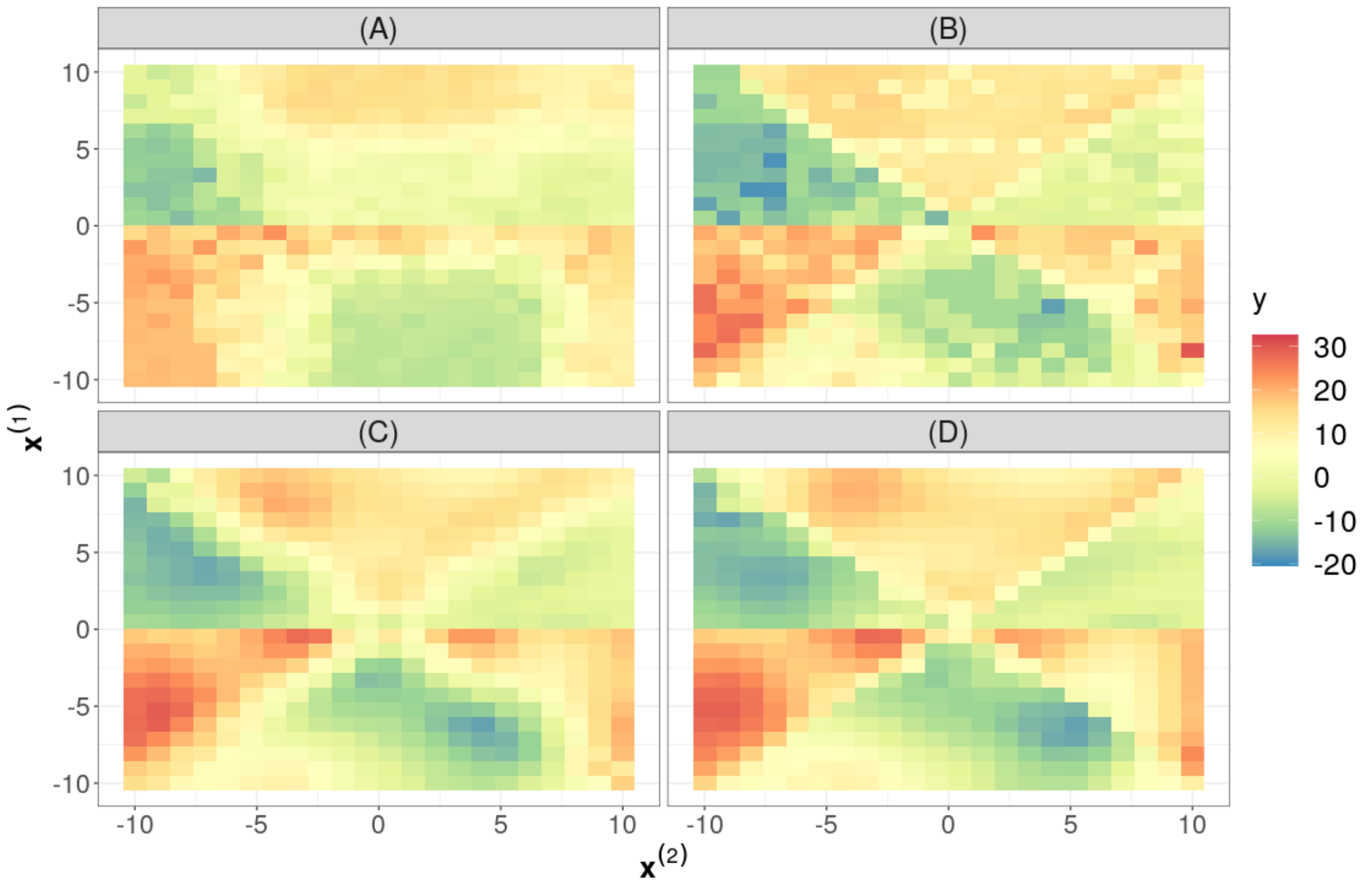}
    \caption{Comparison between the predicted surfaces under the different versions of GP-BART for the $n=500$ simulated data over one randomly chosen test repetition. The surface for (\textbf{D}), the standard version of GP-BART, is qualitatively close to the observed data in the second panel of Figure \ref{fig:simulated_data}.}
     \label{fig:surface_paramter_gpbart_500}
\end{figure}

\begin{figure}[H]
    \centering
    \includegraphics[width = \textwidth]{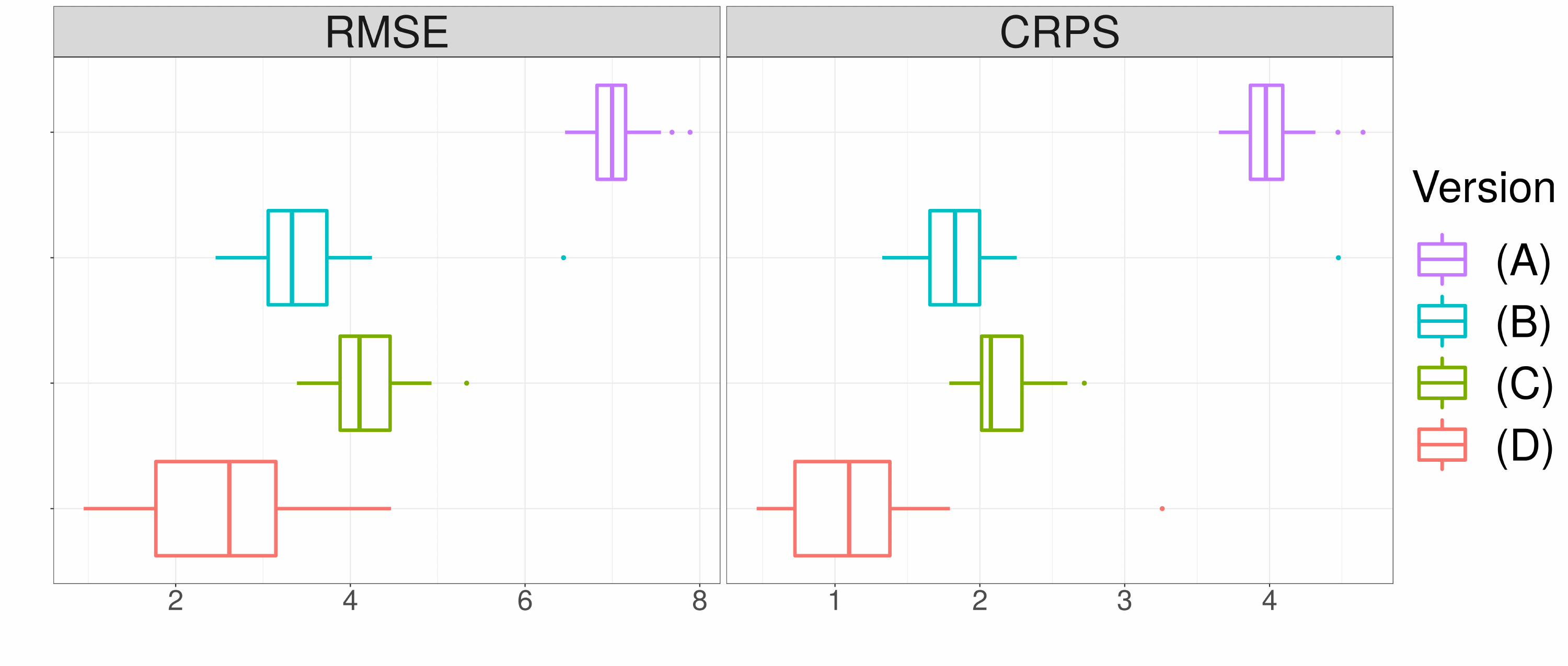}
    \caption{Boxplots of the RMSE  (left) and CRPS (right) values across the different versions of the GP-BART model for the $n=500$ simulated data. The standard GP-BART (\textbf{D}) has the best performance in terms of both RMSE and calibration.\label{fig:both_rmse_compare_gpbart_paramter_500}}
\end{figure}

To highlight the effect of the proposed moves and the use of GPs over the terminal nodes, four different, restricted versions of GP-BART are~compared:\smallskip
\begin{enumerate}
    \item[(\textbf{A})] without any rotated moves or GPs (i.e., the standard BART model);
    \item[(\textbf{B})] without GPs, but with the new rotated `grow' and `change' moves;
    \item[(\textbf{C})] without the new rotated moves, but with GPs;
    \item[(\textbf{D})] the standard GP-BART with both rotated splitting rules and GPs.
\end{enumerate}
We defer the results for the other sample sizes, which lead to similar conclusions, to \ref{sec::appendix_c}, along with an evaluation of the acceptance rates for the tree-proposal moves under version (\textbf{D}), and consider only the $n=500$ setting here,
for brevity. This comparison is summarised~in Figures \ref{fig:surface_paramter_gpbart_500} and \ref{fig:both_rmse_compare_gpbart_paramter_500}, in which the letters above are used to distinguish the model versions. As before, results based on one randomly chosen repetition of the $5$-fold cross-validation are used to construct Figure \ref{fig:surface_paramter_gpbart_500}.
The prediction surface (\textbf{A}) in Figure \ref{fig:surface_paramter_gpbart_500} suggests BART cannot adequately capture different behaviours in the terminal node regions due to the lack~of smoothness and non-linearity compared with GP-BART. Panels (\textbf{B}) and (\textbf{C}) both compare reasonably well with (\textbf{D}), which highlights the benefits of the rotated split rules and use of GPS, respectively. However, there is an apparent lack of smoothness in the terminal node regions of (\textbf{B}), and visible blurriness in the areas where the data splits in (\textbf{C}). Ultimately, it is evident that combining both innovations in (\textbf{D}) yields the best performance. 

This conclusion is reinforced by Figure \ref{fig:both_rmse_compare_gpbart_paramter_500}, which indicates the superior performance of version (\textbf{D}). Despite the larger variance in the RMSE, the standard GP-BART obtains the lowest median value of both metrics shown. \mbox{Versions} (\textbf{B}) and (\textbf{C)}, which respectively incorporate rotated splits and GPs, yield similar yet slightly inferior RMSE values to (\textbf{D}), but their performance in terms of uncertainty calibration as measured by median CRPS differs more substantially. Notably, the standard BART model (\textbf{A}) is unsatisfactory from both points of view. 

\subsection{Friedman data}\label{sec:friedman}

In this scenario, we consider the Friedman equation \cite{friedman1991multivariate}:
\[y_{i} = 10\,\sin\left(\pi x_i^{(1)} x_i^{(2)}\right) + 20\left(x_i^{(3)} - 0.5\right)^{2} + 10 x_i^{(4)} + 5 x_i^{(5)} + \epsilon_{i},\:i = 1, \ldots, n,\]
where $x_i^{(j)} \sim \mbox{Unif}(0,1)\;\forall\: j = 1, \ldots, p$ and $\epsilon_{i} \sim \mbox{N}(0, \tau^{-1})$. This equation is used for benchmarking tree-based methods using synthetic data, and has been examined in many other papers, e.g., \cite{chipman2010bart, linero2018bayesian}. For these data, we compare GP-BART to its explicitly tree-based competitors, namely BART, SoftBART, and tGP. Though there are no spatial features here, we still anticipate that incorporating GPs and rotated splits will help as there are non-linear smooth interactions in these data.

\begin{figure}[H]
    \centering
    \includegraphics[width = \textwidth]{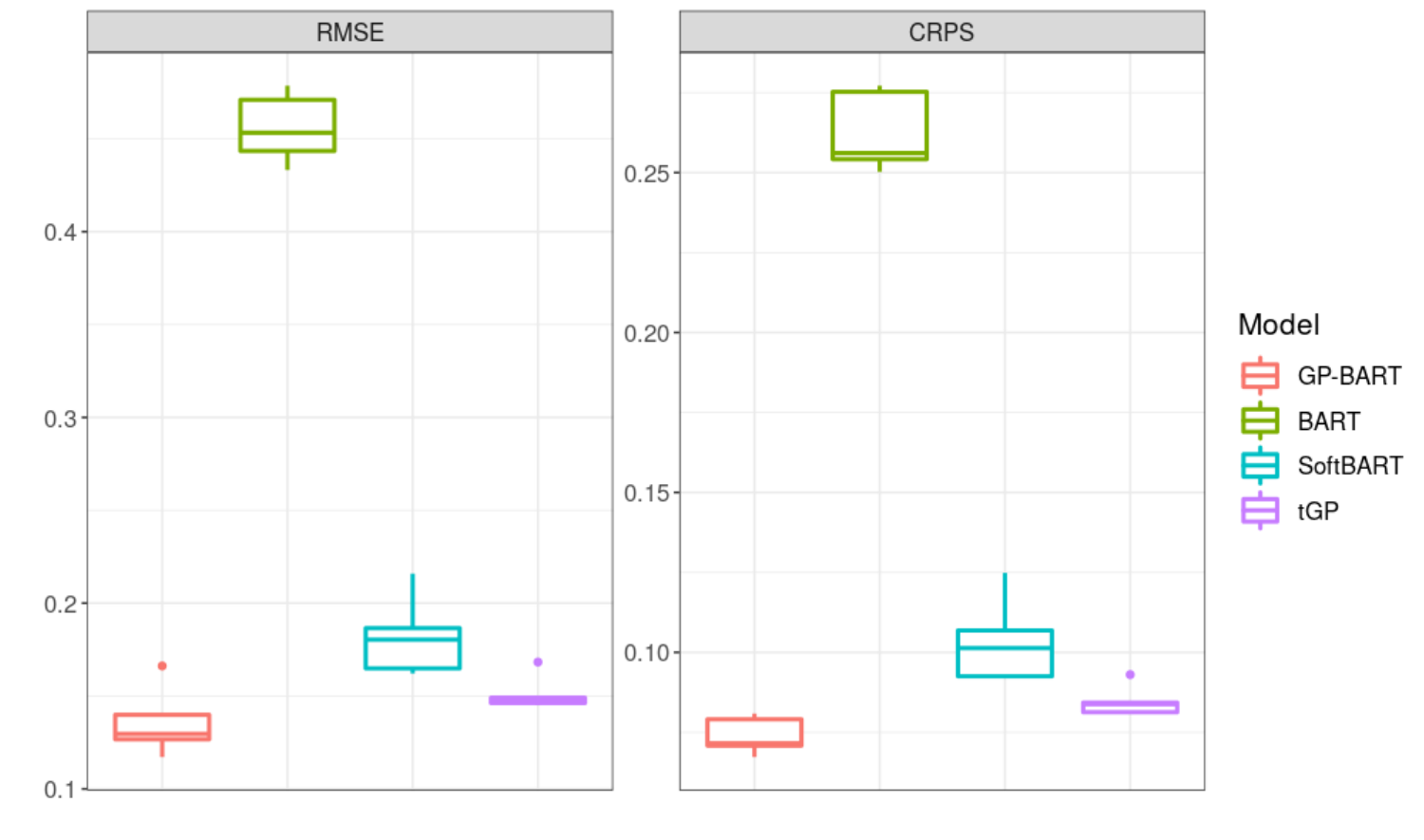}
    \caption{Comparison of the RMSE and CRPS over the test set in the $25$ folds from $5$ repetitions of $5$-fold cross-validation for the Friedman data set with $n=500$ and $p=5$.}
    \label{fig:friedman500_nonoise}
\end{figure}

\begin{figure}[H]
    \centering
    \includegraphics[width = \textwidth]{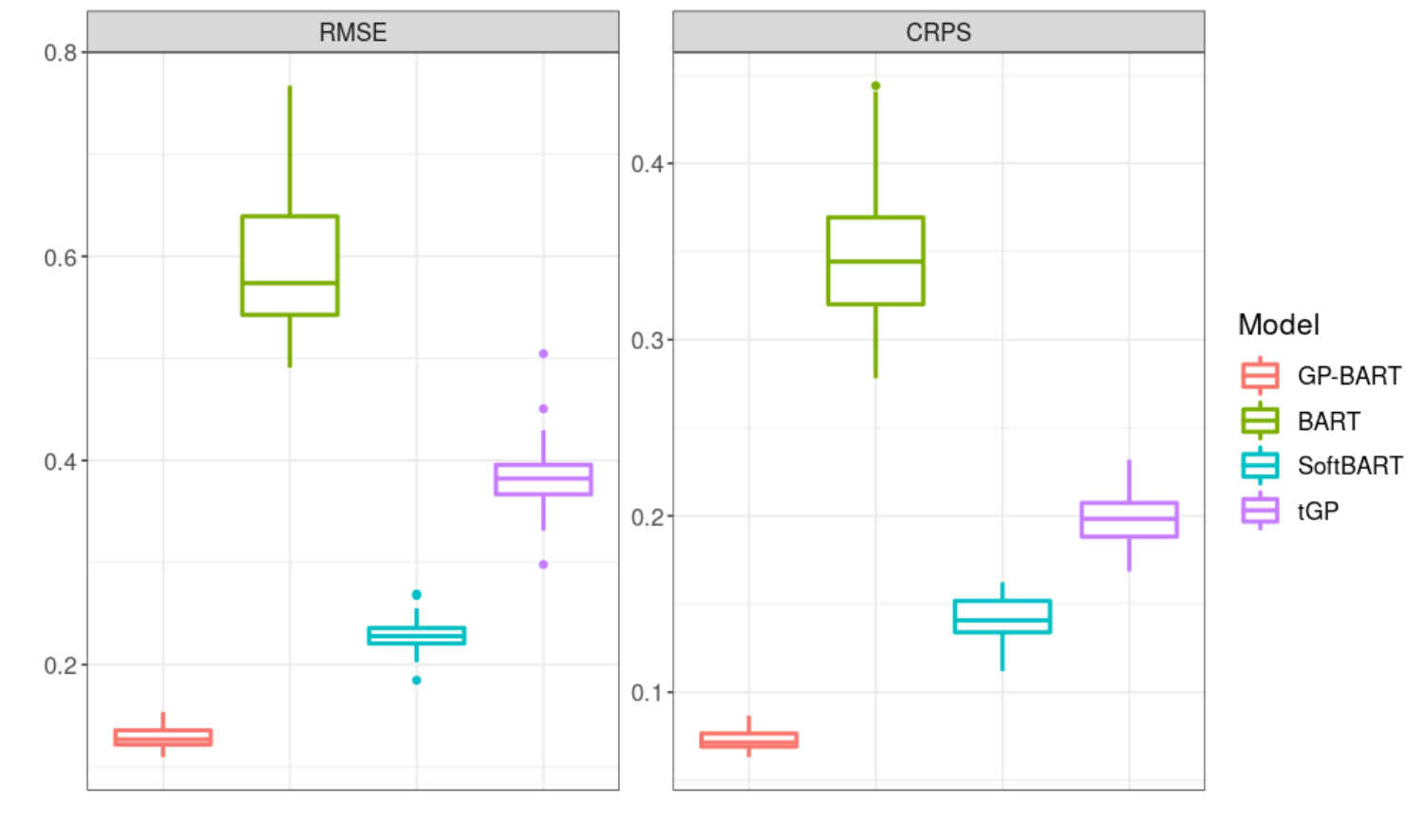}
    \caption{Comparison of the RMSE and CRPS over the test set in the $25$ folds from $5$ repetitions of $5$-fold cross-validation for the Friedman data set with $n=500$ and $p=10$, i.e. with $5$ additional noise variables.}
    \label{fig:friedman500}
\end{figure}

Here, we specify $\tau=100$, $n=500$, and consider two versions of the same data; firstly with $p=5$ and secondly with $p=10$ features, of which the first $5$ are those from the first scenario. As the Friedman equation uses only $5$ covariates to generate the response, the additional five predictors in the second scenario are uninformative noise variables with no effect on~$y_i$. Figure \ref{fig:friedman500_nonoise} shows that GP-BART outperforms the other methods and presents good performance in terms of predictive accuracy and uncertainty calibration, using the RMSE and CRPS metrics as above. Subsequently, Figure \ref{fig:friedman500} shows the same comparison, this time with the additional $5$ noise variables.
\vspace{-1em}\enlargethispage{\baselineskip}
The latter comparison with additional noise variables in Figure \ref{fig:friedman500} is also favourable to GP-BART. In particular, these results show that the uninformative variables do not have a detrimental effect on its performance. This can be attributed to the {discrete prior} assumed for the $\phi_{tj}$ parameters automatically diminishing their influence on the kernels of the GPs. Conversely, the adverse effects of such variables on the RMSE and CRPS values under BART, SoftBART, and tGP are more readily apparent, when one compares Figure \ref{fig:friedman500_nonoise} and Figure \ref{fig:friedman500}. The deterioration is especially notable for tGP.

Table \ref{tab:phi_min_table} further demonstrates the effectiveness of the ARD by examining mean values of the minimum values of $\phi_{tj}$ for each variable (in both the $p=5$ and $p=10$ scenarios) over all trees and all accepted MH proposals in the retained posterior samples across each repetition of $5$-fold cross-validation. In the first scenario ($p=5$), the fourth and fifth variables, which are merely related linearly to the response, are shown to be associated with moderately higher values. In the $p=10$ scenario, the model selects substantially larger values for the $5$ extra noise variables which are unrelated to the response, whereas small values are selected for all of the informative predictors, such that they contribute meaningfully to the GPs.\vspace{-1ex}\enlargethispage{0.5\baselineskip}
\begin{table}[H]
\centering
\caption{Means and standard deviations (in parentheses) of the minimum value for $\phi_{tj}$ for each variable over all trees and all accepted MH proposals in the retained posterior samples across each repetition of $5$-fold cross-validation on the Friedman data set. The first row shows the $p=5$ scenario and the subsequent rows show those same $5$ variables and the $5$ additional noise variables in the $p=10$ scenario.}
\label{tab:phi_min_table}
\begin{tabular}{lccccc}
\hline
\textbf{Friedman data} & \multicolumn{5}{c}{\textbf{Mean (Standard Deviation)}}\\
\hline
\multirow{2}{*}{\shortstack[l]{Without noise\\($p=5$)}} & $0.45$ & $0.45$ & $0.20$ & $0.49$ & $0.55$\\
&($0.13$) & ($0.13$) & ($0.17$) & ($0.08$) & ($0.16$)\\
\hline
\multirow{4}{*}{\shortstack[l]{With noise\\($p=10$)}} & $0.46$ & $0.42$ & $0.39$ & $0.50$ & $0.58$\\
&($0.12$) & ($0.16$) & ($0.18$) & ($0.01$) & ($0.18$) \\
&$47.4$ & $47.2$ & $47.7$ & $48.2$&$43.2$\\
&($10.78$)&($11.09$) & ($10.06$) & ($8.98$) & ($16.93$)\\
\hline
\end{tabular}
\end{table}

\subsubsection{Computational performance and cost considerations}
\label{sec::computationalcost}

While GP-BART exhibits superior performance compared to its tree-based competitors, it is important to acknowledge its additional computational costs. This chiefly arises from its composition as a sum of GPs, which incurs a computational complexity of $\mathcal{O}(n_{t\ell}^3)$ within each terminal node. Although the model shows favorable outcomes in the simulations presented thus far, it is important to weigh this against computational efficiency. While the the aforementioned $\mathcal{O}(n_{t\ell}^3)$ costs can be reduced by encouraging deeper trees \emph{a priori}, the relevant hyperparameters of Equation \eqref{eq:nonterminal_prob} should be handled with care. We continue to adopt the default values of $\alpha=0.95$ and $\beta=2$ as modifying them can decrease run times but comes at the expense of worse predictive performance. See \ref{sec::appendix_d} for further details.

To asses the algorithm's computational demands, the \texttt{microbenchmark} \textsf{R} package \citep{micropackage} was used to obtain accurate measurements of the run times for GP-BART, tGP, and SoftBART, with five replications for each method. These competing treed models were specifically chosen due to their substantial computational requirements, and both were applied using their default settings. For GP-BART, our own \textsf{R} package based on C\texttt{++} code was used. All computations were performed using \textsf{R} version 4.2.1 on a MacBookPro laptop, equipped with a 2.3 GHz Dual-Core Intel Core i5 processor and 8GB of RAM. The experiments were conducted on the Friedman data set with noise variables (i.e., $p = 10$), while varying the training sample size ($n_{tr}$) among $\{50, 100, 500\}$ and keeping the testing sample size fixed at $n_{te}=50$.\enlargethispage{\baselineskip}

The findings are summarized in Table \ref{tab:computational_cost}, which shows that both GP-BART and tGP experience a rapid escalation in computational time as the training sample size ($n_{tr}$) increases. Notably, GP-BART exhibits the highest computational burden in the comparison. Indeed, the run times with $n_{tr}=500$ suggest that GP-BART would need to be run on a dedicated machine or server for feasible modelling of larger datasets. However, it is noteworthy that despite the greater run times required by GP-BART, its timings remain comparable to those of tGP, particularly when considering the ratio of GP-BART's timings to the number of trees ($T=20$ in its default setting).

\begin{table}[H]
\centering
\caption{Computational time statistics for the $p=10$ Friedman data in seconds, across five runs of each implementation for GP-BART and two tree-based competitors.\label{tab:computational_cost}}
\begin{tabular}{llrrr}
\hline
\textbf{Method} & \textbf{Metric} & $\mathbf{n_{tr}=50}$ & $\mathbf{n_{tr}=100}$ & $\mathbf{n_{tr}=500}$\\
\hline
\multirow{3}{*}{\textbf{GP-BART}} & Min. & $150.8$ & $442.9$ & $37633.9$ \\
 & Mean & $167.1$ & $458.4$ & $39360.5$ \\
 & Max. & $176.4$ & $482.2$ & $40187.2$ \\
\hline
\multirow{3}{*}{\textbf{tGP}} & Min. & $5.1$ & $20.6$ & $2062.8$ \\
 & Mean & $6.2$ & $21.6$ & $2119.9$ \\
 & Max. & $6.5$ & $23.8$ & $2177.0$ \\
\hline
\multirow{3}{*}{\textbf{SoftBART}} & Min. & $10.7$ & $12.5$ & $43.2$ \\
 & Mean & $12.8$ & $15.8$ & $44.0$ \\
 & Max. & $13.6$ & $21.7$ & $44.8$ \\
\hline
\end{tabular}
\end{table}
 
\section{Applications}\label{sec:real_data}

In this Section, we appraise the predictive performance of GP-BART compared to BART, SoftBART, tGP, kriging, and INLA on diverse real data sets, as a larger and more challenging test of GP-BART's capabilities. For illustration, we use four public data sets containing spatial features; i.e., with inherent dependence over the observations. These data sets are:\medskip

\begin{enumerate}
    \item The \textit{Auckland} data; consisting of $166$ {observations} describing infant mortality in Auckland, with two spatial covariates and the target variable \citep{bivand2018spdep}.\medskip
    \item The \textit{Baltimore} data; comprising $221$ observations of house sales prices, two spatial features, and $13$ other covariates, not all of which are continuous \citep{bivand2018spdep}.\medskip
    \item The \textit{Boston} data; containing $506$ observations of the median values of owner-occupied suburban homes, two spatial features, and $13$ other covariates, not all of which are continuous. We model a corrected version \cite{boston_housing_correct} of the original data \cite{boston_housing}.\medskip
    \item \textit{Swmud}; a data set of seabed mud content in the southwest Australia Exclusive Economic Zone with $177$ observations of two sets of spatial coordinates and mud content as the target variable \cite{li2010predicting}.\medskip
\end{enumerate}%
Our implementations of each algorithm follow their respective default~\mbox{settings}, including those previously described in Sections \ref{sec:gp_bart} and \ref{sec:gp_bart_sample} for GP-BART. As before, $5$ repetitions of $5$-fold cross-validation are used to evaluate~\mbox{performance}. Categorical features cannot be formally accommodated in the GPs under the present parameterisation of GP-BART's kernel function. Hence, for the \textit{Baltimore} and \textit{Boston} data sets, we restrict the GPs to include the continuous and integer-valued covariates only. However, categorical features are still used to form splitting rules for GP-BART, as described in Section \ref{sec:tree_structure}. All other methods accommodate categorical features using dummy variable representations. In each case, the strictly spatial continuous features represent the exact coordinates of the instances.

\begin{figure}[H]
     \centering
     \includegraphics[width=\textwidth]{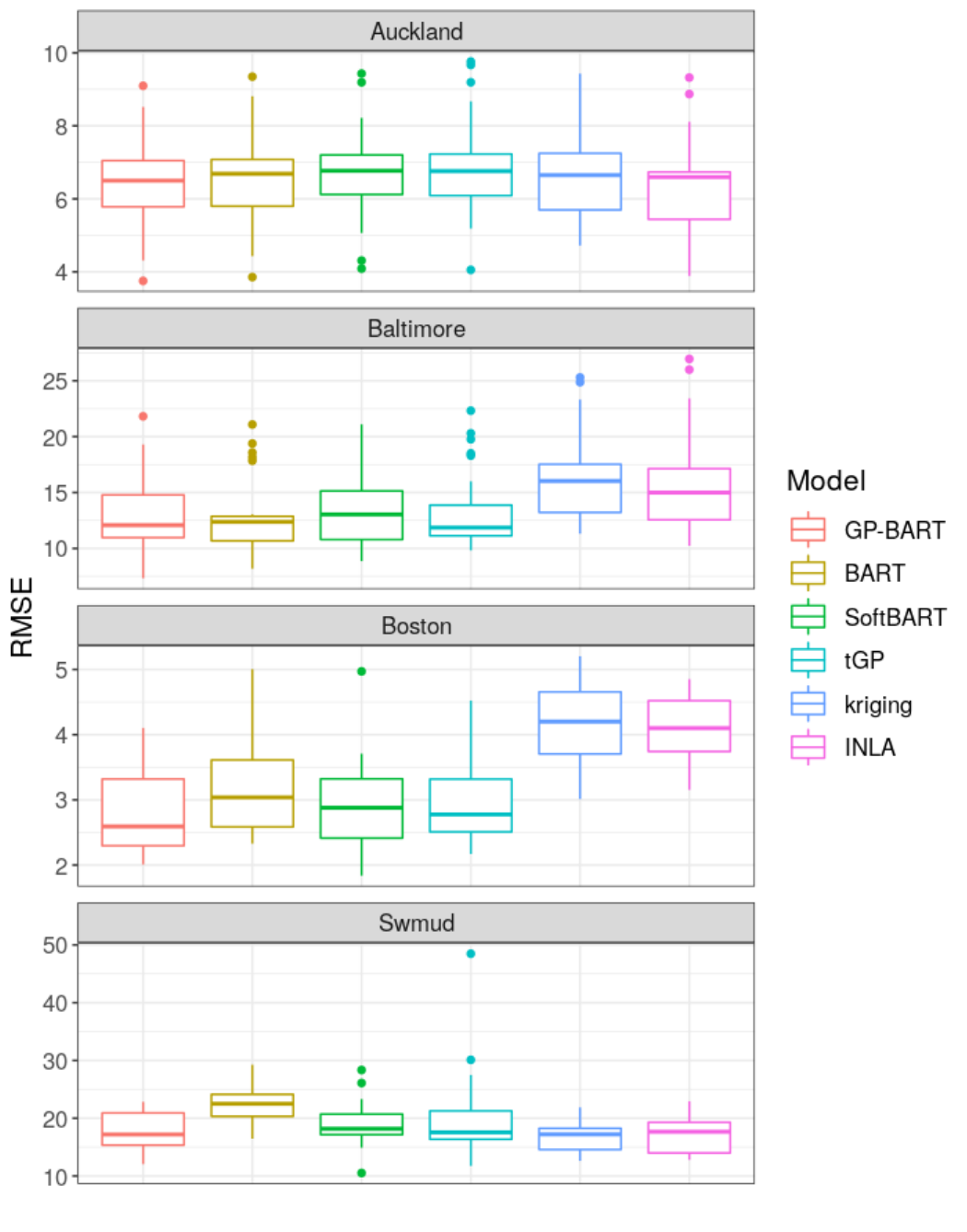}
     \caption{Comparison between the RMSE values for the benchmarking data sets across the six competing methods using $5$ repetitions of $5$-fold cross-validation.}
     \label{fig:rmse_bench}
 \end{figure}

\begin{figure}[H]
     \centering
     \includegraphics[width=\textwidth]{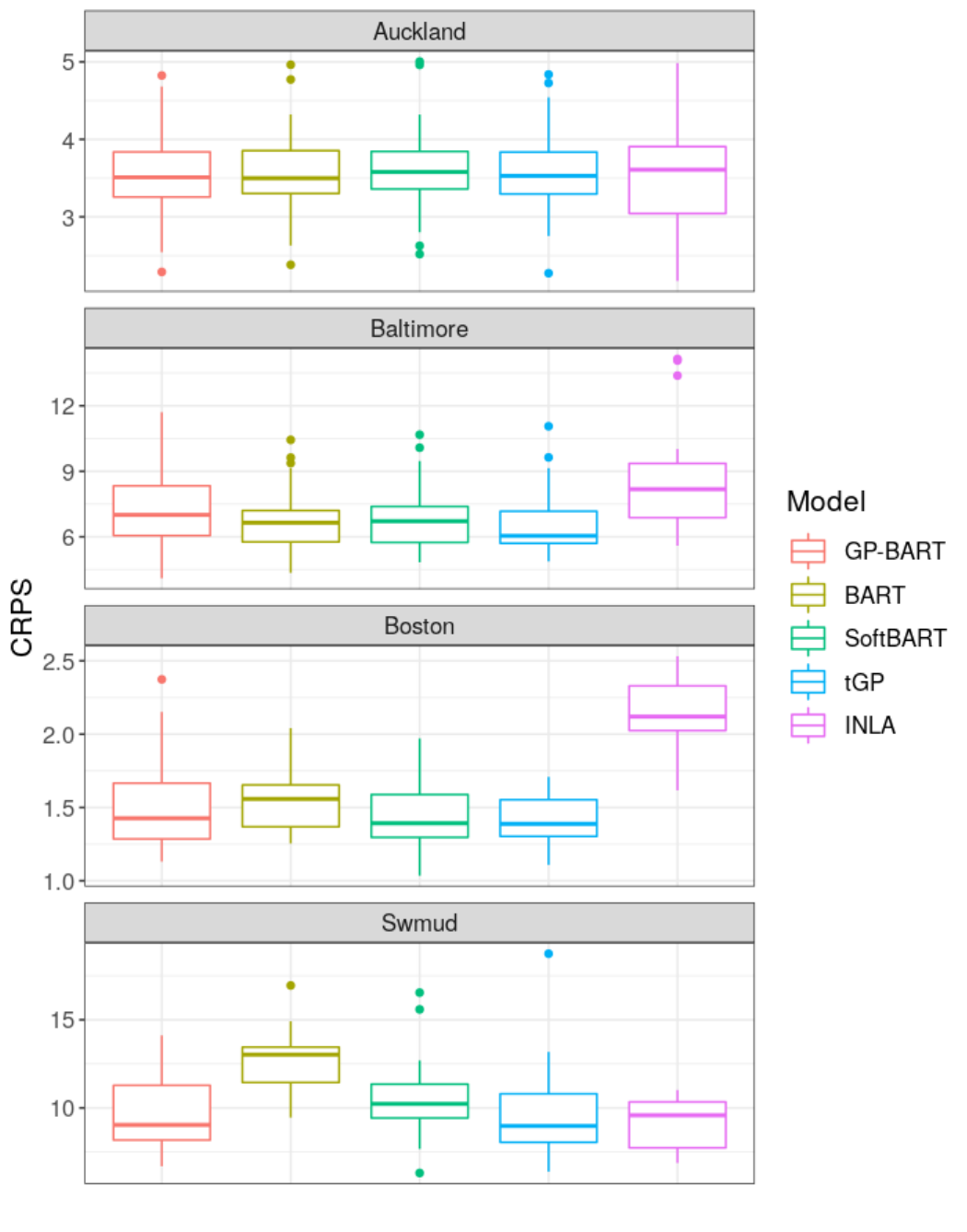}
     \caption{Comparison between CRPS values for the benchmarking data sets across five of the six competing methods using $5$ repetitions of $5$-fold cross-validation.}
     \label{fig:crps_bench}
 \end{figure}

The results are summarised in Figure \ref{fig:rmse_bench} and Figure \ref{fig:crps_bench}, which show the RMSE and CRPS, respectively, for each data set over all folds. According to Figure \ref{fig:rmse_bench}, GP-BART presents the lowest median RMSE for the \textit{Auckland}, \textit{Boston}, and \textit{Swmud} data sets. The difference is most pronounced for the \textit{Boston} data, for which kriging and INLA perform notably worse than all~tree-based methods. For the \textit{Baltimore} data, it ranks second among all methods.

Figure \ref{fig:crps_bench} shows that the CRPS values produced by GP-BART are similarly favourable when compared with the performance of the other algorithms, with GP-BART having the lowest or second-lowest median CRPS values for all but the \textit{Baltimore} data set. Note that boxplots of the CRPS values for kriging are omitted from Figure \ref{fig:crps_bench} for the sake of visual clarity, as they are well outside the range of those for the other models in the comparison. Jointly considering both the predictive accuracy and the uncertainty calibration, GP-BART was able to consistently yield superior or competitive predictions.

\begin{figure}[H]
     \centering
     \includegraphics[width=\textwidth]{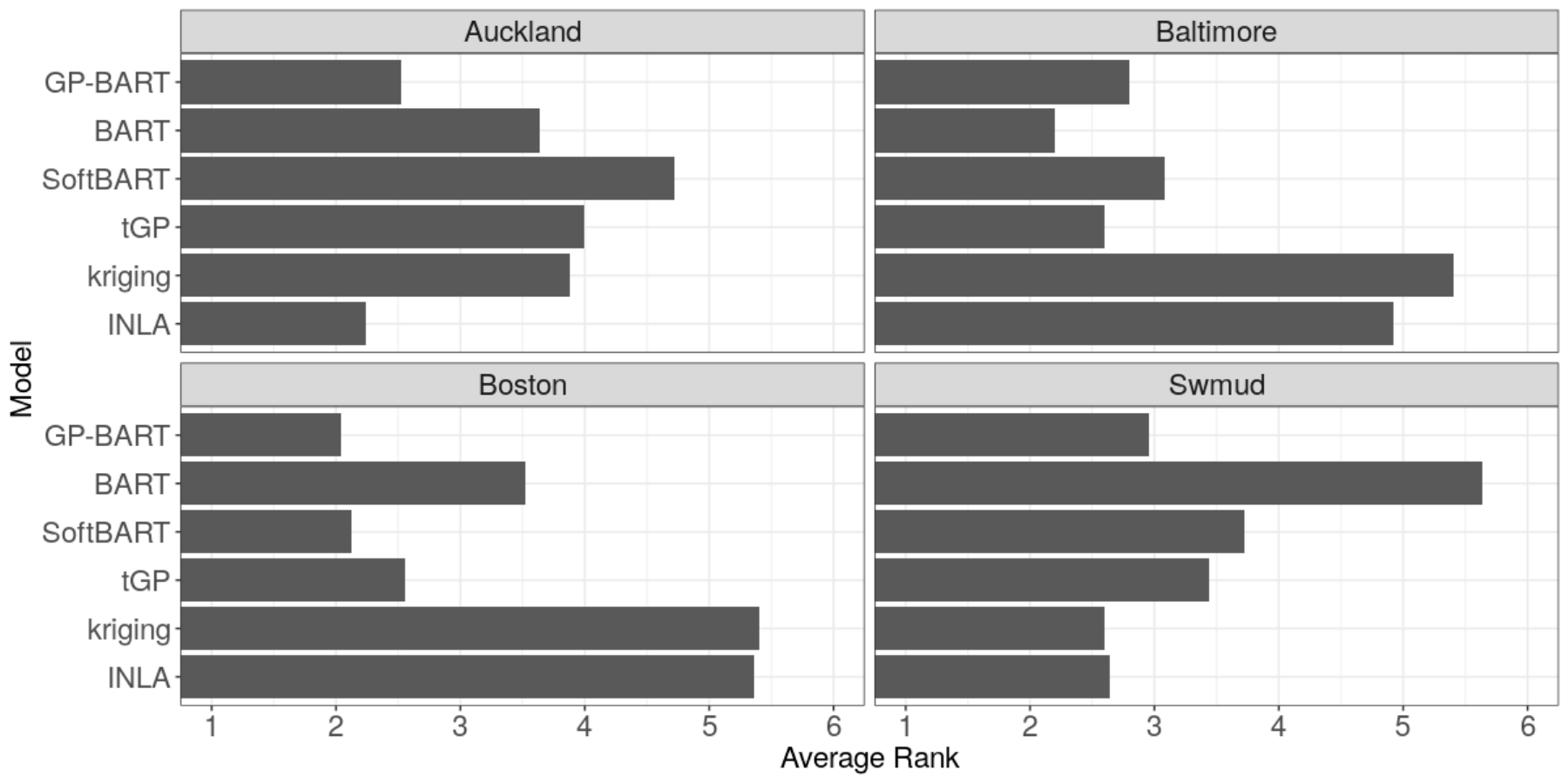}
     \caption{RMSE ranks for all six competing models over the four benchmark data sets, averaged over all five repetitions of the $5$-fold cross validation. The ranks range from $1$ to $6$, with lower ranks being associated with lower mean RMSE values.}
     \label{fig:rank_rmse}
 \end{figure}

Given the variability in these boxplots, another aspect of performance evaluation for each model and data set is illustrated in Figure \ref{fig:rank_rmse}, which presents the average RMSE rank for each of the $25$ test partitions from the repeated cross-validations. Ranks are defined here such that the model yielding the lowest mean RMSE is given a rank of $1$,~while~the one with the worst prediction performance is given the~highest possible rank of $6$, for each test partition. From Figure \ref{fig:rank_rmse}, we can see that GP-BART has  the lowest average RMSE rank for the \textit{Boston} data set, particularly compared to the standard BART model. For the \textit{Auckland} data, INLA's performance in this regard is also the best followed right after by GP-BART, where both jointly outperform the other methods. For the \textit{Swmud} data, GP-BART presents the lowest average ranking among treed methods, losing only to the traditional spatial methods. Finally, GP-BART's performance on the \textit{Baltimore} data is competitive with respect to other methods based on trees and GPs and superior to the traditional spatial methods.

\begin{figure}[H]
     \centering
     \includegraphics[width=\textwidth]{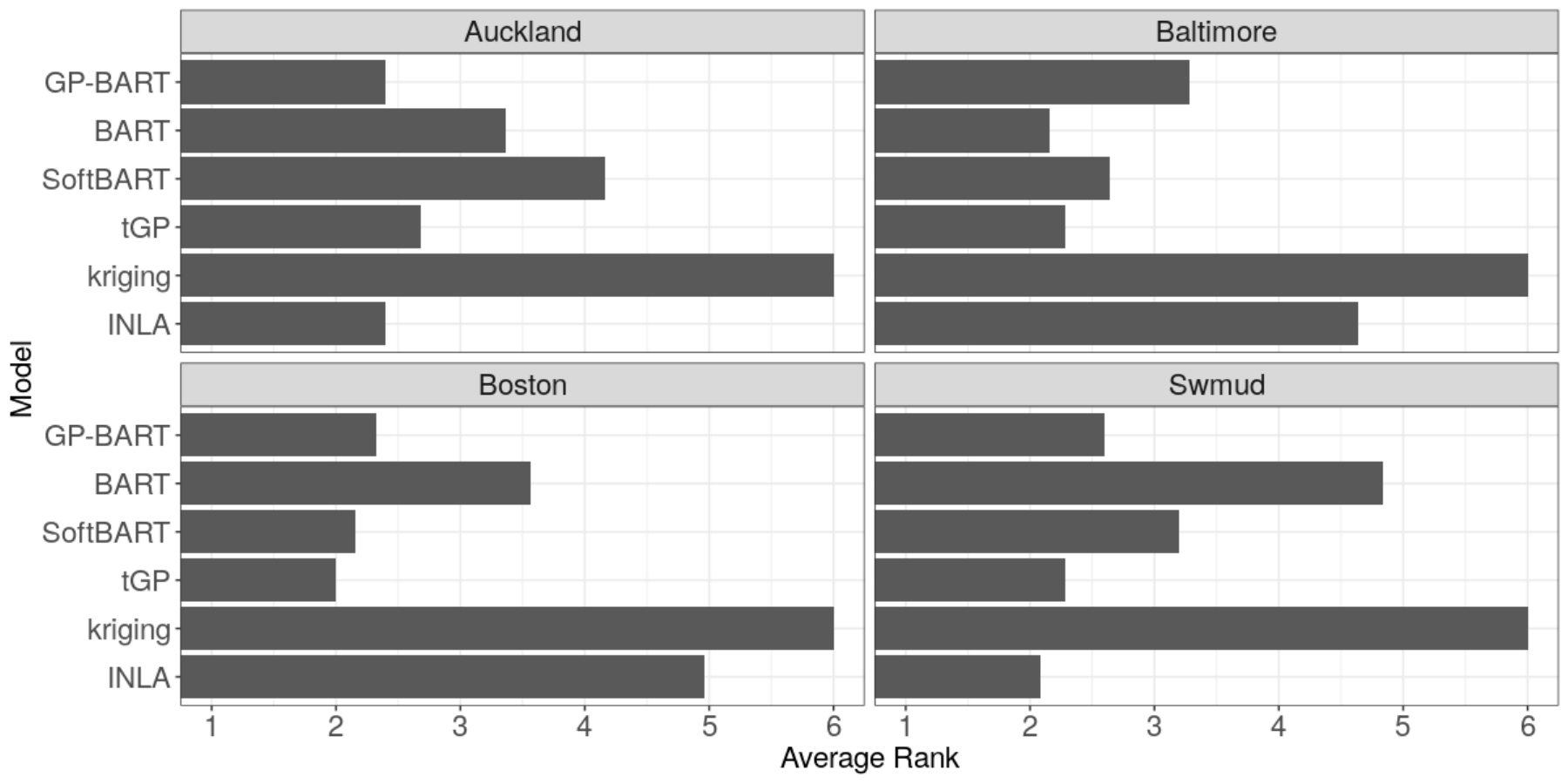}
     \caption{CRPS ranks for all six competing models over the four benchmark data sets, averaged over all five repetitions of the $5$-fold cross validation. The ranks range from $1$ to $6$, with lower ranks being associated with lower mean CRPS values.}
     \label{fig:rank_crps}
 \end{figure}

Figure \ref{fig:rank_crps} also relies on average ranks, though here using CRPS as the metric of comparison in order to evaluate uncertainty quantification. As per Figure \ref{fig:rank_rmse}, GP-BART performs best among the treed methods for the \textit{Auckland} data and performs better than the traditional spatial methods for the \textit{Baltimore} data. For the \textit{Boston} data, SoftBART and tGP surpass all other methods, but GP-BART achieves the next-lowest mean rank. Finally, GP-BART remains competitive for the \textit{Swmud} data, notably outperforming the standard BART. Following its omission from Figure \ref{fig:crps_bench}, kriging's CRPS performance is by far the worst for each data set.

\section{Discussion}\label{sec:discussion}

In this paper, we proposed GP-BART as an extension to the standard BART model. We used Gaussian processes (GPs) to make observation-specific predictions at the terminal node level, and thus are able to capture non-linear relations and spatial dependence through the covariance structure of the GPs. In addition, our novel model allows the use of rotated splitting rules to build rotated partitions, which enable more flexibility in the tree representations.

The performance of GP-BART was evaluated over a number of simulated scenarios, where the model outperformed BART, restricted versions of GP-BART itself without the use of GPs and/or novel rotated splitting rules, and another unrelated BART extension. Our benchmarking studies also highlighted GP-BART's superior performance relative to some spatial models, namely basic kriging and INLA. Our second simulation setting, using data generated according to the well-known Friedman equation, without explicit spatial components, was also favourable to GP-BART over other tree-based methods. In particular, these results demonstrated GP-BART's insensitivity to the inclusion of noise variables through the use of ARD.

When tested on real applications, using out-of-sample data via $5$ repetitions of $5$-fold cross-validation, GP-BART displayed competitive predictive capabilities, beating many of the established methods. We also compared the calibration properties of our method using CRPS; again, GP-BART performed as well or better than competing methodologies. Overall, in terms of predictive accuracy and uncertainty quantification, GP-BART consistently showed promising performance from both perspectives.

There are several potential issues remaining with the model and the sampling algorithm, which may provide opportunities for future research {and further performance improvements}:
\begin{itemize}
    \item Careful choices have been made regarding the specification of prior distributions for the model parameters because the trees and the GPs can compete to explain the variability in the data. We have endeavoured to set sensible default parameters throughout.
    However, a more substantial study might suggest general rules as to how these parameters might be elicited in light of certain data set properties. In some simpler scenarios, reparameterising the kernel functions to specify the length parameters at the tree-level only (i.e., no longer adopting variable-specific $\phi_{tj}$) may be appropriate, and would significantly speed-up the algorithm by reducing the number of likelihood evaluations involved in learning these parameters via MH. However, predictive performance may deteriorate as a result of this simplification in the presence of uninformative variables or other cases where variables contribute unequally to the GPs. {Alternatively, block updates of $\phi_{tj}$ would also reduce the computational burden, though designing an efficient proposal distribution for simultaneously sampling an adequate set of parameter values is not a trivial task.}
    
    \item The model can be computationally challenging to fit for larger data~sets, since the calculation of each terminal node's contribution to the overall likelihood involves inverting each associated covariance matrix, though the cost is reduced from $\mathcal{O}(n^3)$ under a single GP to ${\mathcal{O}(n_{t\ell}^3)}$ \emph{per node}, given the partitioning introduced by the tree structure. Marginalising the GP mean parameters also speeds up the algorithm. Potential strategies for further speeding up the algorithm fall into two categories. 
    \begin{enumerate}
    \item Regarding the necessary matrix computations, scalable, sparse, greedy approximations for GPs --- e.g., the Nystr{\"o}m method \citep{williams2002observations} or the methods of \citet{quinonero2007approximation}, \citet{rahimi2007random}, and \citet{wilson2020efficiently} --- may also be advantageous in future work. However, such approximations may compromise model performance compared to our present MCMC implementation.\smallskip
    
    \item Incorporating warm-up procedures to initialise GP-BART could be another viable strategy. For instance, XBART \cite{xbart2021} employs recursive partitioning and other modifications to the standard BART to rapidly find large trees which fit the data well; by ensuring that its draws are in high-probability regions of the BART posterior, this approach greatly reduces burn-in times. Seeding GP-BART in a similar fashion would allow convergence to be achieved more rapidly. However, it would be crucial to carefully design this initialisation process to align with the specific GP-BART setting, as a faster initialisation may potentially result in local minima or wasteful iterations and trees in high-probability regions of the BART posterior may not be well-suited to GP-BART.
    \end{enumerate}
    
    \item {In general, determining variable importance in GP-BART is difficult as variables may contribute to both the GPs and/or the splits.} Though the ARD appears to adequately {capture relevant variables and} account for irrelevant variables in the applications considered herein, there is further scope for {re-calibrating the discrete prior and proposal distributions for the length parameters in cases where there is prior knowledge about the relative importance of specific predictors, as well as scope for} exploiting variable-selection from the BART component. At present, all continuous predictors used to construct the trees are used in the GPs, which need not be the case. It may be beneficial to restrict the GPs only to the variables used to define splits along the given branch, though this would come with significant additional computational costs. 
    
    \item In the applications herein, we have focused on the use of GP-BART for spatial data sets, but there is nothing to prohibit the model being used in generic machine learning tasks. However, we have restricted the GPs to be covariance-stationary through our use of anisotropic exponentiated-quadratic kernels, which are governed only by scalar rate and tree-level, variable-specific length parameters. A superior approach may introduce non-stationarity to the autocovariance and hence produce more flexible GP surfaces. Relatedly, recall that tGP incorporates non-stationarity in its single `treed-GP'. Doing so for GP-BART may result in our model demonstrating even further performance improvements over tGP in the applications, but it would come with more computational challenges.
    
    Indeed, though the model outperforms its competitors in all simulation experiments and on most of the real data sets analysed above, the underlying exponentiated-quadratic kernel functions used in our parameterisation of the GP components may be inappropriate in other settings. Investigating alternative kernel functions to further improve GP-BART's performance is of great interest for future work. For instance, kernels capable of accommodating the non-continuous features we discarded in our analysis of the \textit{Baltimore} and \textit{Boston} data sets would also be of particular interest. However, this would not be immediately straightforward, given that changing the kernel necessitates specifying priors appropriately and deriving posterior distributions from scratch for sampling parameters with each new kernel and that more sophisticated kernels may further increase the computational burden.
\end{itemize}
We hope to report on these developments as part of our future research plans. 

\section*{Acknowledgements}
\small
Mateus Maia's work was supported by Science Foundation Ireland Career Development Award grant number 17/CDA/4695 and SFI research centre award 12/RC/2289P2. Andrew Parnell's work was supported by: a \mbox{Science} Foundation Ireland Career Development Award (17/CDA/4695); an investigator award (16/IA/4520); a Marine Research Programme funded by the Irish Government, co-financed by the European Regional Development Fund (Grant-Aid Agreement No. PBA/CC/18/01); European Union's Horizon 2020 research and innovation programme under grant agreement No. 818144; SFI Centre for Research Training 18CRT/6049, and SFI Research Centre awards 16/RC/3872 and 12/RC/2289P2.
\normalsize

\bibliographystyle{elsarticle-harv}\interlinepenalty=10000
\bibliography{cas-refs}







\newpage
\section*{Appendices}\medskip
\renewcommand{\thesubsection}{Appendix \Alph{subsection}}

\subsection{Tree likelihood}\label{sec::appendix_a}\bigskip

In general terms, following the initial formulation of the GP-BART model described in Section \ref{section:mu_gp_prior},
the posterior distribution of the residuals for a terminal node $\ell$ in tree $t$ is given by
\[\mathbf{R}_{t\ell} \given \mathcal{T}_{t}, \mu_{t\ell}, \boldsymbol{\phi}_{t}, \nu,
\tau \sim \mbox{MVN} \left(\boldsymbol{\mu}_{t\ell}=\mu_{t\ell}\mathbf{1}_{n_{t\ell}}, \tau^{-1}\bm{\mathcal{I}}_{n_{t\ell}} + \boldsymbol{\Omega}_{t\ell}\right).\]
However, in writing this likelihood, we can marginalise out the terminal-node mean parameters $\mu_{t\ell}\given \mathcal{T}_{t}, \tau_{\mu} \sim \mbox{N}(0, \tau_{\mu}^{-1})$ as follows
\begin{align*}
    \pi \left(\mathbf{R}_{t\ell} \given  \mathcal{T}_{t}, \boldsymbol{\phi}_{t},\nu,
\tau \right) &= \int \pi \left(\mathbf{R}_{t\ell} \given \mu_{t\ell},\boldsymbol{\phi}_{t},\nu,
\tau \right) \pi\left(\mu_{t\ell}\right) \partial \mu_{t\ell}\\
    &\propto \lvert\boldsymbol{\Gamma}_{t\ell}\rvert^{-1/2} \exp\left\{ -\frac{1}{2}\left(\mathbf{R}_{t\ell}-\boldsymbol{\mu}_{t\ell}\right)^{\top}\boldsymbol{\Gamma}_{t\ell}^{-1}\left(\mathbf{R}_{t\ell}-\boldsymbol{\mu}_{t\ell}\right)\right\}\times \\
    & \phantom{\propto\,}~\tau_{\mu}^{-1/2} \exp\left\{-\frac{\tau_{\mu}}{2} \mu_{t\ell}^{2}\right\},
\end{align*}
where $\boldsymbol{\Gamma}_{t\ell} = \tau^{-1}\bm{\mathcal{I}}_{n_{t\ell}}+\boldsymbol{\Omega}_{t\ell}.$
After further calculations, letting
\[v_{t\ell} =  \mathbf{1}_{n_{ t\ell}}^\top\boldsymbol{\Gamma}_{t\ell}^{-1}\mathbf{1}_{n_{ t\ell}}+\tau_{\mu},\]
applying the log, and then summing over the terminal nodes, we obtain
\begin{align*}
   \log\pi\left(\mathbf{R}_t\given \mathcal{T}_{t}, \boldsymbol{\phi}_t,\nu,\tau\right)&= \log\mathcal{C} -\frac{1}{2}\sum_{\ell}^{b_t}\log{v}_{t\ell} -\frac{1}{2}\sum_{\ell}^{b_t}\log{|\boldsymbol{\Gamma}_{t\ell}|}\\
   &\phantom{=}-\frac{1}{2}\sum_{\ell}^{b_t}\mathbf{R}_{t\ell}^{\top}\boldsymbol{\Gamma}_{t\ell}^{-1}\mathbf{R}_{t\ell} + \frac{1}{2}\sum_{\ell}^{b_t}v^{-1}_{t\ell}\boldsymbol{1}^{\top}_{n_{t\ell}}\boldsymbol{\Gamma}_{t\ell}^{-1}\mathbf{R}_{t\ell}\mathbf{R}_{t\ell}^{\top}\boldsymbol{\Gamma}_{t\ell}^{-1}\boldsymbol{1}_{n_{t\ell}},
\end{align*}
where $\mathcal{C}$ is a constant of proportionality. Recalling $\boldsymbol{\Lambda}_{t\ell}=\tau_\mu^{-1}\mathbf{1}_{n_{t\ell}}\mathbf{1}_{n_{t\ell}}^\top + \boldsymbol{\Omega}_{t\ell}$, this expression can be further simplified with the constant $\mu_{t\ell}$ parameters explicitly absorbed into the kernel of the GP. This yields the following distribution for the partial residuals
\[\mathbf{R}_{t\ell} \given \mathcal{T}_{t}, \boldsymbol{\phi}_{t},\nu,
\tau \sim \mbox{MVN}\left(\boldsymbol{0}_{n_{t\ell}}, \tau^{-1}\bm{\mathcal{I}}_{t\ell} + \boldsymbol{\Lambda}_{t\ell}\right),\]
which circumvents the need to sample the $\mu_{t\ell}$ parameters and encourages better mixing.

\subsection{Performance evaluation with varying residual precision on the benchmarking experiments}\bigskip
\label{sec::appendix_b}

\renewcommand{\thefigure}{\Alph{subsection}.\arabic{figure}}
\setcounter{figure}{0}
\renewcommand{\thetable}{\Alph{subsection}.\arabic{table}}
\setcounter{table}{0}

To assess the model's performance across different levels of noise, we replicated the experiments from Section \ref{sec:benchmarking}, varying the residual precision parameter $\tau$ at three levels --- specifically $\tau = \left\{1, 0.1, 0.01 \right\}$ --- and compared GP-BART with its competitors in each case, as before. Recall that the results shown throughout Section \ref{sec:benchmarking} are based on $\tau=10$ only. The results now indicate that even with increasing noise (i.e., lower precision), GP-BART maintains consistent performance and continues to exhibit the lowest median RMSE and CRPS values, though the variability of both metrics does increase as $\tau$ decreases. 

The results for $\tau=0.1$, $\tau=0.1$, and $\tau=0.01$ are presented in \ref{sec::appendix_b_one}, \ref{sec::appendix_b_two}, and \ref{sec::appendix_b_three}, respectively. As per Section \ref{sec:benchmarking}, we show in each case the simulated data surface for the given $\tau$ value with sample sizes of $n=100$, $n=500$, and $n=1000$ and then show the predicted surfaces according to GP-BART and its competitors BART, SoftBART, tGP, kriging, and INLA at each sample size. Finally, we show boxplots of the RMSE and CRPS values obtained by the competing methods on the data generated with the respective $\tau$ value.

\subsubsection{Residual precision \texorpdfstring{$\tau = 1$}{tau = 1}}\bigskip\label{sec::appendix_b_one}

\begin{figure}[H]
     \centering
     \includegraphics[width=\textwidth]{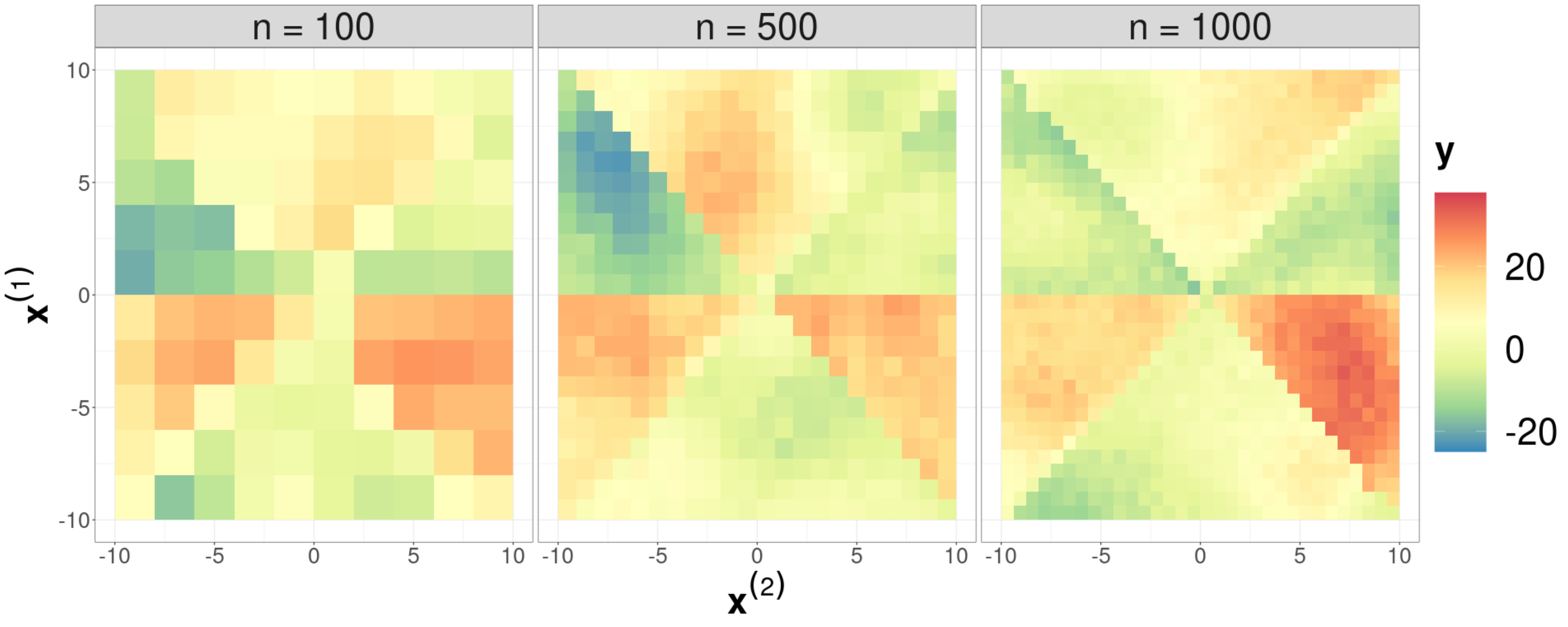}
     \caption{Simulated data with $n=\left\{100,500,1000 \right\}$ observations, respectively, and residual precision of $\tau=1$.}
     \label{fig:simulated_data_tau_one}
 \end{figure}

\begin{figure}[H]
     \centering
     \includegraphics[width=\textwidth]{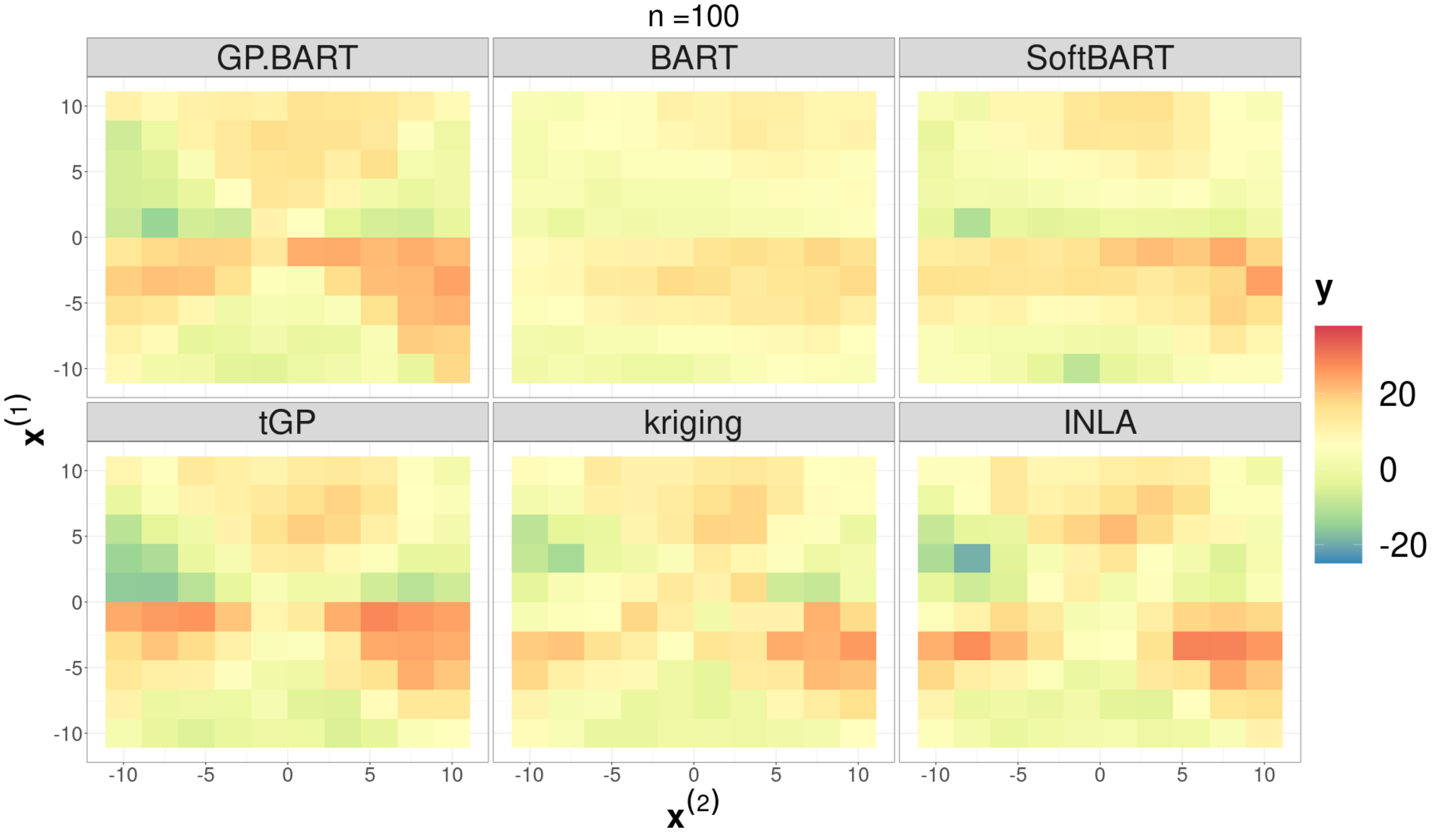}
     \caption{Predicted surfaces for the simulated scenario with $n=100$ observations from the first panel of Figure \ref{fig:simulated_data_tau_one} using different methods over one randomly chosen test repetition. The residual precision is $\tau = 1$.}
     \label{fig:sim_surface_100_tau_one}
 \end{figure}

\begin{figure}[H]
     \centering
     \includegraphics[width=\textwidth]{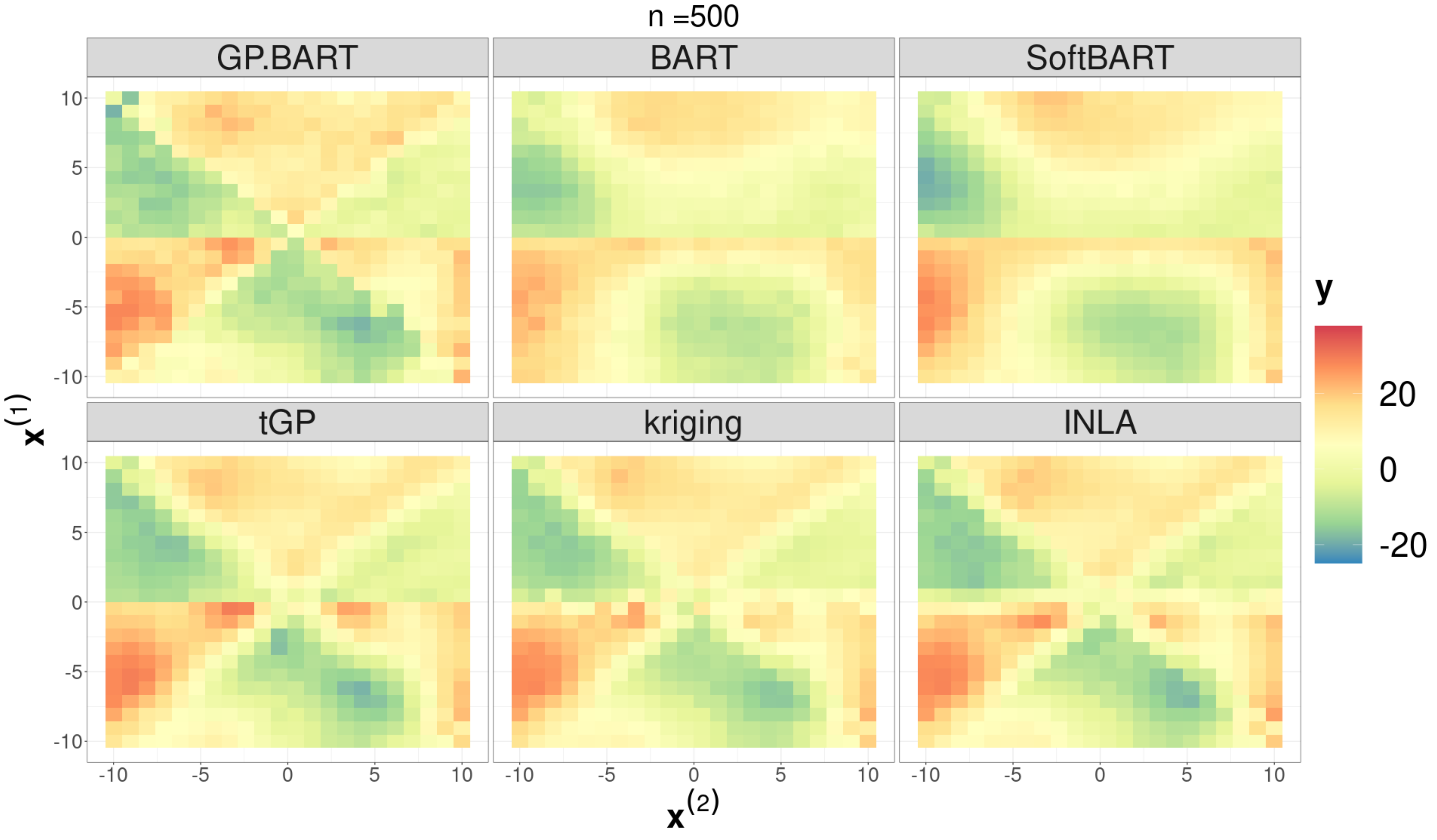}
     \caption{Predicted surfaces for the simulated scenario with $n=500$ observations from the second panel of Figure \ref{fig:simulated_data_tau_one} using different methods over one randomly chosen test repetition. The residual precision is $\tau = 1$.}
     \label{fig:sim_surface_500_tau_one}
 \end{figure}

\begin{figure}[H]
     \centering
    \includegraphics[width=\textwidth]{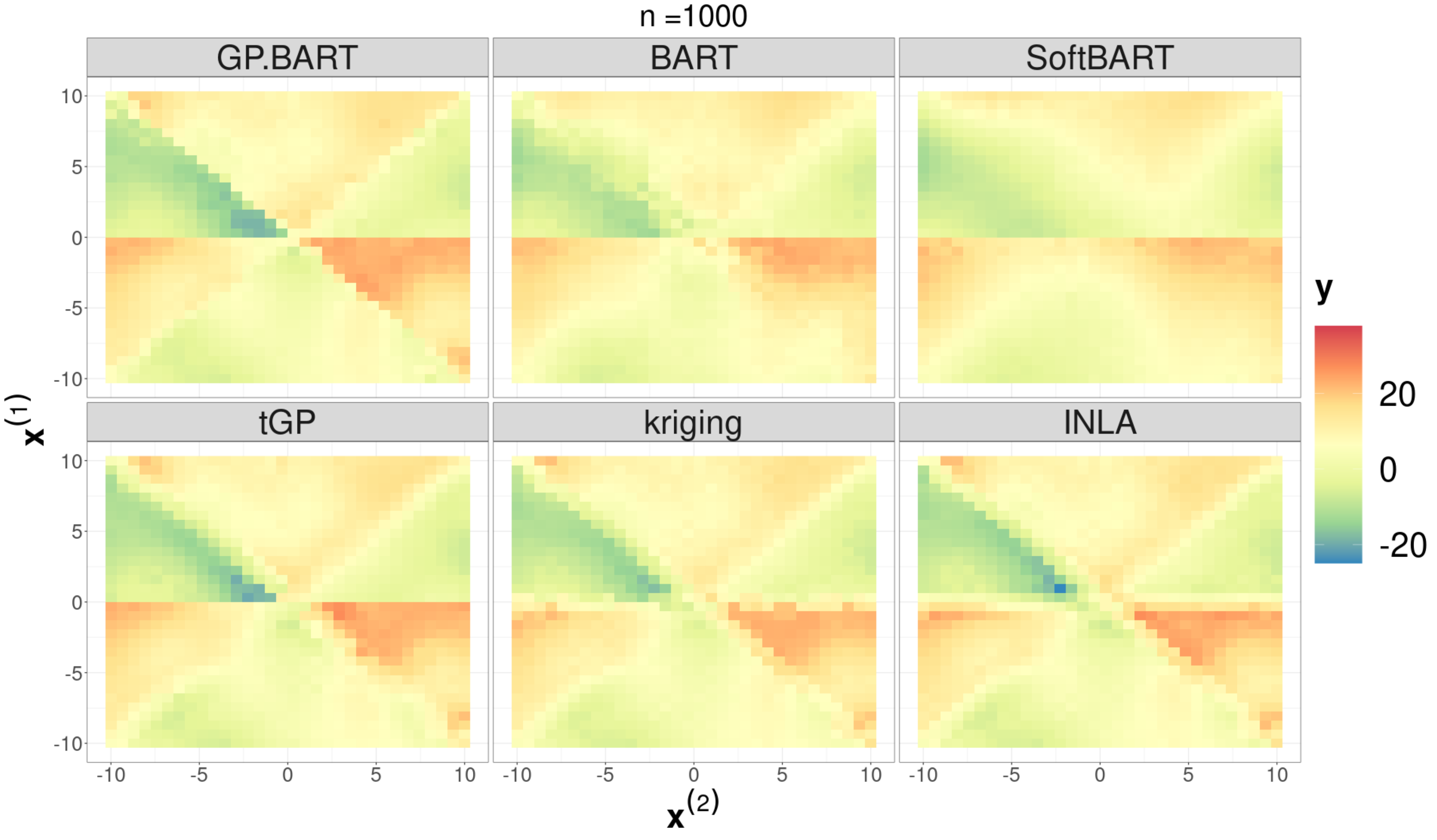}
     \caption{Predicted surfaces for the simulated scenario with $n\!=\!1000$ observations from~the third panel of Figure \ref{fig:simulated_data_tau_one} using different methods over one randomly chosen test repetition. The residual precision is $\tau = 1$.}
     \label{fig:sim_surface_1000_tau_one}
 \end{figure}

\begin{figure}[H]
     \centering
     \includegraphics[width=\textwidth]{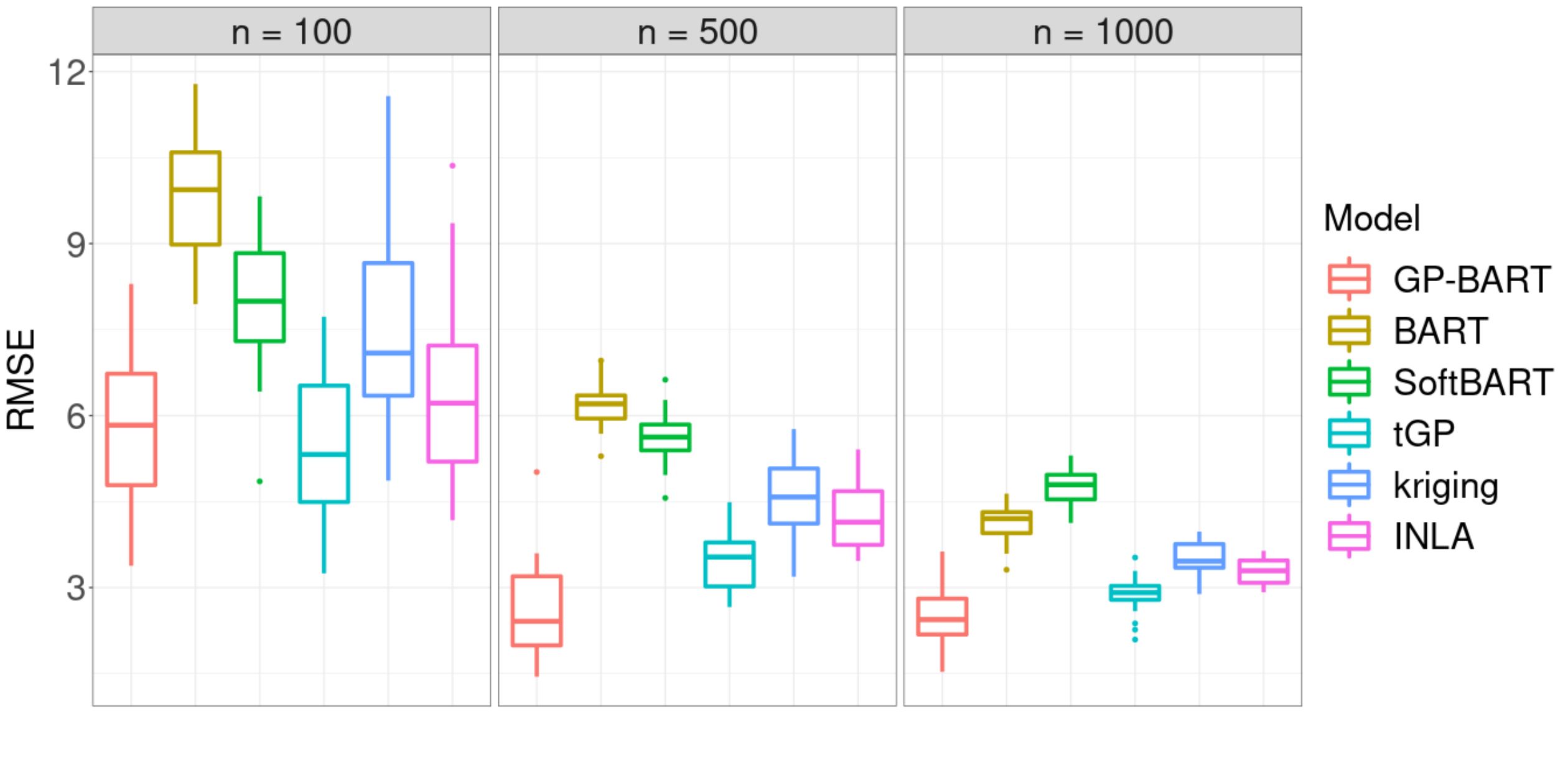}
     \caption{Comparisons between the RMSE obtained by the competing models for the simulated data using $5$ repeated $5$-fold cross validation over different sample sizes, and $\tau=1$. Based on the results, it is evident that GP-BART consistently delivers the best performance on average, as it encompasses assumptions of spatial dependence, smoothness, and allows for rotated splits.}
     \label{fig:sim_rmse_boxplot_tau_one}
 \end{figure}

\begin{figure}[H]
     \centering
     \includegraphics[width=\textwidth]{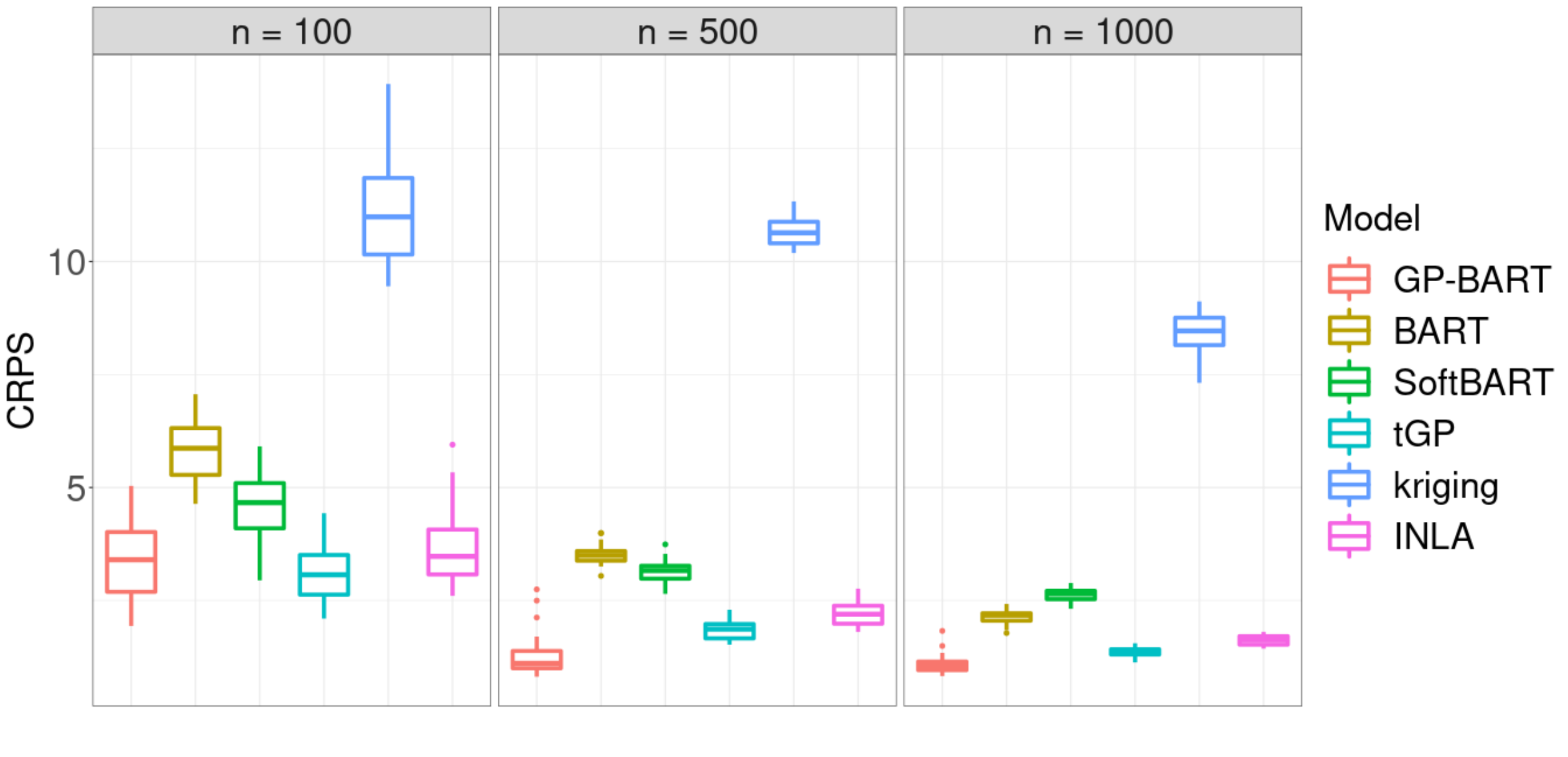}
     \caption{Comparisons between the CRPS values obtained by the competing models for the simulated data using $5$ repeated $5$-fold cross validation over different sample sizes, and $\tau=1$. Based on the results, it is evident that GP-BART consistently delivers the best performance on average, as it encompasses assumptions of spatial dependence, smoothness, and allows for rotated splits.}
     \label{fig:sim_pi_boxplot_tau_one}
 \end{figure}

The simulated data surfaces considering the residual precision $\tau=1$ for different samples sizes $n=\{100,500,1000\}$ are shown in Figure \ref{fig:simulated_data_tau_one}. Figure \ref{fig:sim_surface_100_tau_one}, Figure \ref{fig:sim_surface_500_tau_one}, and Figure \ref{fig:sim_surface_1000_tau_one} show the corresponding predicted surfaces from one randomly chosen repetition of the repeated 5-fold cross-validation for each respective sample size. As before, GP-BART's predicted surfaces more closely resemble the signal from the original data depicted in Figure \ref{fig:simulated_data_tau_one} in every instance, when compared with its competitors. The quantitative comparison is summarised via boxplots of RMSE and CRPS values in Figure \ref{fig:sim_rmse_boxplot_tau_one} and  Figure \ref{fig:sim_pi_boxplot_tau_one}, respectively. These boxplots reflect the conclusions draw from previous plots where, in general, GP-BART presents the lowest median values for RMSE and CRPS across all scenarios.

 \vspace{-1em}\enlargethispage{\baselineskip}


\subsubsection{Residual precision \texorpdfstring{$\tau = 0.1$}{tau = 0.1}}\bigskip\label{sec::appendix_b_two}

\begin{figure}[H]
     \centering
     \includegraphics[width=\textwidth]{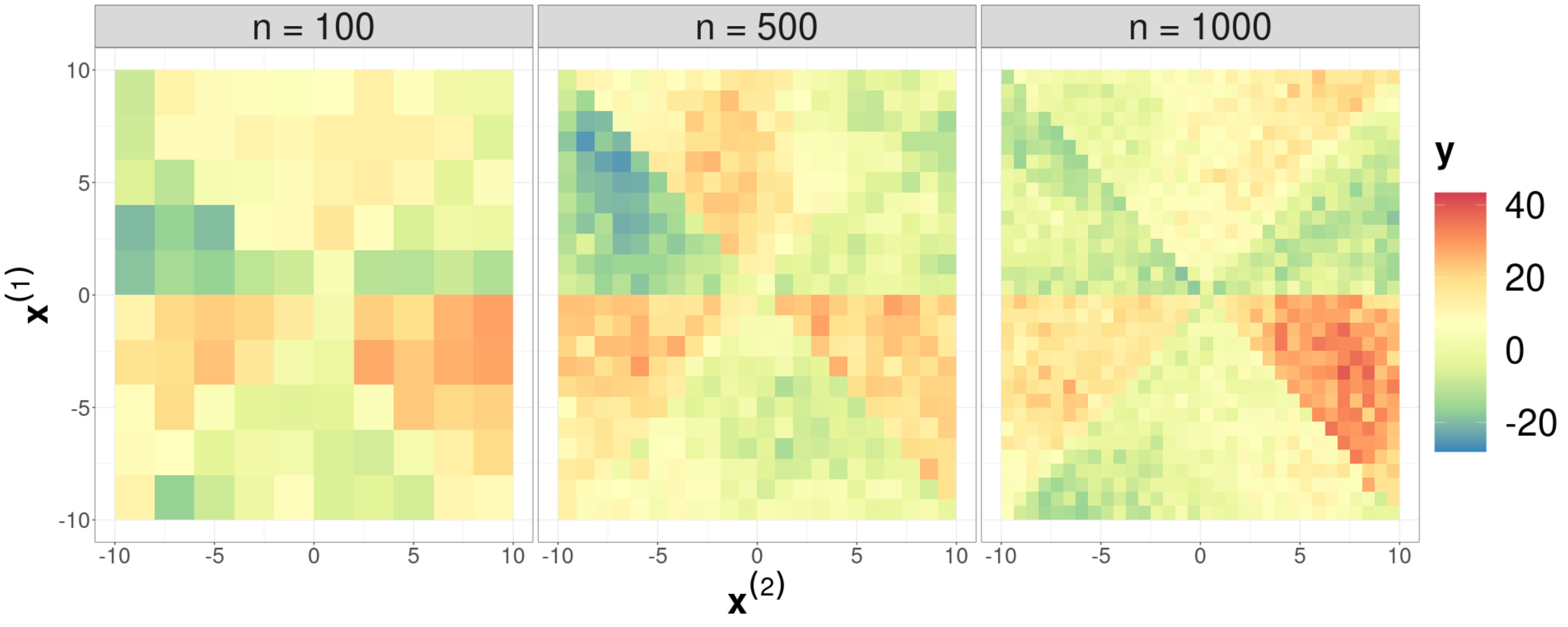}
     \caption{Simulated data with $n=\left\{100,500,1000 \right\}$ observations, respectively, and residual precision of $\tau=0.1$.}
     \label{fig:simulated_data_tau_zero_one}
 \end{figure}

\begin{figure}[H]
     \centering
     \includegraphics[width=\textwidth]{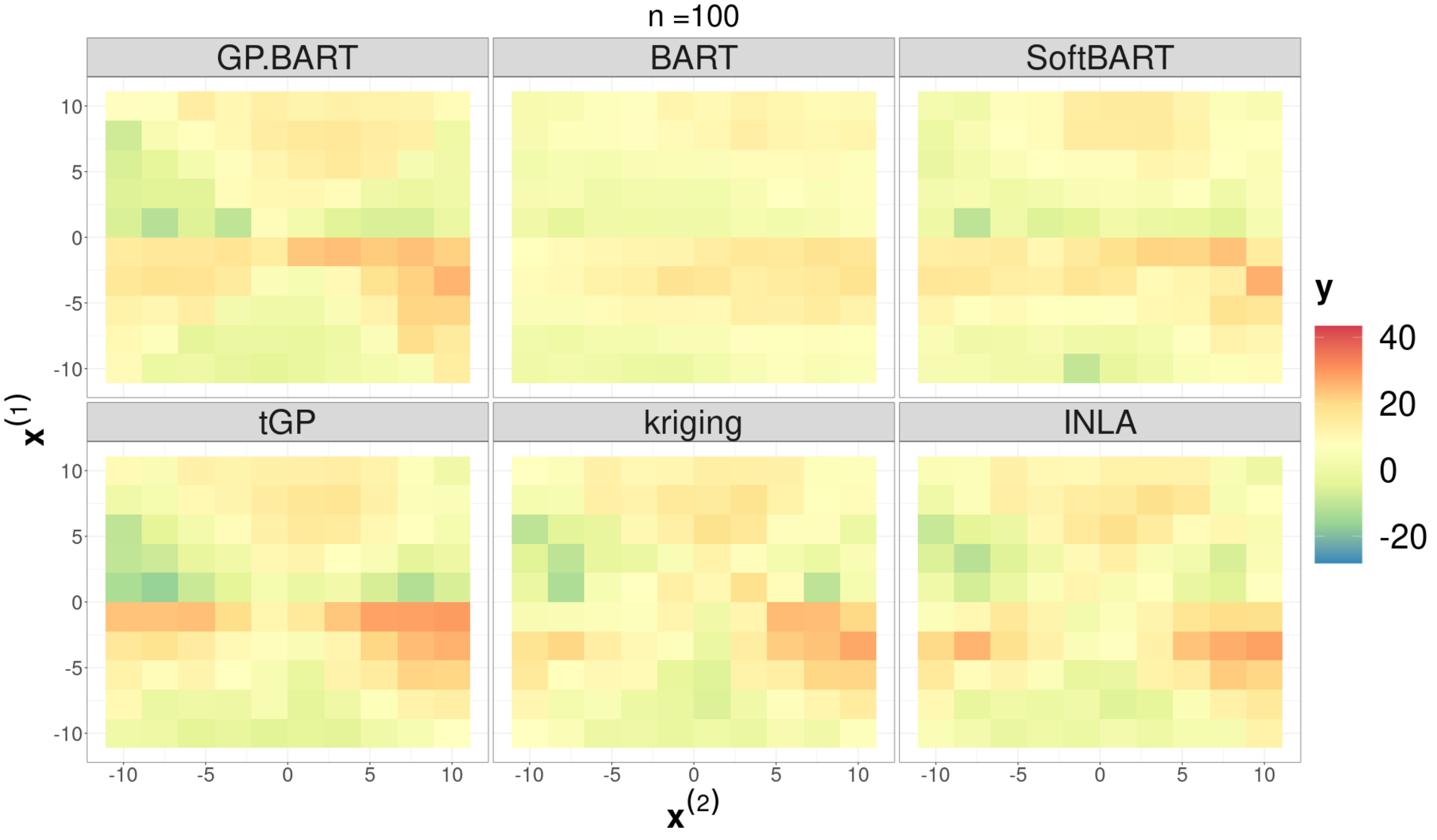}
     \caption{Predicted surfaces for the simulated scenario with $n=100$ observations from the first panel of Figure \ref{fig:simulated_data_tau_zero_one} using different methods over one randomly chosen test repetition. The residual precision is $\tau = 0.1$.}
     \label{fig:sim_surface_100_tau_zero_one}
 \end{figure}

\begin{figure}[H]
     \centering
     \includegraphics[width=\textwidth]{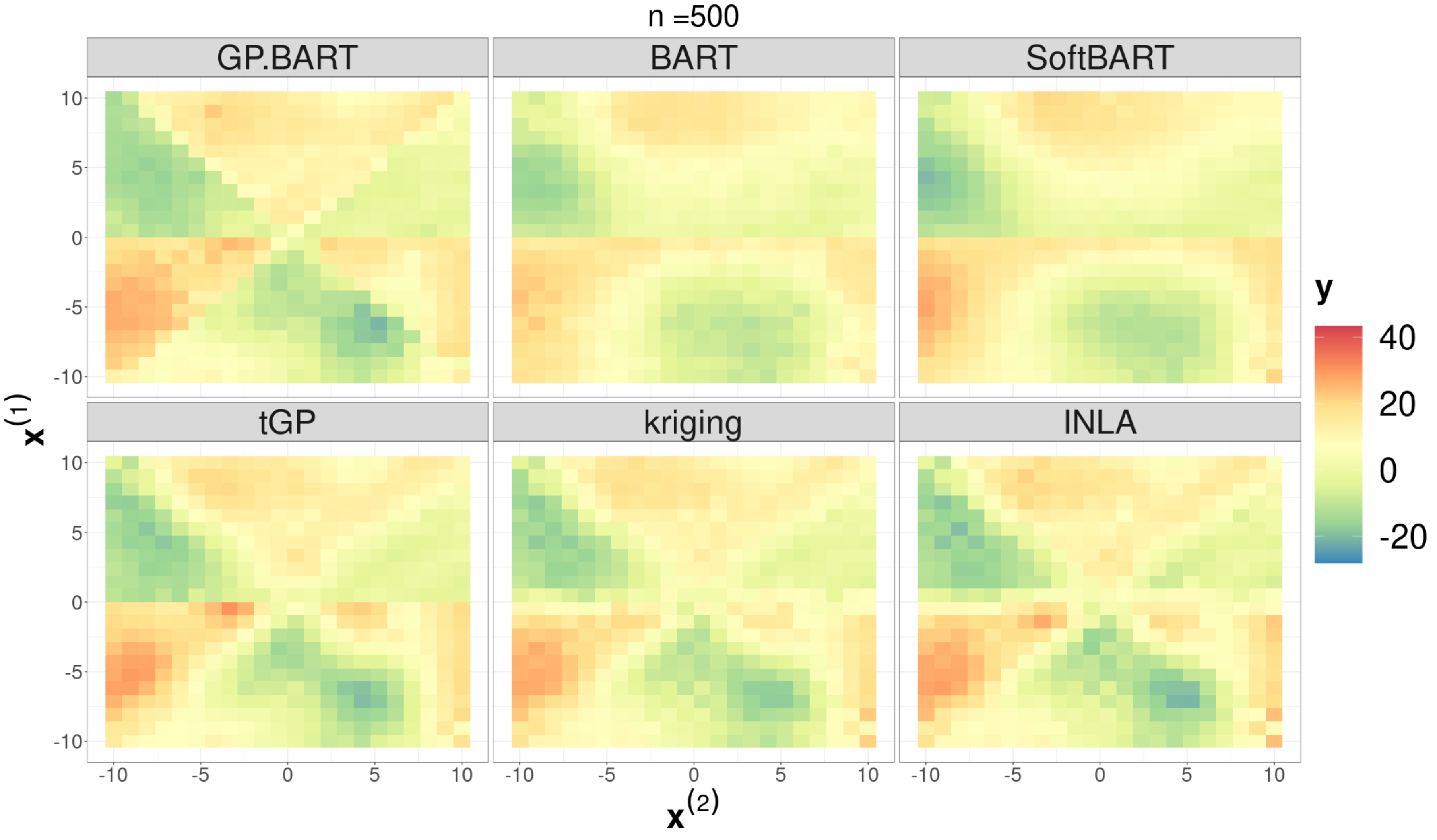}
     \caption{Predicted surfaces for the simulated scenario with $n=500$ observations from the second panel of Figure \ref{fig:simulated_data_tau_zero_one} using different methods over one randomly chosen test repetition. The residual precision is $\tau = 0.1$.}
     \label{fig:sim_surface_500_tau_zero_one}
 \end{figure}

\begin{figure}[H]
     \centering
    \includegraphics[width=\textwidth]{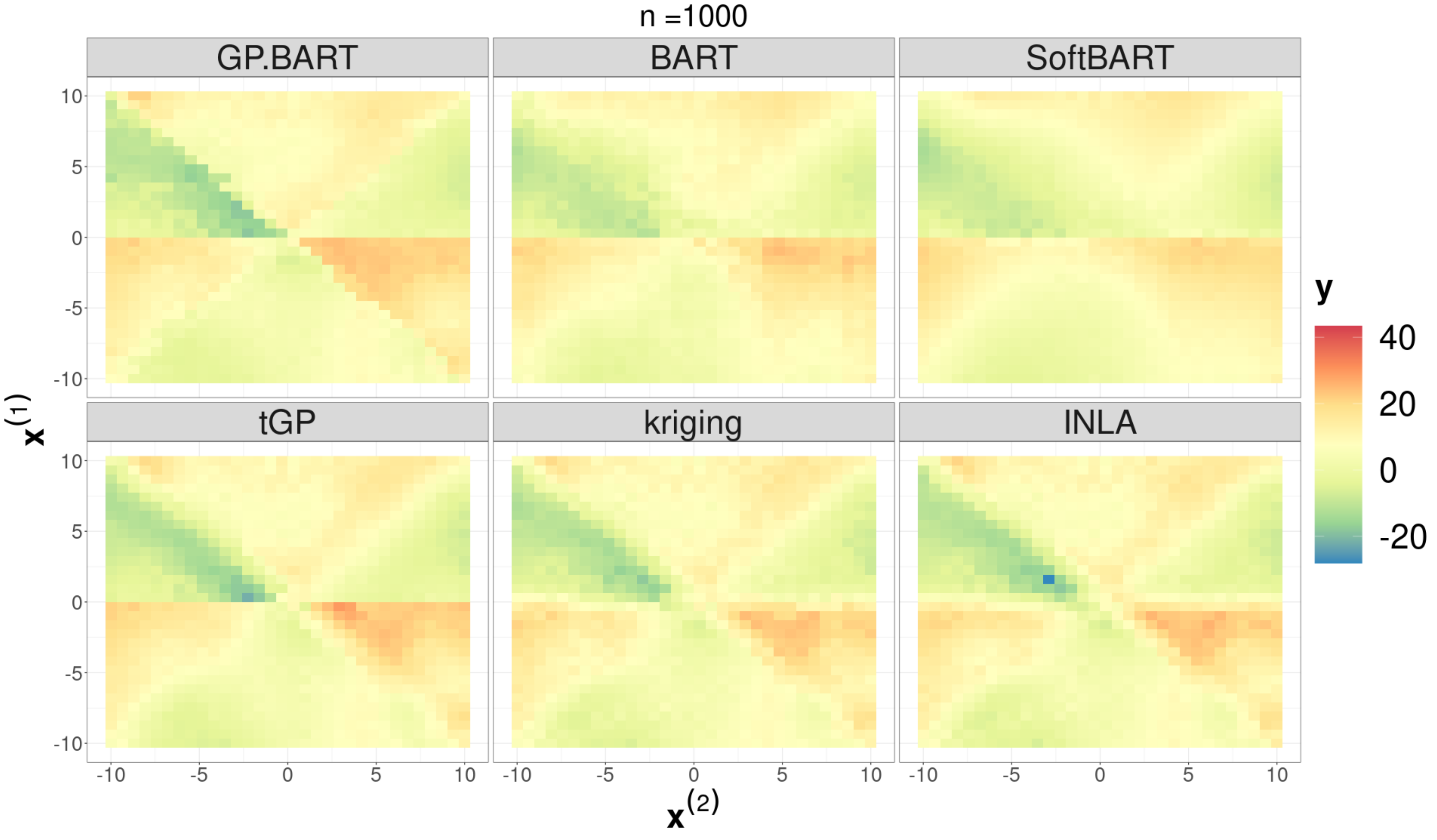}
     \caption{Predicted surfaces for the simulated scenario with $n\!=\!1000$ observations from~the third panel of Figure \ref{fig:simulated_data_tau_zero_one} using different methods over one randomly chosen test repetition. The residual precision is $\tau = 0.1$.}
     \label{fig:sim_surface_1000_tau_zero_one}
 \end{figure}

\begin{figure}[H]
     \centering
     \includegraphics[width=\textwidth]{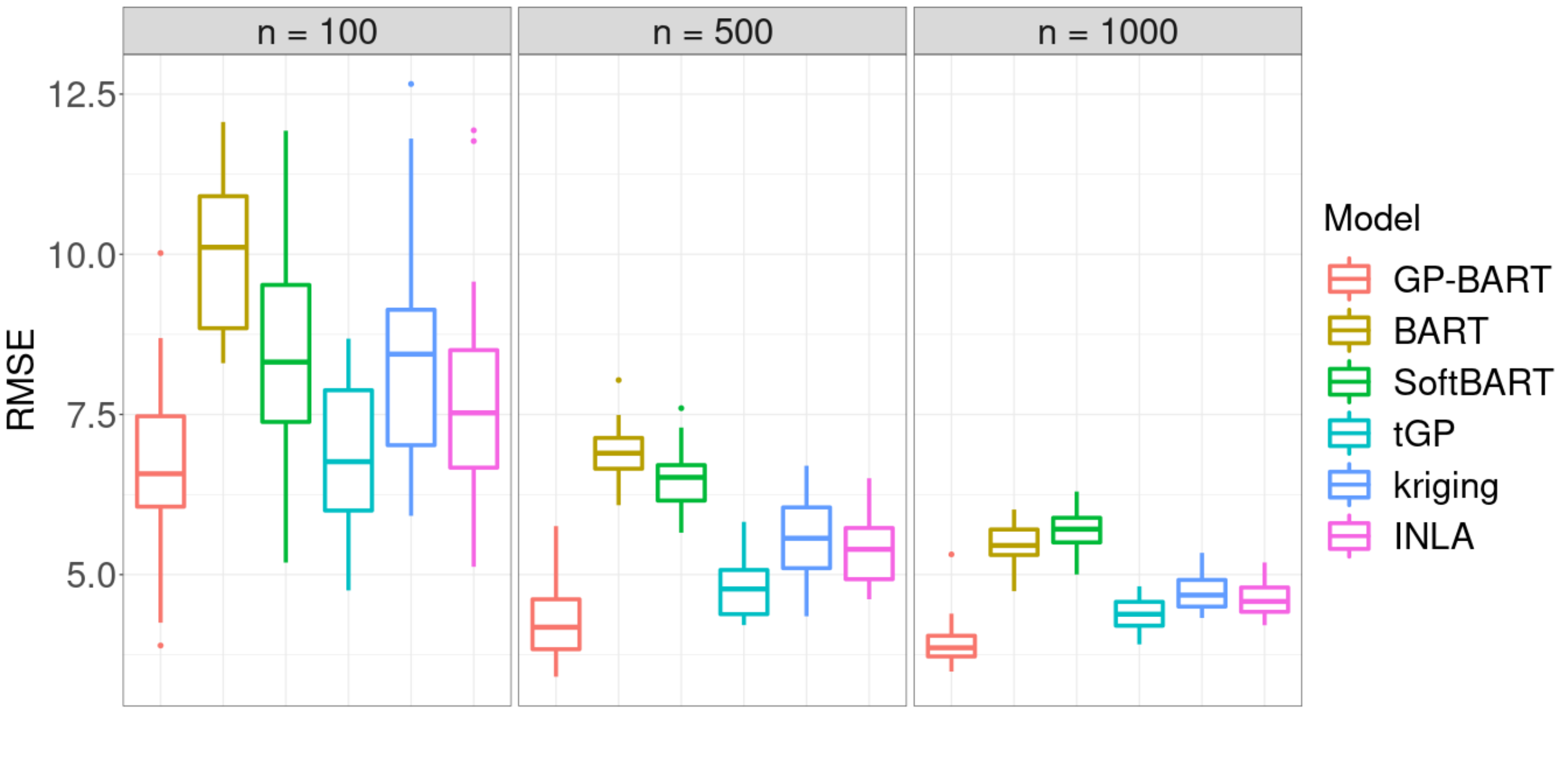}
     \caption{Comparisons between the RMSE obtained by the competing models for the simulated data using $5$ repeated $5$-fold cross validation over different sample sizes, and $\tau=0.1$. Based on the results, it is evident that GP-BART consistently delivers the best performance on average, as it encompasses assumptions of spatial dependence, smoothness, and allows for rotated splits.}
     \label{fig:sim_rmse_boxplot_tau_zero_one}
 \end{figure}

\begin{figure}[H]
     \centering
     \includegraphics[width=\textwidth]{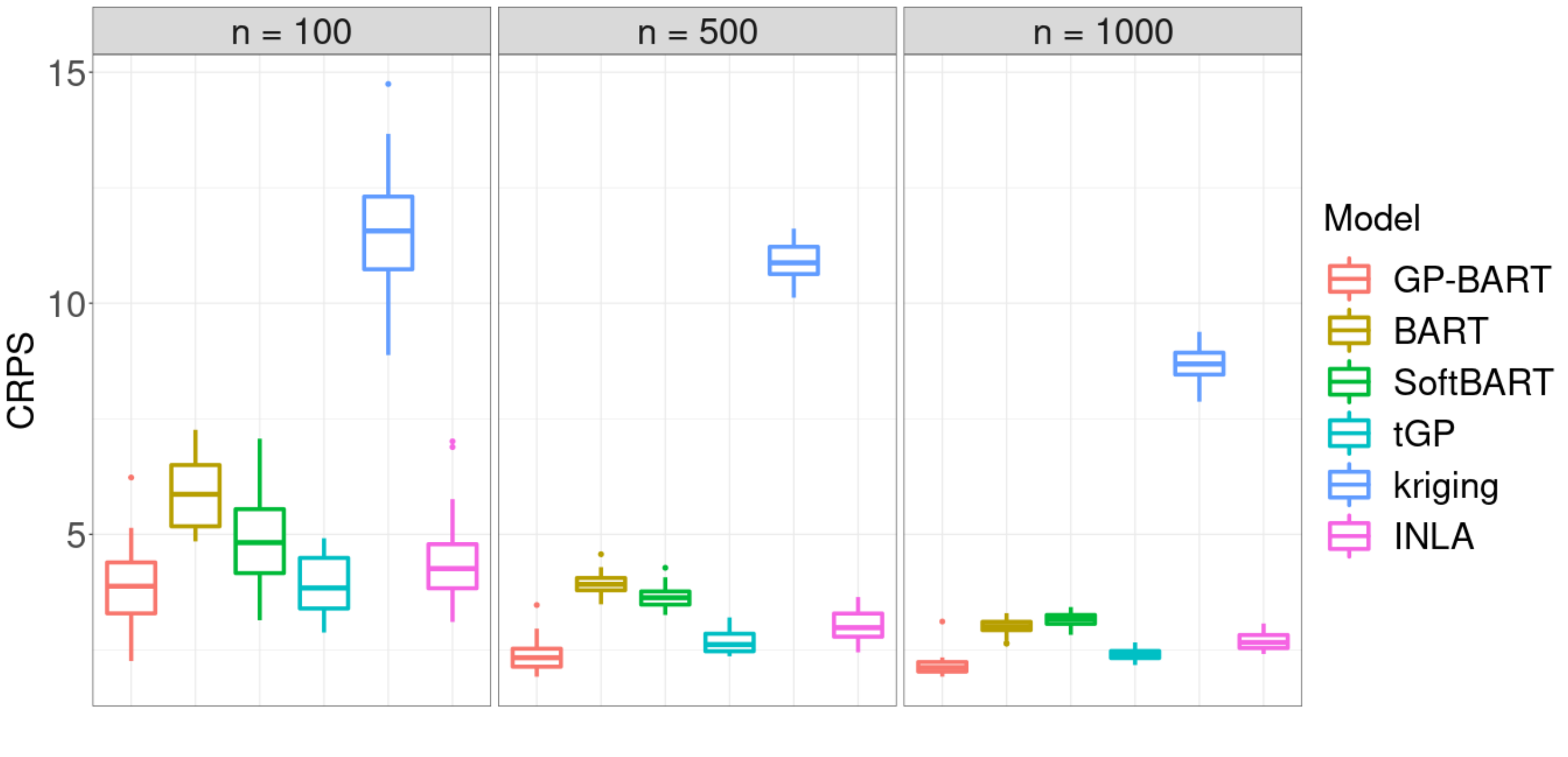}
     \caption{Comparisons between the CRPS values obtained by the competing models for the simulated data using $5$ repeated $5$-fold cross validation over different sample sizes, and $\tau=0.1$. Based on the results, it is evident that GP-BART consistently delivers the best performance on average, as it encompasses assumptions of spatial dependence, smoothness, and allows for rotated splits.}
     \label{fig:sim_pi_boxplot_tau_zero_one}
 \end{figure}

The simulated data surfaces considering the residual precision $\tau=0.1$ for different samples sizes $n=\{100,500,1000\}$ are shown in Figure \ref{fig:simulated_data_tau_zero_one}. Figure \ref{fig:sim_surface_100_tau_zero_one}, Figure \ref{fig:sim_surface_500_tau_zero_one}, and Figure \ref{fig:sim_surface_1000_tau_zero_one} show the corresponding predicted surfaces from one randomly chosen repetition of the repeated 5-fold cross-validation for each respective sample size. As before, GP-BART's predicted surfaces more closely resemble the signal from the original data depicted in Figure \ref{fig:simulated_data_tau_zero_one} in every instance, when compared with its competitors. The quantitative comparison is summarised via boxplots of RMSE and CRPS values in Figure \ref{fig:sim_rmse_boxplot_tau_zero_one} and  Figure \ref{fig:sim_pi_boxplot_tau_zero_one}, respectively. These boxplots reflect the conclusions draw from previous plots where, in general, GP-BART presents the lowest median values for RMSE and CRPS across all scenarios.

 \vspace{-1em}\enlargethispage{\baselineskip}
 \vspace{-1em}\enlargethispage{\baselineskip}

 \subsubsection{Residual precision \texorpdfstring{$\tau = 0.01$}{tau = 0.01}}\bigskip\label{sec::appendix_b_three}

\begin{figure}[H]
     \centering
     \includegraphics[width=\textwidth]{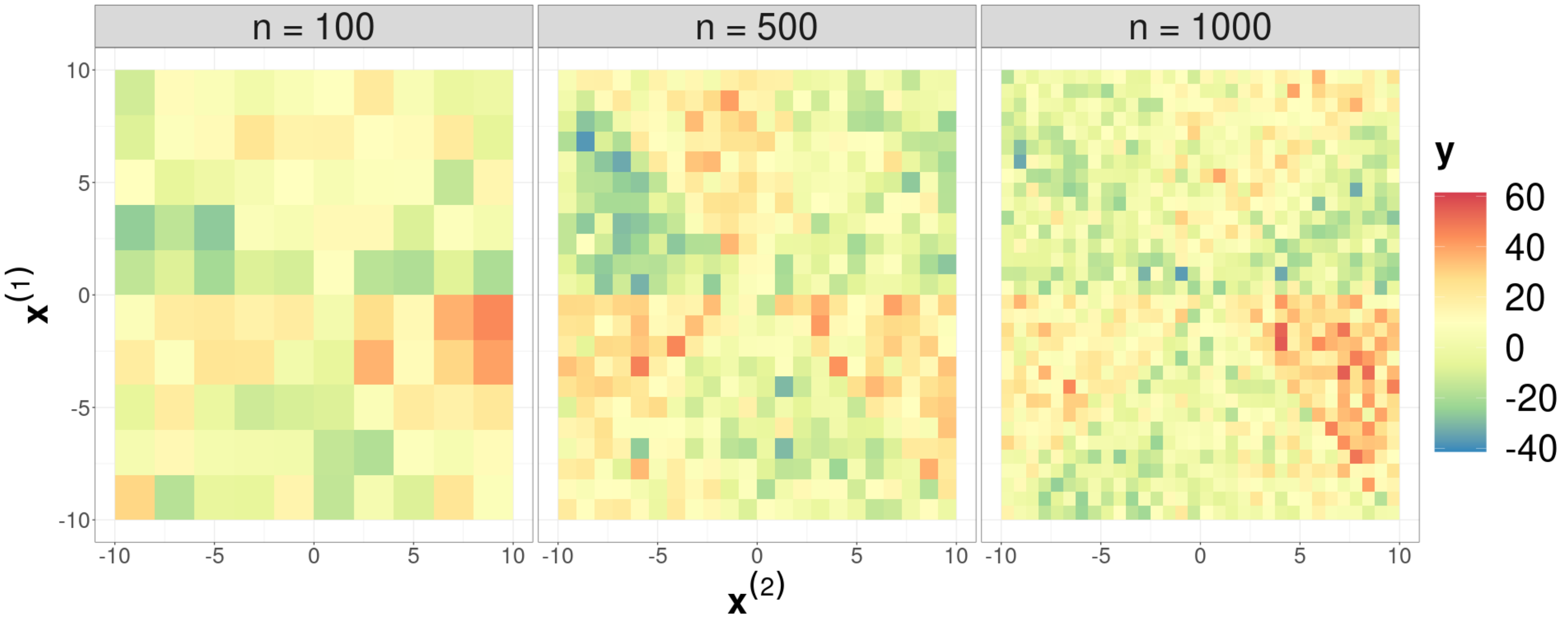}
     \caption{Simulated data with $n=\left\{100,500,1000 \right\}$ observations, respectively, and residual precision of $\tau=0.01$.}
     \label{fig:simulated_data_tau_zero_zero_one}
 \end{figure}

\begin{figure}[H]
     \centering
     \includegraphics[width=\textwidth]{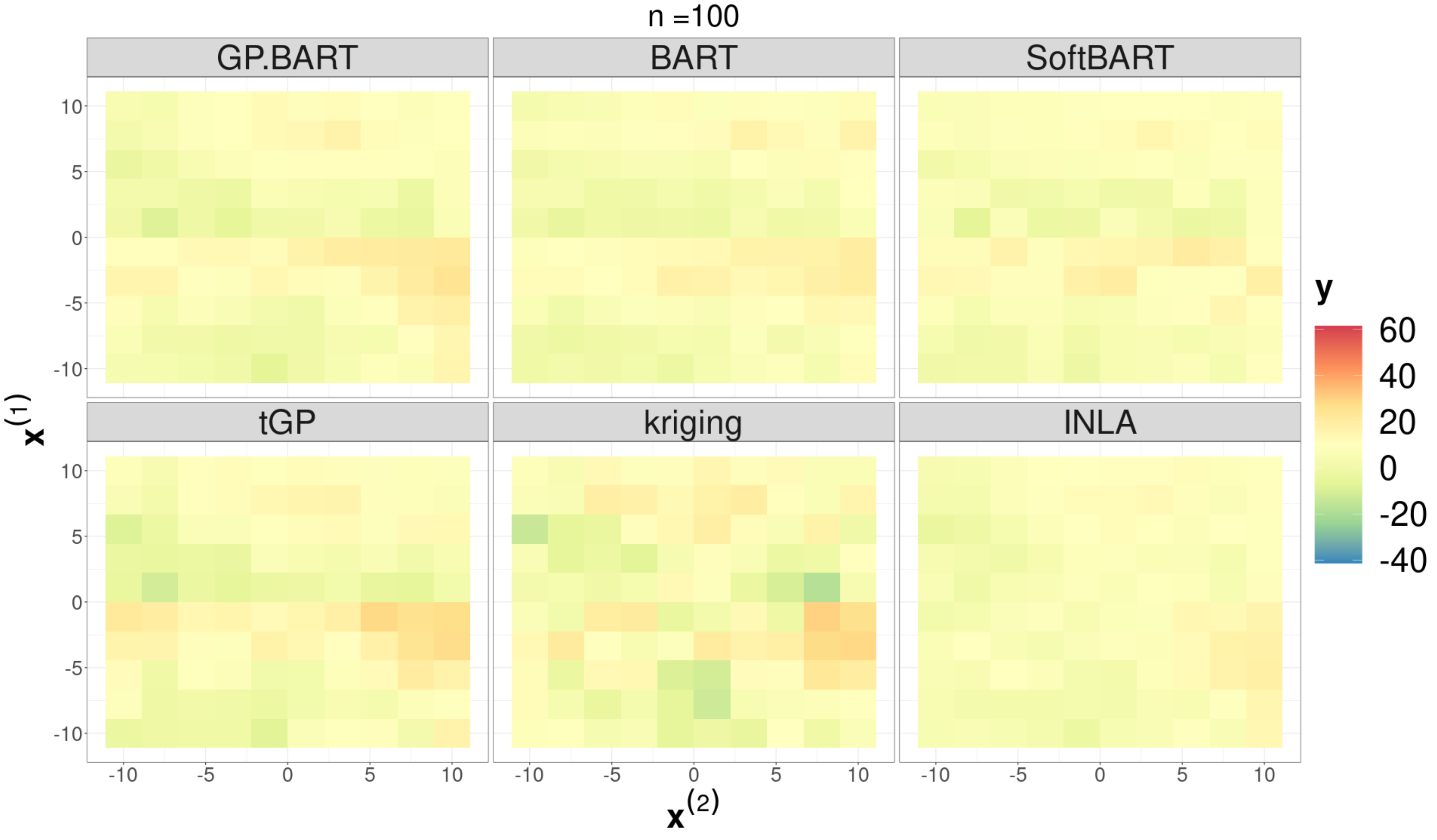}
     \caption{Predicted surfaces for the simulated scenario with $n=100$ observations from the first panel of Figure \ref{fig:simulated_data_tau_zero_zero_one} using different methods over one randomly chosen test repetition. The residual precision is $\tau = 0.01$.}
     \label{fig:sim_surface_100_tau_zero_zero_one}
 \end{figure}

\begin{figure}[H]
     \centering
     \includegraphics[width=\textwidth]{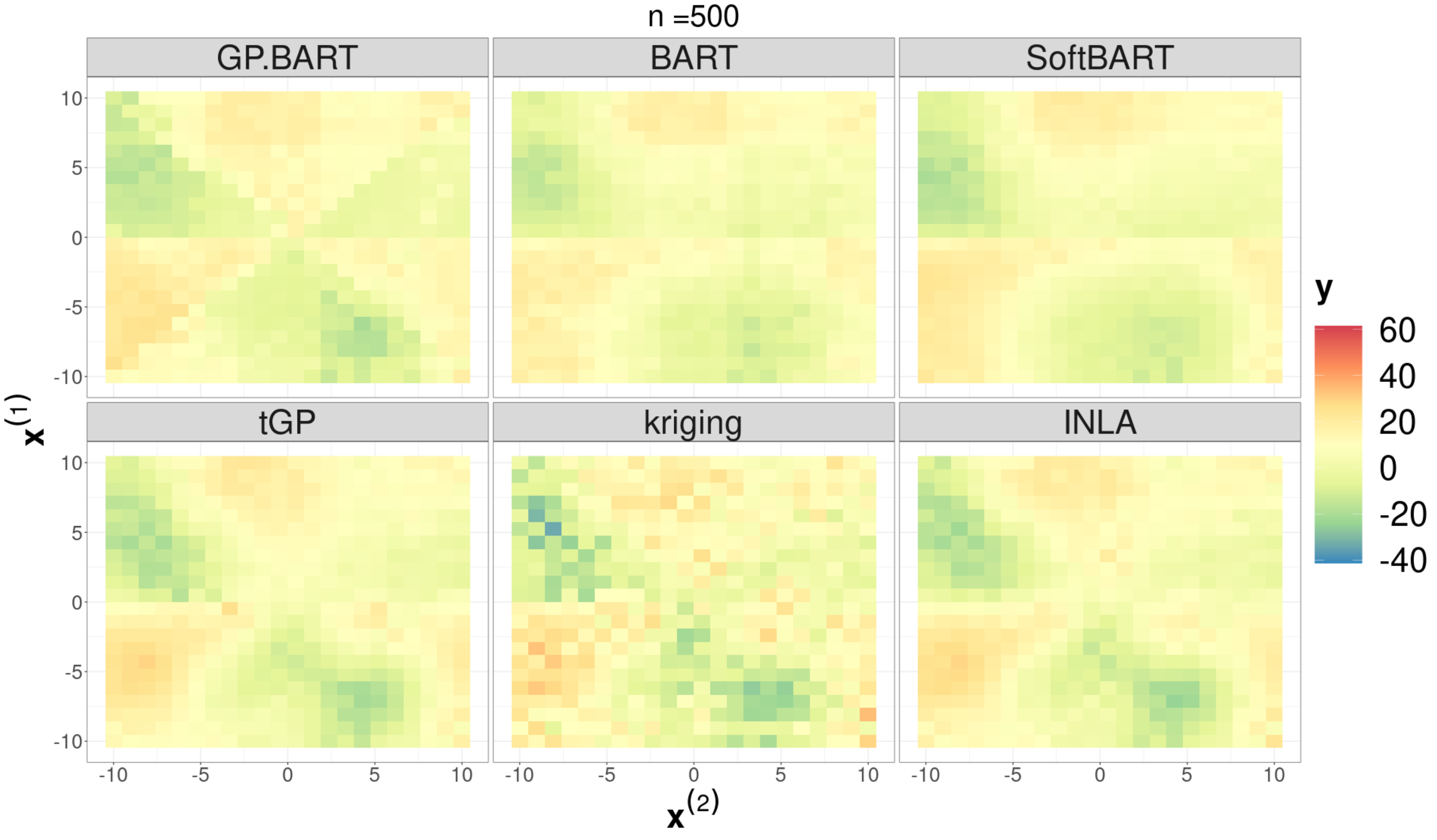}
     \caption{Predicted surfaces for the simulated scenario with $n=500$ observations from the second panel of Figure \ref{fig:simulated_data_tau_zero_zero_one} using different methods over one randomly chosen test repetition. The residual precision is $\tau = 0.01$.}
     \label{fig:sim_surface_500_tau_zero_zero_one}
 \end{figure}

\begin{figure}[H]
     \centering
    \includegraphics[width=\textwidth]{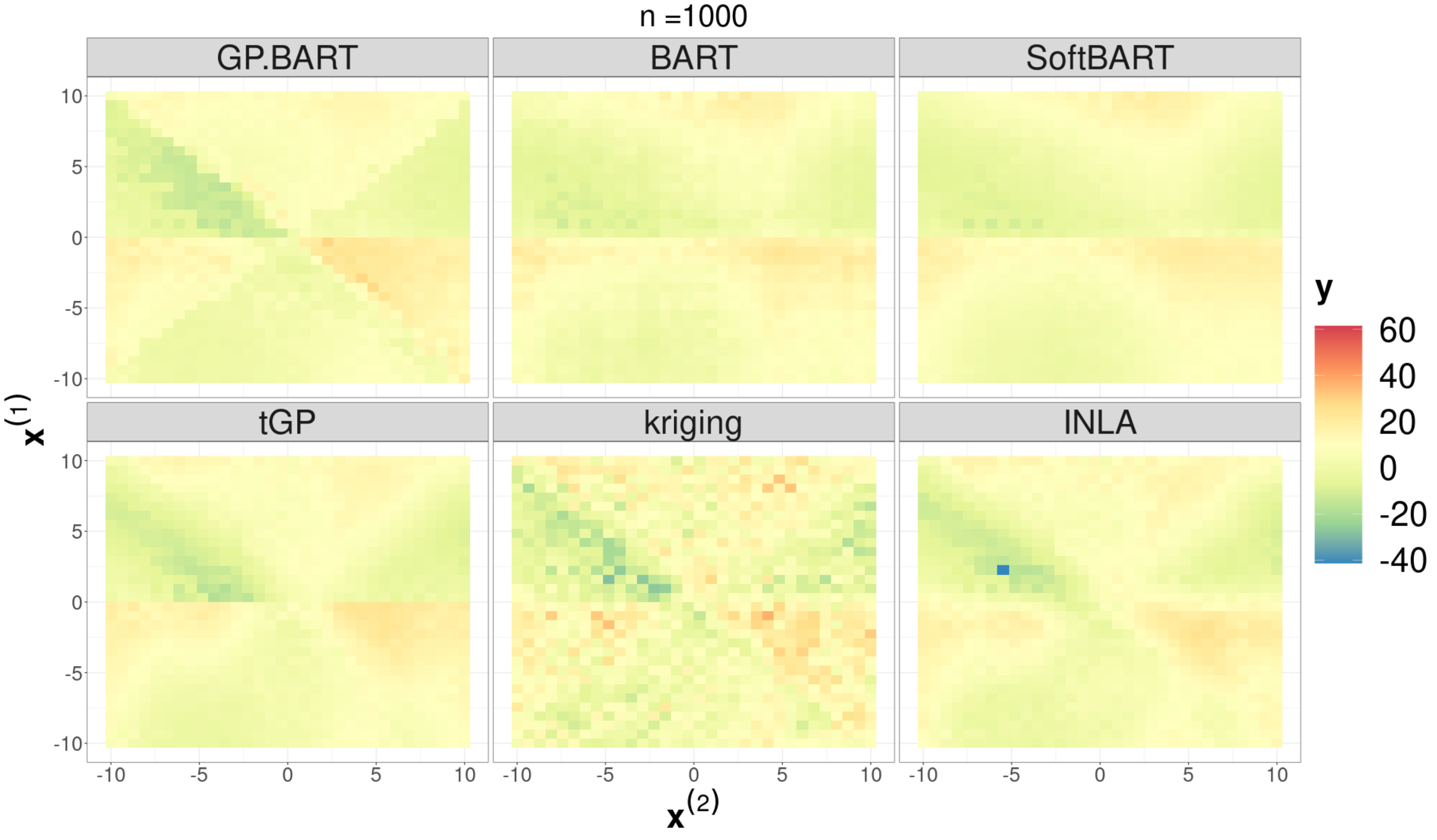}
     \caption{Predicted surfaces for the simulated scenario with $n\!=\!1000$ observations from~the third panel of Figure \ref{fig:simulated_data_tau_zero_zero_one} using different methods over one randomly chosen test repetition. The residual precision is $\tau = 0.01$.}
     \label{fig:sim_surface_1000_tau_zero_zero_one}
 \end{figure}

\begin{figure}[H]
     \centering
     \includegraphics[width=\textwidth]{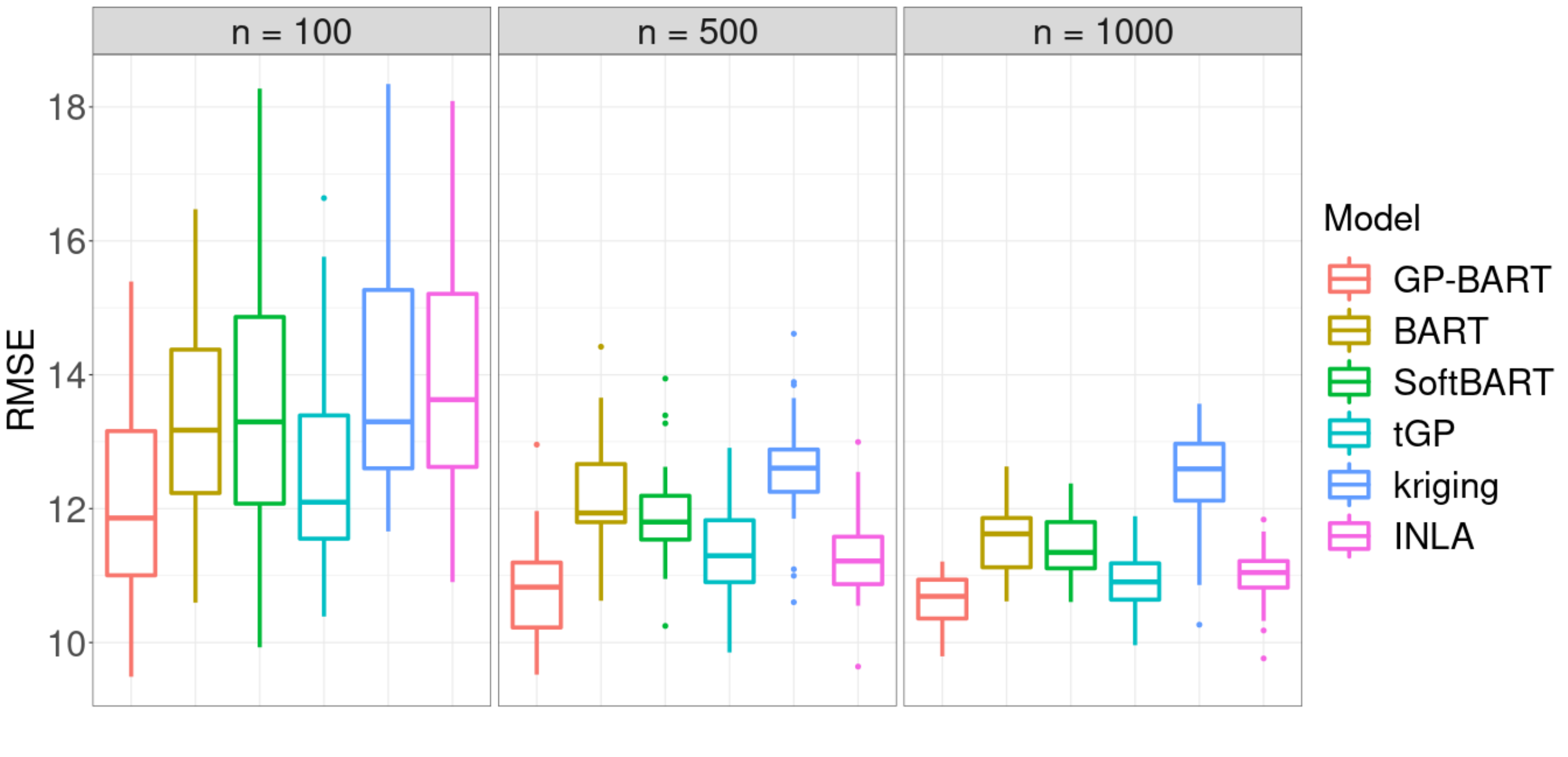}
     \caption{Comparisons between the RMSE obtained by the competing models for the simulated data using $5$ repeated $5$-fold cross validation over different sample sizes, and $\tau=0.01$. Based on the results, it is evident that GP-BART consistently delivers the best performance on average, as it encompasses assumptions of spatial dependence, smoothness, and allows for rotated splits.}
     \label{fig:sim_rmse_boxplot_tau_zero_zero_one}
 \end{figure}

\begin{figure}[H]
     \centering
     \includegraphics[width=\textwidth]{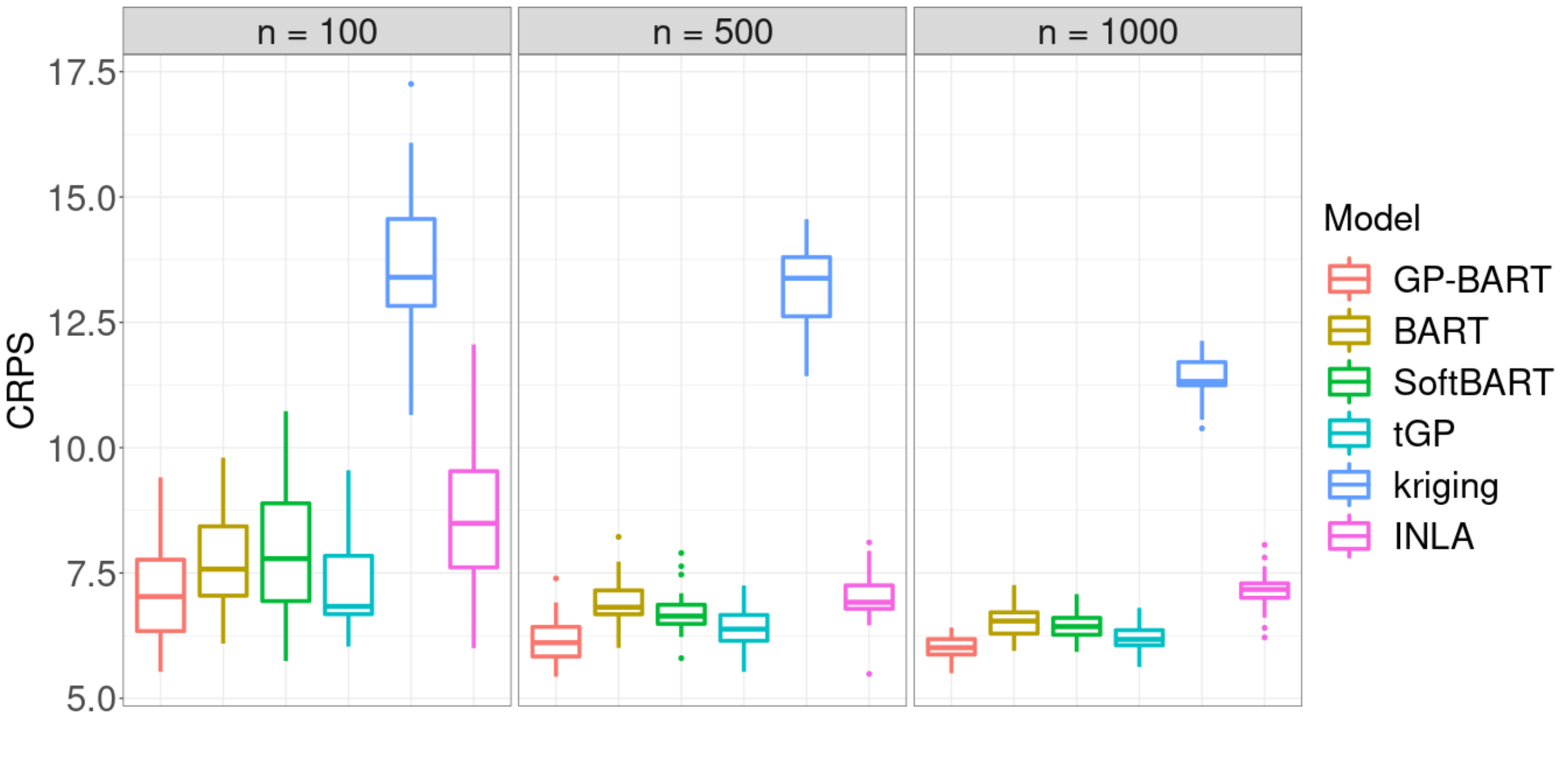}
     \caption{Comparisons between the CRPS values obtained by the competing models for the simulated data using $5$ repeated $5$-fold cross validation over different sample sizes, and $\tau=0.01$. Based on the results, it is evident that GP-BART consistently delivers the best performance on average, as it encompasses assumptions of spatial dependence, smoothness, and allows for rotated splits.}
     \label{fig:sim_pi_boxplot_tau_zero_zero_one}
 \end{figure}

The simulated data surfaces considering the residual precision $\tau=0.01$ for different samples sizes $n=\{100,500,1000\}$ are shown in Figure \ref{fig:simulated_data_tau_zero_zero_one}. Figure \ref{fig:sim_surface_100_tau_zero_zero_one}, Figure \ref{fig:sim_surface_500_tau_zero_zero_one}, and Figure \ref{fig:sim_surface_1000_tau_zero_zero_one} show the corresponding predicted surfaces from one randomly chosen repetition of the repeated 5-fold cross-validation for each respective sample size. As before, GP-BART's predicted surfaces more closely resemble the signal from the original data depicted in Figure \ref{fig:simulated_data_tau_zero_zero_one} in every instance, when compared with its competitors. The quantitative comparison is summarised via boxplots of RMSE and CRPS values in Figure \ref{fig:sim_rmse_boxplot_tau_zero_zero_one} and Figure \ref{fig:sim_pi_boxplot_tau_zero_zero_one}, respectively. These boxplots reflect the conclusions draw from previous plots where, in general, GP-BART presents the lowest median values for RMSE and CRPS across all scenarios.

 \vspace{-1em}\enlargethispage{\baselineskip}


\subsection{Performance evaluation for restricted versions of GP-BART}\bigskip\label{sec::appendix_c}

\renewcommand{\thefigure}{\Alph{subsection}.\arabic{figure}}
\setcounter{figure}{0}
\renewcommand{\thetable}{\Alph{subsection}.\arabic{table}}
\setcounter{table}{0}

\begin{figure}[H]
     \centering
     \includegraphics[width=\textwidth]{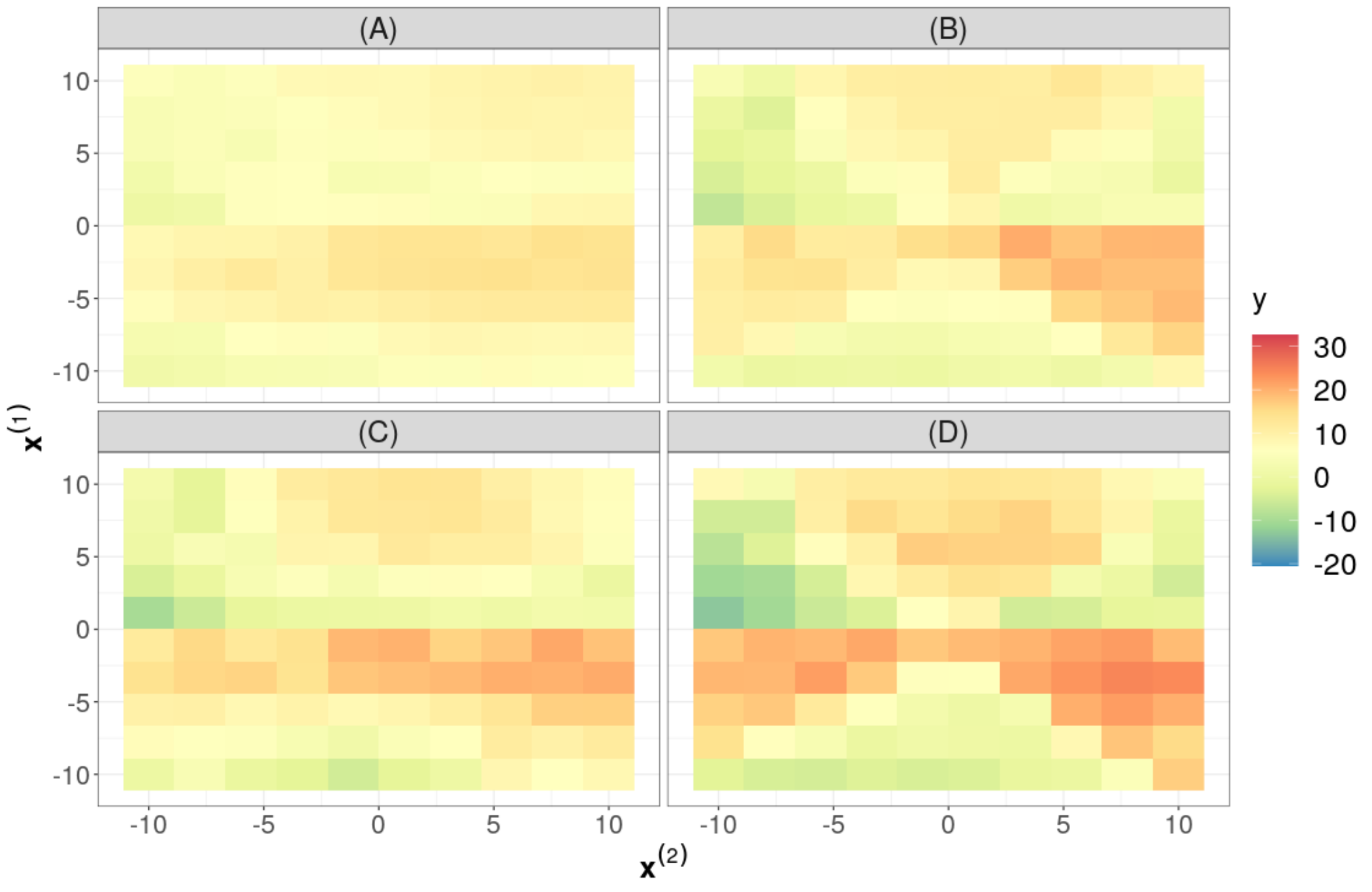}
     \caption{Comparison between the predicted surfaces under the different versions of GP-BART for the $n=100$ simulated data over one randomly chosen repetition. The surface for (\textbf{D}), the standard version of GP-BART, is qualitatively close to the observed data in the first panel of Figure \ref{fig:simulated_data}.}
     \label{fig:surface_paramter_gpbart_100}
 \end{figure}

\begin{figure}[H]
     \centering
     \includegraphics[width=\textwidth]{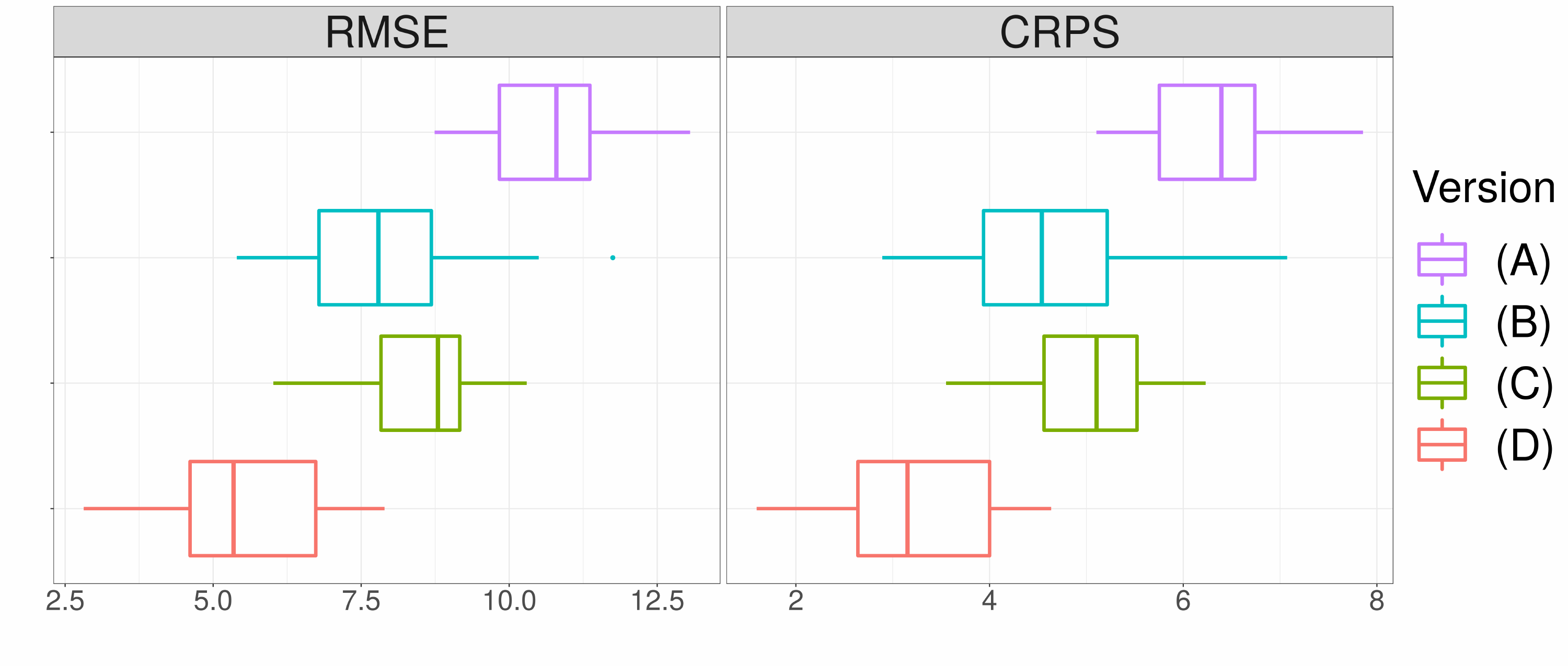}
     \caption{Boxplots of the RMSE  (left) and CRPS (right) values across the different versions of the GP-BART model for the $n=100$ simulated data. The standard GP-BART (\textbf{D}) has the best performance in terms of both RMSE and calibration.}
     \label{fig:both_rmse_compare_gpbart_paramter_100}
\end{figure}

\begin{figure}[H]
     \centering     \includegraphics[width=\textwidth]{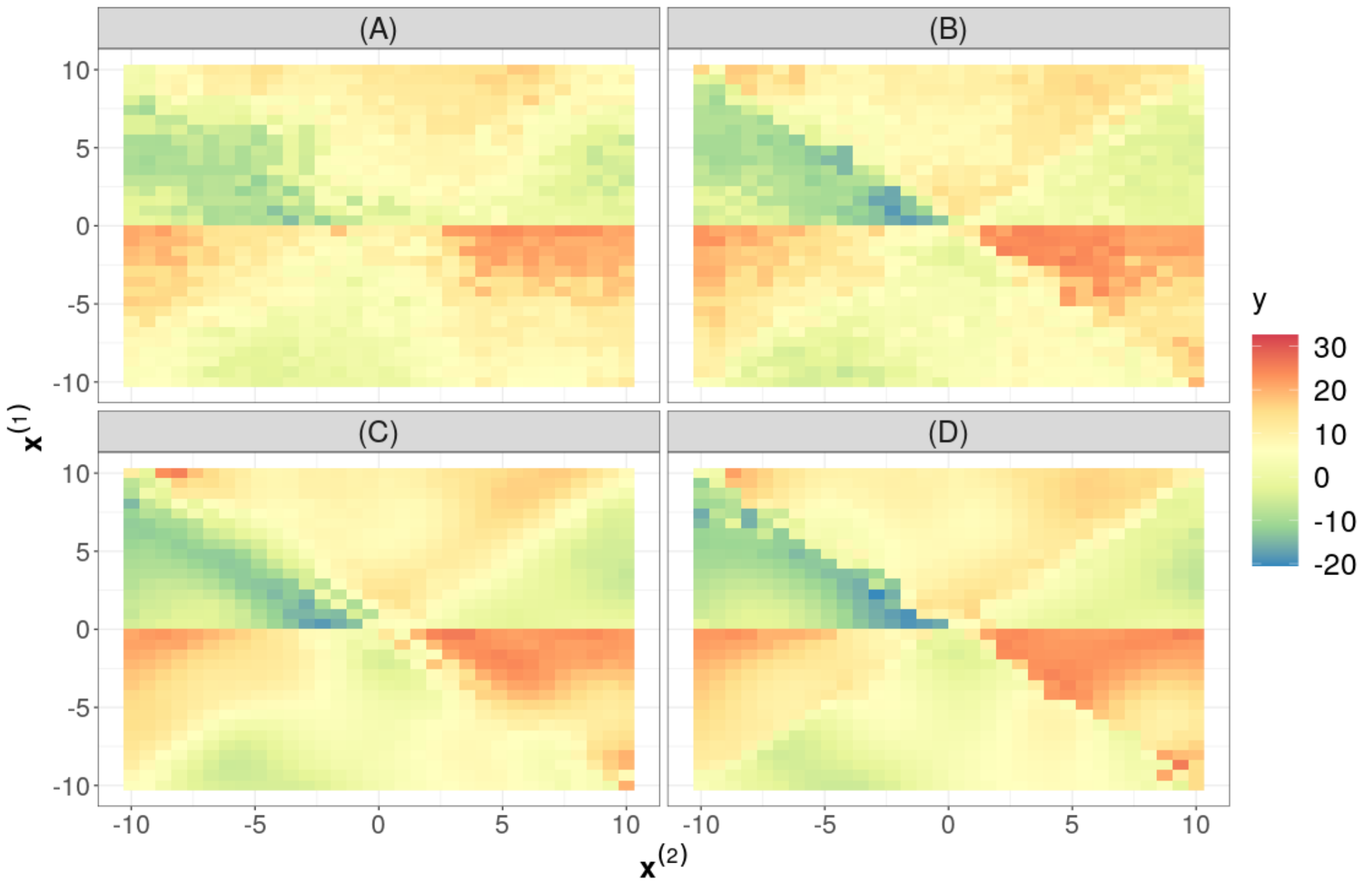}
     \caption{Comparison between the predicted surfaces under the different versions of GP-BART for the $n=1000$ simulated data over one randomly chosen test repetition. The surface for (\textbf{D}), the standard version of GP-BART, is qualitatively close to the observed data in the third panel of Figure \ref{fig:simulated_data}.}
    \label{fig:surface_paramter_gpbart_1000}
 \end{figure}

\begin{figure}[H]
     \centering
     \includegraphics[width=\textwidth]{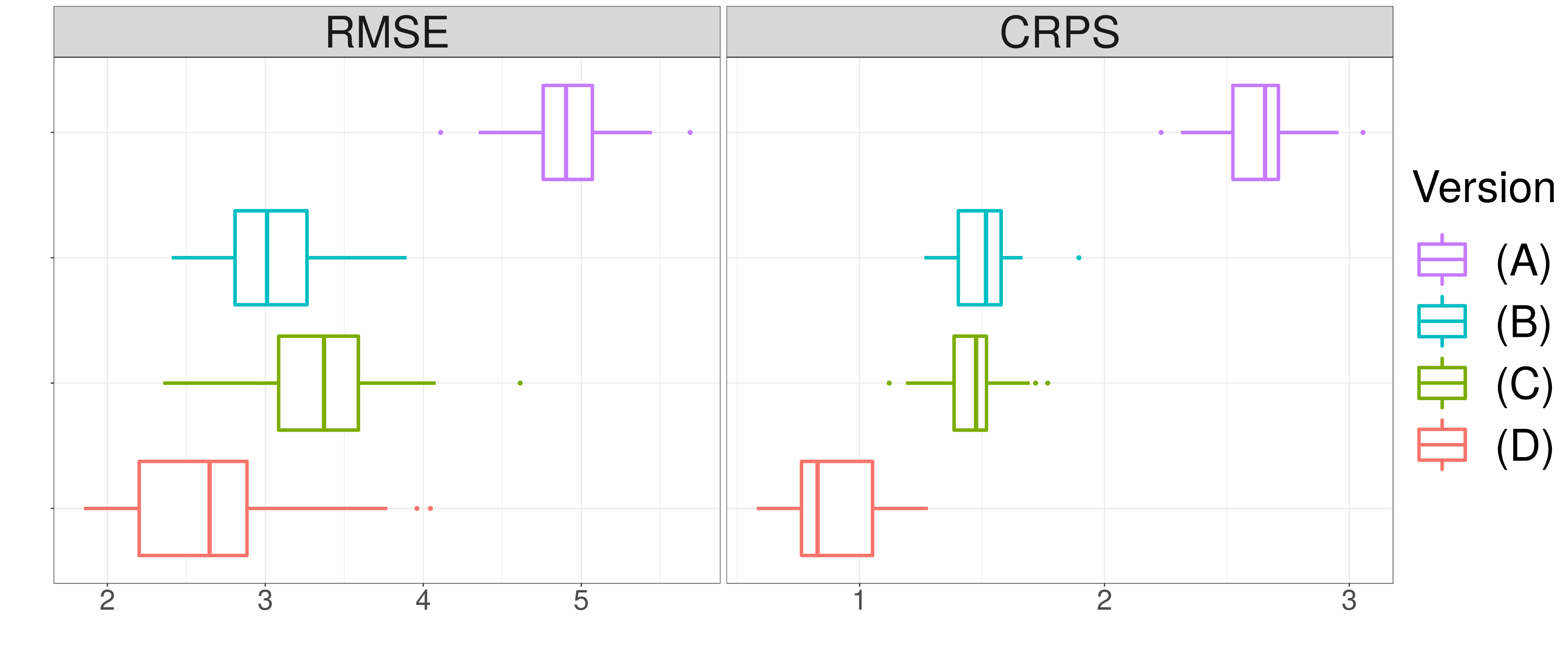}
     \caption{Boxplots of the RMSE  (left) CRPS (right) values across the different versions of the GP-BART model for the $n=1000$ simulated data. The standard GP-BART (\textbf{D}) has the best performance in terms of both RMSE and calibration.}
    \label{fig:both_rmse_compare_gpbart_paramter_1000}
 \end{figure}

The results of a comparison between different versions of GP-BART for simulated data with $n=500$ are illustrated in Figure \ref{fig:surface_paramter_gpbart_500}, showing predicted surfaces, and Figure \ref{fig:both_rmse_compare_gpbart_paramter_500}, showing boxplots of the RMSE and CRPS values. For completeness, we provide here the analogous plots for the other sample sizes considered in the simulation study, with predicted surfaces and~\mbox{boxplots} for the $n=100$ data in Figures \ref{fig:surface_paramter_gpbart_100} and \ref{fig:both_rmse_compare_gpbart_paramter_100}, respectively, and equivalent plots for the $n=1000$ data in Figures \ref{fig:surface_paramter_gpbart_1000} and \ref{fig:both_rmse_compare_gpbart_paramter_1000}. Recall that the restricted versions of GP-BART evaluated here are: (\textbf{A}) without any projection moves or GPs (equivalent to the standard BART model); (\textbf{B}) without GPs, but with the addition of the new rotation moves; (\textbf{C}) without the new moves, but with GPs; and (\textbf{D}) the standard GP-BART with both rotated split rules and GPs. Finally, numerical summaries of the median RMSE and CRPS values for all sample sizes across all four versions are summarised in Table \ref{tab:all_benchmarking_medians} and the acceptance rates for the tree-proposal moves under the full GP-BART are summarised in Table \ref{tab:acceptance_verbs}.\enlargethispage{\baselineskip}
 The behaviour of versions (\textbf{B}) and (\textbf{C}) in Figures \ref{fig:both_rmse_compare_gpbart_paramter_100} and \ref{fig:both_rmse_compare_gpbart_paramter_1000} is consistent with the corresponding Figure \ref{fig:both_rmse_compare_gpbart_paramter_500} for the $n=500$ benchmarking experiment in Section \ref{sec:benchmarking}. These versions clearly demonstrate the efficacy of the novel grow-rotate and change-rotate moves and the use of GP priors over terminal nodes, in that they show improved performance relative to the standard BART according to both metrics, but incorporating both innovations under GP-BART (\textbf{D}) yields the best performance. Regarding Figures \ref{fig:surface_paramter_gpbart_100} and \ref{fig:surface_paramter_gpbart_1000}, the predicted surface under GP-BART is the one which is closest to the observed data in each case.
To provide further clarity, Table \ref{tab:all_benchmarking_medians} numerically summarises the median lines of the boxplots from Figure \ref{fig:both_rmse_compare_gpbart_paramter_500}, Figure \ref{fig:both_rmse_compare_gpbart_paramter_100}, and Figure \ref{fig:both_rmse_compare_gpbart_paramter_1000}. All versions present lower values of both metrics as the sample size increases. While (\textbf{B}) and (\textbf{C}) improve on the standard BART (\textbf{A}) in each case, GP-BART remains the superior method from both perspectives at each value of $n$. Though the difference between it and its competitors in terms of RMSE and CRPS becomes less pronounced as $n$ increases, GP-BART remains the best from the points of view {of} prediction accuracy and uncertainty calibration. Interestingly, there is no unanimous tendency for version (\textbf{B}), which adds rotated split rules only, or version (\textbf{C}), which adds GPs only, to be second best; when jointly considering both RMSE and CRPS, (\textbf{C}) outperforms (\textbf{B}) in terms of CRPS at $n=1000$. This reaffirms that combining both innovations is necessary to achieve the best performance.
\begin{table}[H]
\centering
\caption{Summaries of the median RMSE and CRPS values over the $5$ repetitions of $5$-fold cross-validations for the $n=\{100, 500, 1000\}$ simulated data sets from the benchmarking experiments in Section \ref{sec:benchmarking}.\label{tab:all_benchmarking_medians}}
\begin{tabular}{ccccccc}
\hline
\multirow{2}{*}{Version} & \multicolumn{2}{c}{$n=100$} & \multicolumn{2}{c}{$n=500$} & \multicolumn{2}{c}{$n=1000$}\\
\cline{2-7}
& RMSE & CRPS & RMSE & CRPS & RMSE & CRPS\\
\hline
(\textbf{A}) & $10.80$ & $6.39$ & $7.00$ & $3.97$ & $4.90$ & $2.66$\\
(\textbf{B}) & $7.79$ & $4.54$ & $3.33$ & $1.83$ & $3.01$ & $1.52$\\
(\textbf{C}) & $8.80$ & $5.11$ & $4.10$ & $2.08$ & $3.37$ & $1.48$\\
(\textbf{D}) & $5.35$ & $3.15$ & $2.65$ & $1.10$ & $2.61$ & $0.83$\\
\hline
\end{tabular}
\end{table}
Finally, we present the MH acceptance rates of the newly proposed moves used for learning the tree structures under the standard GP-BART (\textbf{D}). Table \ref{tab:acceptance_verbs} shows the proportion of new trees that were accepted after the burn-in phase using each of the three available moves for each simulated data set, over all $25$ folds in total. The acceptance rates of the novel grow-rotate and change-rotate moves highlight their effectiveness.
\begin{table}[H]
\centering
\caption{Acceptance rates for the tree-proposal moves available under GP-BART for the three simulated data sets, obtained by dividing the number of times the given move~was accepted by the total number of trees across all $25$ folds in all retained posterior samples.}\label{tab:acceptance_verbs}
\small
\begin{tabular}{lccc}
 \hline
 Move    & $n=100$ &  $n=500$ & $n=1000$ \\ \hline
grow-rotate             & $0.107$   & $0.052$   & $0.037$    \\
change-rotate           & $0.218$   & $0.053$   & $0.031$    \\
prune                   & $0.109$   & $0.056$   & $0.038$    \\
\hline
\end{tabular}
\end{table}%

\subsection{Examining the effects of the hyperparameters of the tree prior.}\bigskip\label{sec::appendix_d}

\renewcommand{\thefigure}{\Alph{subsection}.\arabic{figure}}
\setcounter{figure}{0}
\renewcommand{\thetable}{\Alph{subsection}.\arabic{table}}
\setcounter{table}{0}

The choice of the tree hyperparameters $\alpha$ and $\beta$ from the tree prior in Equation \eqref{eq:nonterminal_prob} controls the depth of the trees which compose the ensemble. The default choice is $\alpha = 0.95$ and $\beta = 2$, which tends to favour shallow trees. In the GP-BART context, it would appear to be of interest to consider alternative hyperparameter specifications, in order to encourage deeper trees with fewer observations in each terminal node, given the computational complexity of $\mathcal{O}(n_{t\ell}^3)$ per node. However, we show here that doing so comes at the expense of worse predictive performance.

\begin{figure}[H]
        \centering
        \includegraphics[width=\textwidth]{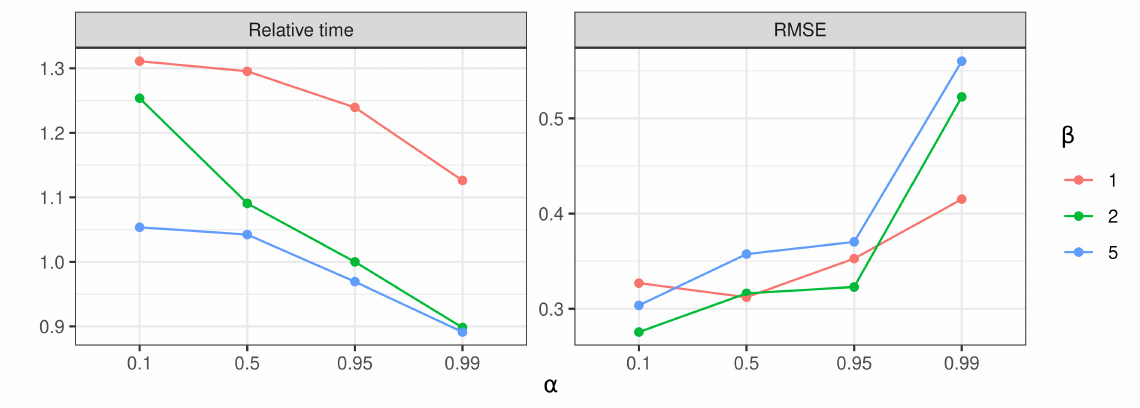}
        \caption{Performance assessment for Friedman data with noise variables and $n=500$, over a range of $\alpha$ and $\beta$ values in the tree prior, in terms of run time (relative to the default parameterisation of $\alpha=0.95$ and $\beta=2$) and RMSE.}
        \label{fig:alpha_time}
\end{figure}

To evaluate the joint effect of alternative specifications of $\alpha$ and $\beta$ on the computational cost and the accuracy of the predictions, we conducted an experiment using data generated via the Friedman equation \cite{friedman1991multivariate}; specifically, we use the same data from Section \ref{sec:friedman} with $p=10$ predictors, of which five are additional noise variables, as an example. In this case, GP-BART was trained with $n_{\text{train}} = 500$ and evaluated with $n_{\text{test}} = 500$. The tree parameters were evaluated over a discrete grid of $\alpha = \{0.1, 0.5, 0.95, 0.99\}$ and $\beta = \{1,2,5\}$. All possible combinations of these parameters were evaluated, constituting a total of 12 different scenarios. All other parameters were set to their default values. The outcomes are summarised in Figure \ref{fig:alpha_time}, in the form of relative run times and RMSE values. The run time of each setting is given relative to the time taken under the defaults of $\alpha = 0.95$ and $\beta = 2$.
From these results, it is evident that setting priors which favour more splits can reduce the computational cost of the model. As the cost of each matrix inversion is at the scale of $\mathcal{O}(n_{t\ell}^3)$, deeper trees with fewer observations in each terminal node reduces the burden of matrix inversion. However, the predictive performance diminishes due to forcing splits that should not exist. Conversely, the few settings which slightly improve the RMSE are substantially slower. Therefore, it remains sensible to adopt the default values for $\alpha$ and $\beta$ from the standard BART are as the default for GP-BART. Indeed, we do so throughout the main body of the paper and note that changing these settings to increase the speed of computations should be done with caution as it can significantly harm predictive performance.

\end{document}